\documentclass[fleqn,usenatbib]{mnras}

\usepackage{newtxtext,newtxmath}

\usepackage[T1]{fontenc}

\DeclareRobustCommand{\VAN}[3]{#2}
\let\VANthebibliography\thebibliography
\def\thebibliography{\DeclareRobustCommand{\VAN}[3]{##3}\VANthebibliography}

\usepackage{graphicx}	
\usepackage{amsmath}	

\usepackage{multicol}
\usepackage{ulem}

\usepackage{multirow}
\usepackage{booktabs}
\usepackage{hyperref}
\usepackage{xcolor}
\definecolor{medium-blue}{rgb}{0,0,1}
\hypersetup{colorlinks, urlcolor={medium-blue}}

\title[Field burial by accretion mounds]{Investigating field burial by magnetically confined accretion mounds on Neutron Stars }

\author[S. Yeole et al.]{
Saurabh Yeole,$^{1}$\thanks{E-mail: saurabh.yeole@iucaa.in}
Dipanjan Mukherjee$^{1}$\thanks{E-mail: dipanjan@iucaa.in} and
Ankush Mandal$^{2,1}$
\\
$^{1}$ Inter-University Centre for Astronomy $\&$ Astrophysics, Post Bag 4, Ganeshkhind, Pune 411007, India\\
$^{2}$ Leibniz-Institut f\"ur Astrophysik Potsdam (AIP), An der Sternwarte 16, 14482, Potsdam, Germany \\
}

\date{Accepted 2025 July 10. Received 2025 July 10; in original form 2025 April 3}

\pubyear{2025}

\usepackage{footnote}
\makesavenoteenv{tabular}
\begin{document}
\label{firstpage}
\pagerange{\pageref{firstpage}--\pageref{lastpage}}
\maketitle

\begin{abstract}
We explore the problem of magnetic confinement of accreted matter forming an accretion mound near the magnetic poles of a neutron star. We calculate the magnetic field geometry of the accreted mound by solving the magnetostatic Grad Shafranov (GS) equation in radially stretched spherical coordinates with high resolution and an extended domain. In this work, we propose a new physically motivated multipolar current free boundary condition at the outer radial boundary. We have evaluated a large suite of GS solutions for different neutron star magnetic fields and mound configurations. We find that with sufficient resolution, the ring-shaped mound profiles spread latitudinally on the neutron star surface, towards the equator, with a potential decline in dipole moment at outer radii, demonstrating the onset of field burial. A higher latitudinal spread towards the equator leads to more effective magnetic field burial. Along with the ring-shaped mound profile on a hard crust majorly used in this work, we also model mounds formed on a pre-existing ocean, which is more physically motivated. Additionally, we explore different GS solutions for a quadru-dipolar surface magnetic field. We find that such configurations lead to asymmetric polar mounds. We discuss the validity of such solutions for different relative strengths of the quadrupole and dipolar components.
\end{abstract}




\section{Introduction}

Neutron stars are known to have two distinct populations in terms of their surface magnetic fields, viz. low magnetic fields of $\sim 10^{8}-10^{10}$ G such as in milli-second pulsars, and those with higher magnetic fields ($ \gtrsim 10^{12}$ G), comprising of the bulk of known neutron star population \citep[see ][ and references therein]{konar2017magnetic}. An accretion induced reprocessing scenario is often invoked to explain both the higher spin frequencies and lower magnetic field in millisecond pulsars. Although long-term accretion driven spin up has now been generally accepted as the cause of origin of such sources \citep{smarr1976,radhakrishnan1984,Alpar1982}, the reason for low magnetic fields in such systems has not been understood yet. Several mechanisms have been proposed for this, such as surface magnetic field burial in the crust due to the accreted matter \citep{Romani1990,cumming2001magnetic,melatos2001hydromagnetic,konar2002,konar2004,payne2004burial,payne2007burial}, high crust resistivity due to accretion induced heating \citep{urpin95,urpin97,konar97,konar99a}, fluxoid-vortex migration to the crust \citep{muslimov,srinivasan1990,jahanmiri,konar99b}, crustal magneto-thermal evolution \citep{Blondin1986}, crustal tectonic motions \citep{ruderman1,ruderman2}, vortex destruction due to accretion induced currents \citep{istomin2016} and screening of core magnetic field generating currents by accretion induced currents \citep{aronslea1980}. Magnetic field burial due to accreted matter has been one of the widely studied problems explored previously by various authors without a resolution to the issue.

Matter accreted on the neutron star could also contribute to the emission of continuous gravitational waves (CGW) \citep{Melatos_2005,singh_2020,sur_2021b,Rossetto_2025}, which has served as another strong motivation for studying the feasibility of forming magnetically confined mounds\footnote{Instead of "mounds", the term "mountains" have also been used in literature to describe the magnetically confined accreted matter on neutron stars.} on neutron stars. Neutron star deformation due to magnetic stresses can also lead to the emission of CGW \citep{bonazzola1996gravitational,mastrano2015,chatterjee2021,surhaskell2021a}. The various other sources of CGW from isolated neutron stars could be from quakes due to cooling and cracking of the crust \citep{pandharipande1976,kerin2022}, changing centrifugal stress induced by stellar spindown \citep{ruderman1969,baym1969,fattoyev2018,giliberti2022}, non-axisymmetric distribution of magnetic energy trapped beneath the crust \citep{zimmermann1978,cutler2002}, pinned neutron superfluid component in the star's interior \citep{jones2010,melatos2015,haskell2022}, r-modes \citep{mytidis2015,alford2014,owen2010}, high frequency f-modes \citep{chandrashekhar1970,friedman1978,lindblom1995} or excitation of r-modes due to accretion \citep{bildsten1998}. See \citep{riles2023} for a recent review on efforts to detect CGW from rotating neutron stars.

\citet{1974bisnov,1979blandford,1986taam} were some of the first few proponents of the idea of diamagnetic screening due to accreting matter. Later, brief semi-analytical calculations were presented by \cite{hameury1983magnetohydrostatics,brown1998ocean,melatos2001hydromagnetic,konar2002}.  
One of the first detailed self-consistent solutions to demonstrate field reduction due to magnetic confinement of accreted matter was by \cite{payne2004burial}. The authors developed a numerical method to solve a dimensionless Grad Shafranov (hereafter GS) equation for solutions of the hydromagnetic equilibria of accreted matter on the neutron star. Using an isothermal equation of state (hereafter EOS\footnote{In this work, the term equation of state or EOS will be used to refer to the thermodynamic description of the accreted matter, and not the compositional description of matter interior to the neutron star crust used to solve for neutron star structures \citep{lattimer,haensel2007neutron}.}) for the accreted plasma, the authors obtained masses as large as $10^{-4}$ M$_{\odot}$. These solutions were improved by \cite{priymak2011quadrupole,priymak2014cyclotron} who tested the mounds for an adiabatic EOS and $B=10^{12.5}$ G. Similar to earlier results, they showed that the characteristic mass at which the outer radius dipole moment reaches $50 \%$ of the surface dipole moment is $5\times10^{-4}$ M$_{\odot}$ for isothermal EOS. However\textcolor
{magenta}{,} solutions with a more physically motivated Fermi gas EOS yielded much lower limits such as $10^{-7}$ M$_{\odot}$ for relativistic degenerate electron EOS and $3\times10^{-8}$ M$_{\odot}$ for non-relativistic degenerate electron EOS. A limitation of these works is that the assumed mound profile loads matter onto the magnetic field lines from the pole to the equator. In a realistic system, the accreted matter should only be loaded near the magnetic pole up to the magnetic field line attached to the inner cut-off radius of the accretion disk. 

GS solutions with mounds restricted to a polar cap radius defined by Alfv\'en radius were performed by \citet{mukherjee2012phase,mukherjee2013mhd2d,muk2017revisit}. These solutions demonstrated much lower confined masses ($\lesssim 10^{-12} M_\odot$) beyond which viable converged solutions were not obtained due to the appearance of closed field loops in the compute domain \citep[also reported in ][]{hameury1983magnetohydrostatics,payne2004burial}. \cite{muk2017revisit} provided a brief review of the work on field burial till $2017$. 

All of these works above have certain limitations (refer to the limitations in \cite{muk2017revisit}). One of the deficiencies was the boundary condition at the outer radius of the simulation domain. Previous works have considered either a free boundary condition \citep{payne2004burial,vigelius2010sinking,priymak2011quadrupole,suvorov2020multi}, a fixed boundary condition \citep{mukherjee2012phase,muk2017revisit} or a reduced dipole boundary condition \citep{rossetto23} at the outer radial boundary. However, a boundary fixed to the dipolar value as used in \citet{mukherjee2012phase} is very restrictive and is not able to probe diamagnetic screening by design. On the other hand, a free boundary as in \citet[][and later]{payne2004burial} leads to radial magnetic fields which may not be physical. The reduced dipolar boundary in \citep{rossetto23} addresses these issues to some extent, but still is limiting in nature as it suppresses other multi-poles from freely evolving. 

Another limitation of the previous papers is that all the works discussed above had calculated solutions for high magnetic field pulsars. Recently, \cite{fujisawa2022magneticallymulti} solved the Grad Shafranov equilibria in spherical coordinates (similar to \cite{payne2004burial}) that included the effect of an axisymmetric toroidal field, with an arbitrary smooth source function profile. 
The authors numerically solved the integral form of the Grad Shafranov equation (accounting for the force free nature of the magnetic field at the outer boundary) for a range of neutron star magnetic fields, and presented changes in mass-ellipticity and multipole moments of the magnetostatic mounds. However, a self-consistent evolution of the profile function duly accounting for the change in Alfv\'en radius due to varying magnetic fields was not considered.

Analysis of recent NICER observations suggests that pulsars have a complex nondipolar magnetic field profile \citep{Bilous_2019,Chen_2020,Kalapotharakos_2021,Riley_2021}. \cite{fujisawa2022magneticallymulti} calculated solutions for an initial dipolar field, an initial dipolar $+$ quadrupolar field and an initial dipolar $+$ octupolar field. Similar to the inferences of \cite{suvorov2020multi}, \cite{fujisawa2022magneticallymulti} found that for an initial dipolar field, the solutions show a buried dipole field and an increase in multipolar components. \cite{fujisawa2022magneticallymulti} also found that for an initial dipolar and strong multipole fields ($10$ times the dipolar field at the surface), the multipolar fields are buried and transformed into negative dipolar components. However, the authors considered only a single value for surface quadrupole to dipole fraction. Additionally, the authors did not account for a change in the Alfv\'en radius due to the new quadrudipolar magnetic field.

In this work, we address some of the limitations described above. We propose a new current free boundary condition for the outer radial boundary and compare the results with other traditional boundary conditions. The solutions here have been carried out with improved EOS \citep[][]{paczynski1983models,muk2017revisit} that better describes the pressure over a broader range of densities, than a polytropic EOS used in previous works \citep[e.g.][]{priymak2011quadrupole,mukherjee2012phase} for the degenerate plasma. Using the above, we have computed a suite of GS solutions for different mound properties and neutron star conditions. The section-wise break up is as follows. Section \ref{accrmnd} describes the Grad Shafranov equation form solved here, the numerical method used to solve it, the EOS used, the boundary conditions for the simulation domain and the parameter space explored here. Section \ref{SEC3} describes the form of the solutions for the ring-shaped mound profile and the effect of the boundary condition at the outer radius on the solution. We have calculated the ellipticity and dipole moments of the mounds for a range of magnetic fields (low field $10^{9}$ G to high field $10^{12}$ G pulsars) with a mound profile duly accounting for the change in the Alfv\'en radius. We have also calculated the same for $B_{d}=10^{12}$ G and an arbitrary truncation angle $50^{0}$. Section \ref{largeconfines} presents solutions and an analysis of the results for three different profile functions. Section \ref{SEC4} describes the form of the solutions for the ring-shaped mound profile for a quadru-dipolar inner boundary. The final section summarizes all the results of the paper and discusses their implications. Thus, we present high resolution simulations of accretion mounds with a larger domain, an improved multipolar current free boundary condition (CFB), new profiles and solutions for an initial quadru-dipolar field.

\section{Accretion mounds}\label{accrmnd}

\subsection{Grad Shafranov equation}
Let the magnetic field vector, plasma pressure, plasma density and gravitational potential be denoted by \textbf{B}, p, $\rho$ and $\phi_{g}$ respectively.
From force balance in a neutron star plasma, we have
\begin{equation}\label{forcebal}
    \frac{(\nabla\times\mathbf{B})\times\mathbf{B}}{4\pi} = \nabla p + \rho \nabla \phi_{g}.
\end{equation}
Assuming axisymmetry and a zero toroidal magnetic field, we have
\begin{equation}\label{magvec}
    \mathbf{B} = \frac{\nabla \psi(r,\theta)\times\hat{\phi}}{r\sin \theta},
\end{equation}
where $\psi = r\sin \theta A_{\phi}$. Here ($r,\theta,\phi$) are spherical coordinates, $\psi$ is the magnetic poloidal flux function and $A_{\phi}$ is the toroidal component of the magnetic vector potential. We assume the gravitational potential to be $\phi_{g}=gr$, where $g=GM_{\ast}/R_{\ast}^{2}$ is the constant gravitational acceleration due to the neutron star ($G$ is the gravitational constant, $M_{\ast}$ and $R_{\ast}$ are the mass and radius of the neutron star respectively). Using the EOS, we can write $\nabla p = \rho \nabla G(\rho)$. Using equation \ref{magvec}, equation \ref{forcebal} can be rewritten as
\begin{equation}\label{fb2}
    -\frac{\Delta^{2}\psi}{4\pi r^{2} \sin^{2}\theta} \nabla \psi = \rho \nabla (G(\rho) + \phi_{g})
\end{equation}
where
\begin{equation}\label{delsq}
    \Delta^{2} = \frac{\partial^{2}}{\partial r^{2}} + \frac{\sin \theta}{r^{2}} \frac{\partial}{\partial \theta}\left(\frac{1}{\sin \theta} \frac{\partial}{\partial \theta}\right).
\end{equation}
Defining $r_{0}$ as a function of $\psi$ such that
\begin{equation}\label{grho}
    \rho g \nabla r_{0}(\psi) = \rho \nabla (G(\rho) + \phi_{g}).
\end{equation}
Rearranging, we get
\begin{equation*}
    \rho g \nabla (r_{0}(\psi)-r) = \rho \nabla G(\rho)
\end{equation*}
Integrating the above equation, we find
\begin{equation}\label{inv}
    g (r_{0}(\psi)-r) = G(\rho) + C
\end{equation}
We assume $\rho = 0$ at $r=r_{0}(\psi)$ and calculate $C$ accordingly.
$r_{0}(\psi)$ is analytically assigned to be a function of $\psi$ such that it represents the radial extent of the mound for each flux surface. $\rho$ is calculated by inverting equation \ref{inv}. Substituting equation \ref{grho} in equation \ref{fb2}, we get
\begin{equation}
    -\frac{\Delta^{2}\psi}{4\pi r^{2} \sin^{2}\theta} \nabla \psi = \rho g \nabla r_{0}(\psi) = \rho g \frac{dr_{0}}{d\psi} \nabla \psi.
\end{equation}
The Grad Shafranov (hereafter GS) equation for zero toroidal field in spherical coordinates (r,$\theta$) is \citep{payne2004burial,muk2017revisit}
\begin{equation}\label{GS_sph}
    \Delta^{2}\psi = K(\psi,r,\theta),
\end{equation}
where
\begin{equation}\label{source_func}
    K(\psi,r,\theta) = -4\pi r^{2}\sin^{2}\theta \rho g \frac{dr_{0}(\psi)}{d\psi},
\end{equation}
 $r_{0}(\psi)$ is a profile function which defines the shape of flux surfaces and $\Delta^{2}$ is given by Eqn. \ref{delsq}. To better resolve the strong gradients at the base of the mound and in the latitudinal direction, it is convenient to recast the GS equation in terms of new variables: 
\begin{align}
\left(r,\theta\right) &\rightarrow \left(y=\mbox{log}\left(\frac{r - aR_{\ast}}{R_{\ast}(1-a)}\right), \mu = \cos\theta\right). 
\end{align}
The resulting GS equation is
\begin{align}\label{GSeq}
    \frac{\partial^{2} \psi}{\partial y^{2}} - \frac{\partial \psi}{\partial y} &+ \frac{(1-\mu^{2})}{\left(1 + \frac{a}{e^{y}(1-a)}\right)^{2}} \frac{\partial^{2} \psi}{\partial \mu^{2}} \\
    &= R_{\ast}^{2}(1-a)^{2}e^{2y} K(\psi,y,\mu). \nonumber
\end{align}
Here, $R_{\ast}$ is the radius of the neutron star in km and a is a parameter that defines the stretching of the grid. We use a$=0.999$ for all simulations in this work (Appendix \ref{app1}). 

\subsection{Numerical Method}
We solve the Grad-Shafranov equation through an iterative scheme using the successive over-relaxation method (SOR). The solver has been updated to be compatible with MPI (Message Passing Interface), using a red-black parallelization scheme \citep{PresTeukVettFlan92}. The details of the numerical scheme and the comparison with the previous results of \cite{muk2017revisit} are presented in Appendix \ref{app1}. We assume a solid crust with a fixed magnetic field, which is dipolar for the results presented in Section \ref{SEC3} and a mix of dipole and quadrupolar fields in Section \ref{SEC4}. MPI and updated coordinates allow us to present an extended simulation domain and larger grid resolution such as $5000^{2}$ relative to previous works.

\subsection{Equation Of State (EOS)}

In the outer envelope of the neutron star, matter has a composition of degenerate electron gas and strongly coulomb coupled ions. The pressure is dominated by the electron gas \citep{haensel2007neutron}. Plasma is degenerate for $\rho > 10^{3}$ g cm$^{-3}$ for typical HMXB hotspot temperatures of $2\times10^{7}$ K \citep{Coburn_2002}. Thus, Fermi gas EOS for zero temperature gives a fair estimation of the pressure in the mound. Fermi gas is relativistic beyond a density range of $10^{6} - 10^{7}$ g cm$^{-3}$. The densities at the base of the accretion mound may reach values of $\rho \gtrsim 10^{8} - 10^{9}$ g cm$^{-3}$. Thus, to cover the wide density range in such mounds, a simple polytropic approximation for the EOS is insufficient. Hence, we use an empirical function that very closely approximates the pressure of a Fermi gas to an accuracy $1.8 \%$ \citep{paczynski1983models,muk2017revisit}. The Paczynski EOS and a detailed comparison with the often used polytropic asymptotic forms of the Fermi EOS are presented in \cite{muk2017revisit}. The Paczynski EOS is formulated as 
\begin{equation}
    p = \frac{\pi}{3}\frac{m_{e}^{4} c^{5}}{h^{3}} \frac{(8/5)x_{F}^{5}}{\sqrt{1+(16/25)x_{F}^{2}}}
\end{equation}
where
\begin{equation*}
    x_{F} = \frac{1}{m_{e}c} \left(\frac{3h^{3}}{8\pi \mu_{e} m_{p}}\right)^{1/3} \rho^{1/3}.
\end{equation*}
$G(\rho)$ used in equation \ref{grho} for the Paczynski EOS is
\begin{equation}
    G = \frac{m_{e}c^{2}}{8\mu_{e}m_{p}}\left(\frac{(15/2) + (32/5)x_{F}^{2}}{\sqrt{1+(16/25)x_{F}^{2}}}\right)
\end{equation}
We have assumed pure ionized helium matter ($\mu_{e}=2.0$) for all the mound simulations unless specified otherwise.

\subsection{Boundary conditions}

\subsubsection{Inner radial boundary $R=R_{\rm in}$}
The inner radial boundary is assumed to be a fixed dipole on the hard crust of the neutron star, which keeps the mound stable from interchange modes \citep{payne2004burial,vigelius2008three}. However, in a realistic scenario, a mixture of accreted matter and neutron star matter at the inner boundary is expected to sink into the ocean and outer crust due to gravitational compression. However, such a sinking boundary condition \citep[e.g.][]{vigelius2010sinking} has not been directly implemented in this work and will be explored in the future. The impact of accretion on a pre-existing atmosphere has been partly addressed in subsection \ref{sunkmound}, although the inner boundary is assumed to be fixed even in that subsection.

\subsubsection{Outer radial boundary $R=R_{\rm out}$}
As calculated in \cite{payne2004burial}, Greens function solution for the Grad Shafranov operator is given by
\begin{align}
    \quad & \quad \psi(r^{\prime},\theta^{\prime}) = \nonumber \\ & \frac{\psi^{\ast} \sin^{2}\theta^{\prime} R_{\ast}}{r^{\prime}} \nonumber \\ & + \sum_{\ell=1}^{\infty} \frac{\sin \theta^{\prime} P_{\ell}^{1}(\cos \theta^{\prime})}{2\ell(\ell+1)}  \nonumber \\ 
    & \times \left( r^{\prime(-\ell)} \int_{0}^{\pi} \int_{R_{\ast}}^{r^{\prime}} r^{\ell+1} Q(r,\ell) P_{\ell}^{1}(\cos \theta) K(\psi,r,\theta) dr d\theta \right. \nonumber \\ 
    & \left. \quad + \; r^{\prime(\ell+1)} Q(r^{\prime},\ell) \int_{0}^{\pi} \int_{r^{\prime}}^{\infty} r^{-\ell} P_{\ell}^{1}(\cos \theta) K(\psi,r,\theta) dr d\theta \right).
\end{align}
where
\begin{equation*}
    Q(r,\ell) = \left(\left(\frac{R_{\ast}}{r}\right)^{2\ell+1}-1\right)
\end{equation*}
Here $\psi^{\ast} = \frac{1}{2}B_{d}R_{\ast}^{2}$, where $B_{d}$ is the dipolar magnetic field at the poles, and $K(\psi,r,\theta)$ is the source function from equation \ref{source_func}.

Using the Greens function formalism, we update the outer radial boundary value ($R_{\rm out}$) at each iterative SOR step by utilizing the value of $\psi$ at a radius $R_{\rm in}$ just above the maximum height $r_{c}$ of the mound. $\psi$ at the boundary is calculated from
\begin{align} \label{FFeq}
    \psi(R_{\rm out},\theta^{\prime}) = \sum_{\ell=1}^{\ell_{\rm max}} & \left( \frac{(2\ell+1) P_{\ell}^{1}(\cos \theta^{\prime})\sin \theta^{\prime}}{2\ell(\ell+1)} \right. \nonumber \\ & \left. \times \left(\frac{R_{\rm in}}{R_{\rm out}}\right)^{\ell} (1 + (-1)^{\ell+1})\right. \nonumber \\
    & \left. \times \int_{0}^{1} \frac{\psi(R_{\rm in},\cos\theta) P_{\ell}^{1}(\cos \theta)}{\sqrt{1-\cos^{2}\theta}} d(\cos\theta) \right). 
\end{align}
Derivation of this equation is presented in subsection \ref{derivn}, Appendix \ref{app2}. $\ell_{\rm max}$ has been chosen to be $33$ for all simulations by several tests performed in subsection \ref{lmx}, Appendix \ref{app2}. $R_{\rm in}$ needs to be selected at a point in vacuum above the diamagnetic screening currents \citep{payne2004burial}. The above boundary condition has been validated by a comparison with a greens function based boundary condition involving volume terms described in subsection \ref{greensvol}, Appendix \ref{app2}.

\subsubsection{Latitudinal boundaries}
At the magnetic pole (i.e. $\theta=0^{0}$) it is assumed that $\psi = 0$. At the magnetic equator (i.e. $\theta=90^{0}$), we assume a symmetric solution given by $\frac{\partial \psi}{\partial \cos \theta} = 0$.
  
\subsection{Parameter space and Profile function} \label{param}

\begin{table}
\caption{Name of the fixed properties and its values for the calculations performed are noted in this table.}
\label{constpar}
\begin{center}
\begin{tabular}{|c|c|}
\hline
Name & Value \\ 
 \hline\hline
Mass of the Neutron Star ($M_{\ast}$) & $1.4$ M$_{\odot}$ \\
 \hline
Radius Of the Neutron Star ($R_{\ast}$) & $10$ km\\
  \hline
Mass accretion rate ($\Dot{M}$) & $1.94 \times 10^{-8}$ M$_{\odot}$ yr$^{-1}$ \\
  \hline
Outer radius of the simulation domain ($R_{\rm out}$) & $22$ km  \\
 \hline
$\ell_{\rm max}$ (Appendix \ref{app2}) & $33$  \\
 \hline 
\end{tabular}
\end{center}
\end{table}

We assume the values of the properties noted in table \ref{constpar} to be constant for all simulations. A near Eddington accretion rate ($L_{Edd} = \eta\Dot{M}c^{2}$) where $\eta=0.16$ is assumed to find some of the largest stable mound masses possible. The angular extent of the simulation is $\theta \in [0^{0},90^{0}]$ for a dipolar inner boundary while it is $\theta \in [0^{0},180^{0}]$ for a quadru-dipolar case. Each GS solution is further characterized by the following parameters.
\begin{enumerate}
    \item Surface Magnetic field strength $B_{d}$
    \item Maximum Height of the mound $r_{c}$
    \item $\zeta$ = ratio of truncation radius ($r_{t}$) to Alfv\'en radius ($R_{A}$)
    \item Truncation angle $\theta_{t}(\zeta,B_{d},R_{\ast},\Dot{M},M_{\ast})$, as defined in eq.~\ref{TTeq}.
    \item Form of the profile function $r_{0}(\psi,R_{\ast},r_{c},\theta_{t})$ (shape of the density profile, e.g. eq.~\ref{HWeq})
    
\end{enumerate}

 The profile function ($r_{0}(\psi)$) defines the shape of the mound and the radial distribution of matter in the mound. Previously, various forms of this function have been explored to solve the GS equation \citep{hameury1983magnetohydrostatics,brown1998ocean,mukherjee2012phase,mukherjee2013mhd2d,muk2017revisit}. In all such cases, the outer angular edge of the mound corresponds to a truncation angle of $\theta_{t}$, which has been determined by the inner edge of the accretion disk. In this work, we have primarily used the ring-shaped mound profile or the hollow mound profile \citep{mukherjee2012phase,muk2017revisit} along with some other profiles. The ring-shaped mound profile is  
\begin{alignat}{2}\label{HWeq}
    r_{0}(\psi) = R_{\ast} + \frac{r_{c}}{0.25}\left(0.25-\left(\frac{\psi}{\psi_{a}(\theta_{t})}-0.5\right)^{2}\right) \quad \quad \quad 0.0\leq\psi\leq\psi^{\ast} 
\end{alignat}
$r_{0}(\psi)-R_{\ast}$ is negative for $\psi>\psi_a$. Thus $\rho$ which is a function of $r_{0}(\psi)-r$ is defined to be zero for $\psi>\psi_a$. For $\psi>\psi_a$, though $dr_{0}/d\psi$ has some finite value, RHS term or source term $K(\psi,r,\theta)$ is zero due to zero $\rho$. Though an accretion profile depends on complex disk magnetosphere interaction where the matter penetrates through the magnetosphere (based on effective diffusivity) in non-axisymmetric 3D instabilities such as Rayleigh Taylor and Kelvin Helmholtz instabilities \citep{aronslea1976a,aronslea1976b}, such considerations require a complicated and non axisymmetric analysis which is outside the scope of this work. 
 
Ring-shaped mound profile requires two input parameters $-$ maximum height ($r_{c}$) of the mound and the truncation angle ($\theta_{t}$). The truncation angle ($\theta_{t}$) is calculated from the accretion disk truncation radius ($r_{t}$) as
\begin{equation}\label{TTeq}
    \theta_{t} = \sin^{-1}\left(\sqrt{\frac{R_{\ast}}{r_{t}}}\right).
\end{equation}
We assume the truncation radius to be fraction of the classical Afv\'en radius ($R_{A}$) as $r_{t} = \zeta R_{A}$, where the Alfv\'en radius is given by \citep{Psaltis_1999,Mukherjee15a,muk2017revisit}  
\begin{align}\label{RAeq}
    R_{A} = 3.53 \times 10^{3} \mbox{ km} & \left(\frac{B_{d}}{2.0 \times 10^{12} \mbox{ G}}\right)^{\frac{4}{7}} \left(\frac{R_{\ast}}{10 \mbox{ km}}\right)^{\frac{12}{7}} \nonumber \\ & \left(\frac{\Dot{M}}{10^{-9} \mbox{ M}_{\odot} \mbox{yr}^{-1}}\right)^{-\frac{2}{7}} \left(\frac{M_{\ast}}{1.4 \mbox{ M}_{\odot}}\right)^{-\frac{1}{7}}. 
\end{align}
As $R_{\ast}$, $M_{\ast}$ and $\Dot{M}$ are fixed for all models, in this work $\theta_{t}$ primarily depends on $B_{d}$ and $\zeta$. We have performed simulations for different surface magnetic field strengths $B_{d} \in (10^{9},10^{10},10^{11},10^{12})$ G. 
MHD simulations generally find the truncation radius to vary as $r_{t} = 0.5-1.0 R_{A}$ \citep{Long_2005,Romanova_2008,bessolaz2008,zanni08,kulkarni2013,parfrey,parfrey17b}. For each $B_{d}$, we solve the GS equations for five different values of $\zeta$ viz. $\zeta \in ( 0.6, 0.7, 0.8, 0.9, 1.0)$, with $\zeta = 1$ corresponding to $r_{t} = R_{A}$ and hence $\theta_{t} = \theta_{A}$. Furthermore, for a given choice of $\zeta$ and $B_{d}$, various simulations have been performed for different mound heights ($r_{c}$) to scan a wide range of parameter space. In subsection \ref{3.2}, \ref{1cosh}, last part of \ref{3.4}, \ref{sunkmound} and Appendix \ref{app3}, $\theta_{t}$ is arbitrarily chosen to explore hypothetical maximal mass loading, and is not calculated from Equation \ref{TTeq}. In such cases, it has been explicitly specified that an arbitrary $\theta_{t}$ has been selected. The total mass of the mounds near both poles ($M$) is computed by integrating the density over all volume in the domain for all runs, with higher $r_{c}$, yielding larger mound masses.

\section{Dipolar inner boundary for ring-shaped mound profile}\label{SEC3}

This section presents results for a dipolar boundary at the neutron star surface and a multipolar current free boundary condition (CFB) at the outer boundary along with a comparison of CFB to other boundary conditions. The results of GS solutions for a typical magnetically confined mound with a CFB are presented in \ref{3.1}. In subsection \ref{3.2}, the new boundary condition and the effect of the boundary condition at the outer radial boundary have been discussed.  

\subsection{General description of GS solutions of mounds}\label{3.1}
\begin{figure}
\vbox{
\includegraphics[width=\columnwidth]{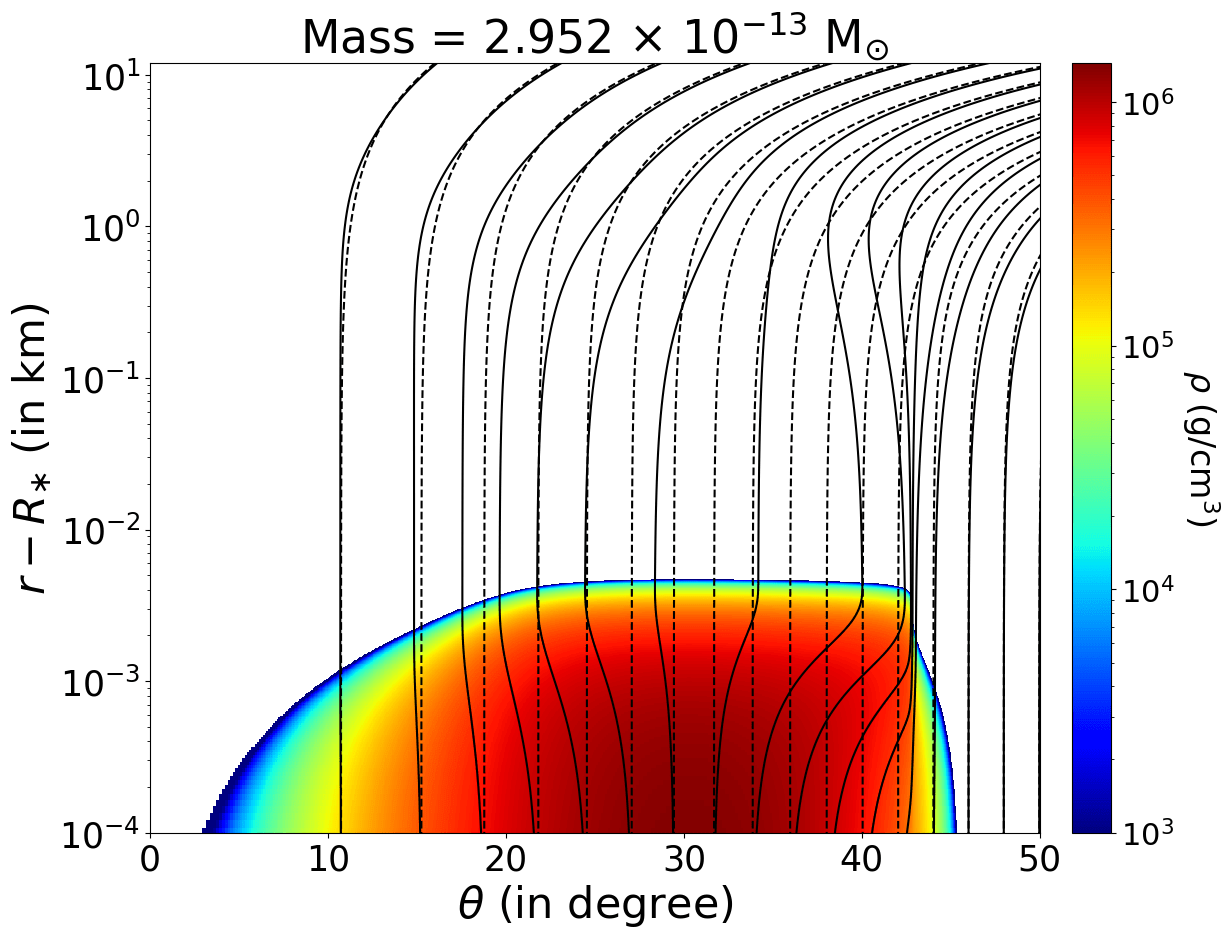}
\includegraphics[width=\columnwidth]{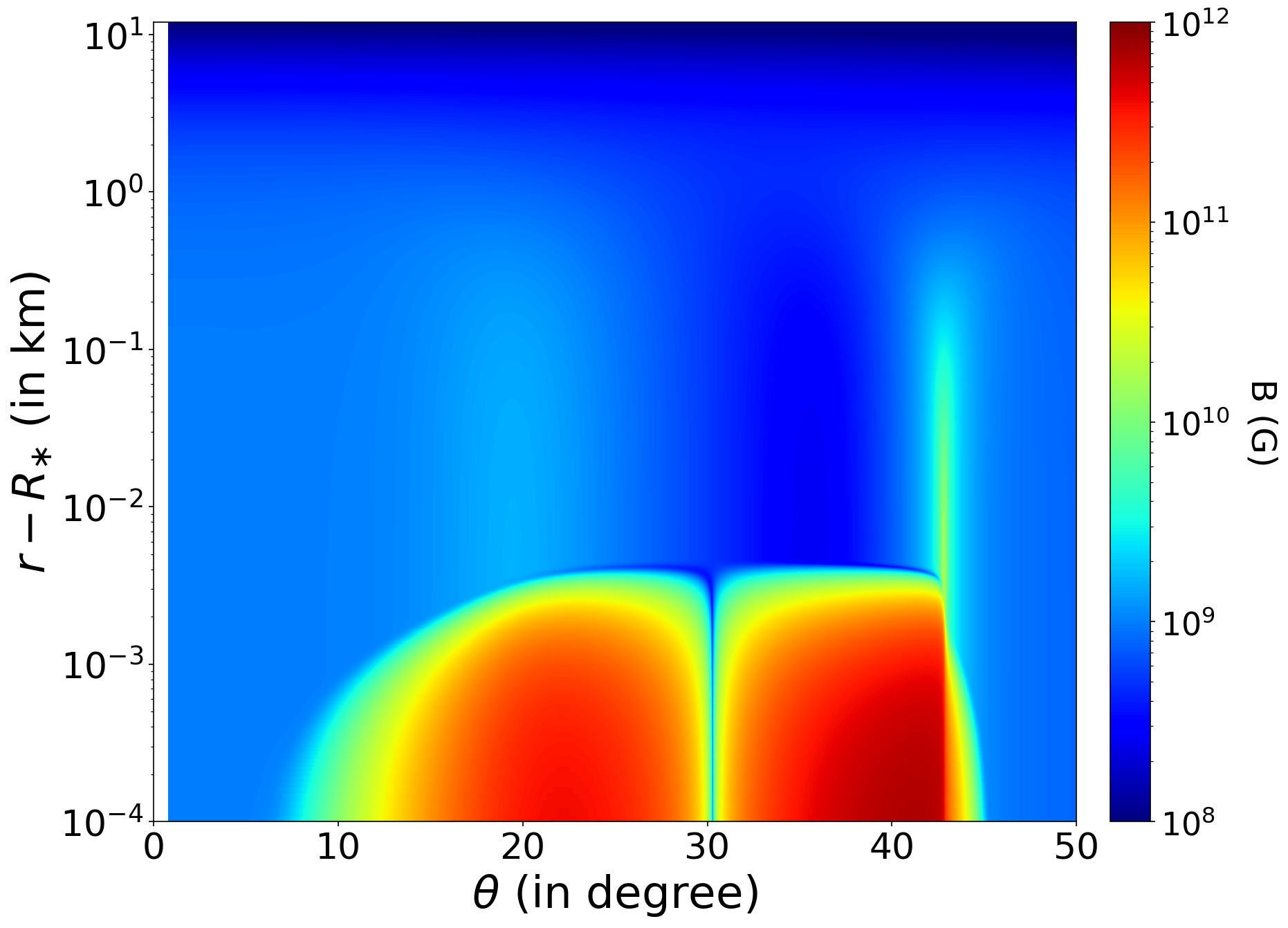}
\includegraphics[width=\columnwidth]{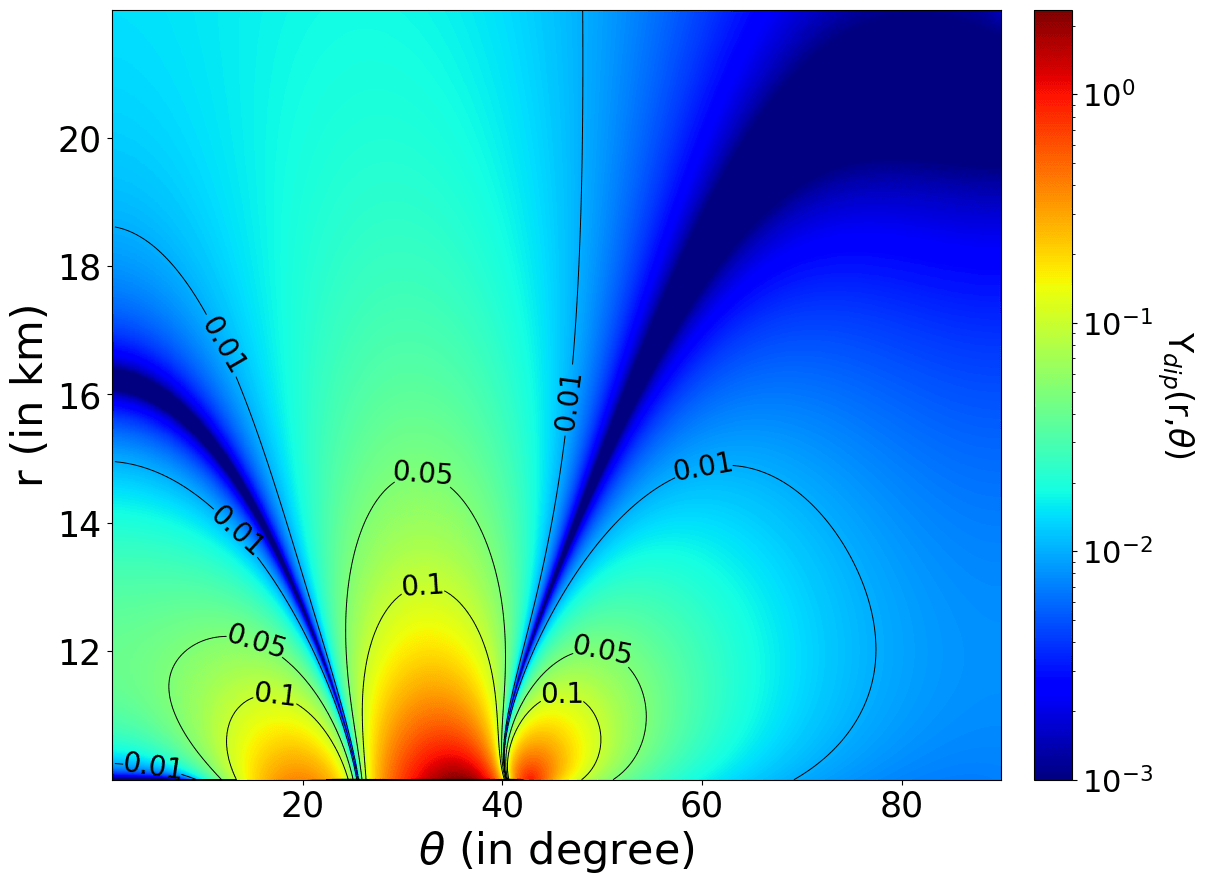}
}
\caption{Plot of Density Profile and Magnetic field lines (solid) for parameters $B_{d} = 10^{9}$ G, $\theta_{t} = 45.5^{0}$, $r_{c} =$ $ 4.7$ m, $M$ $= 2.952 \times 10^{-13}$ M$_{\odot}$ (top), colormap of magnetic field magnitude B (in Gauss) (middle) and colormap of $Y_{\mbox{dip}}$ (Eqn \ref{ydip}) (bottom). Dashed lines in the top figure are dipolar magnetic field lines. The accretion mound changes the magnetic field at the surface of the neutron star to a range of $10^{8} - 10^{11}$ Gauss. Bottom plot has an extended $\theta$ range to show the deviation from the dipolar field in the whole simulation domain. There is a significant deviation of $10\%$ $-$ $5\%$ from a dipolar solution till a height of $3$ $-$ $5$ km above the neutron star surface due to an accretion mound.}
\label{Eg}
\end{figure}

To demonstrate the form of the final solutions using CFB, the results of a $5000\times5000$ simulation of an accretion mound with parameter values of $B_{d} = 10^{9}$ G, $\theta_{t} = \theta_{A} = 45.5^{0}$, $r_{c} =$ $ 4.7$ m, $M$ $= 2.952 \times 10^{-13}$ M$_{\odot}$ are plotted in Figure \ref{Eg}. The plot at the top of Figure \ref{Eg} is a density profile of the mound where the solid lines represent the magnetic field lines while the dashed lines are the dipolar field lines. The magnetic field lines are contours of $\psi$ (placed equidistant in $\psi$) and thus in all the plots of the paper, the tangent to the lines represents the direction of the magnetic field but the density of the magnetic field lines is not equal to the magnetic field magnitude.

The ring-shaped mound profile pushes the magnetic field lines outwards. The magnetic field geometry is similar to the solutions presented before in spherical \citep{muk2017revisit} and cylindrical coordinates \citep{mukherjee2012phase}. 
Inside the mound, $B$ increases to $\sim 10^{11}$ G, two orders of magnitude larger than the starting magnitude of the dipole magnetic field at the base of the mound. Above the mound in vacuum, $B$ increases near the two edges of the mound as the mound tends to stretch in both directions. The value of $B$ is reduced right above the mound to accommodate the stretching. To quantify the deviation from the initial dipolar magnetic field, we present in the lower panel of Figure \ref{Eg} the colormap of $Y_{\mbox{dip}}$, defined as
\begin{equation}\label{ydip}
    Y_{\mbox{dip}}(r,\theta) = \frac{|B(r,\theta) - B_{\mbox{dip}}(r,\theta)|}{B(r,\theta)},
\end{equation}
where $B$ is the magnitude of the magnetic field vector. For this example, we can see that $B$ differs from the dipolar solution by $10\% - 5\%$ for a height of $3 - 5$ km above the neutron star surface. Maximum height above the neutron star surface till which $Y_{\mbox{dip}}(r,\theta)$ has values $0.1$ and $0.05$ have been noted down in Table \ref{Bmag} for different magnetic fields and truncation angle $\theta_{t}=\theta_{A}$ for a given maximum mass allowed by our solutions (Appendix \ref{numapp}). Low magnetic field pulsars have a larger deviation from the dipolar field than high magnetic field pulsars due to relatively higher $\theta_{t}$ values of the former's solutions. For the given range of magnetic fields, $B$ differs from the dipolar field by $5\%$ to heights ranging from $0.7-9.5$ km. This has significant implications for expected radiation from accretion columns and its polarization \citep{Becker_2007,caiazzo_2021}. 

\begin{table}
\caption{The table lists results of GS solutions with different magnetic fields (1st column) and truncation angle (2nd column). The maximum height ($h_{\rm max}=r-R_{\ast}$) where GS solution deviates by 5\% and 10\% from a dipolar field (defined by $Y_{\mbox{dip}}(r,\theta)$ in (Equation \ref{ydip}) are presented in the 3rd and 4th columns respectively. This demonstrates the vertical extent of the non-dipolar distortion for different neutron star and mound parameters. 
}
\label{Bmag}
\begin{center}
\begin{tabular}{|c|c|c|c|}
\hline
 $B_{d}$ (G) & $\theta_{t}$ & $h_{\rm max}$ (km) & $h_{\rm max}$ (km) \\ 
   &  & $\left(Y_{\mbox{dip}}=0.05\right)$ & $\left(Y_{\mbox{dip}}=0.1\right)$ \\
 \hline\hline
$10^{9}$ & $45.5^{0}$ & $9.5$ & $5.2$ \\
 \hline
$10^{10}$ & $21.7^{0}$ & $2.5$ & $1.7$ \\
  \hline
$10^{11}$ & $11.0^{0}$ & $1.14$ & $0.82$ \\
  \hline  
$10^{12}$ & $5.7^{0}$ & $0.7$ & $0.53$ \\
 \hline\hline
\end{tabular}
\end{center}
\end{table}

\subsection{Effects on solution due to outer radial boundary}\label{3.2}

\subsubsection{Current free boundary condition (CFB) v/s Fixed boundary condition}\label{3.2.1}
\begin{figure}
\vbox{
\includegraphics[width=\columnwidth]{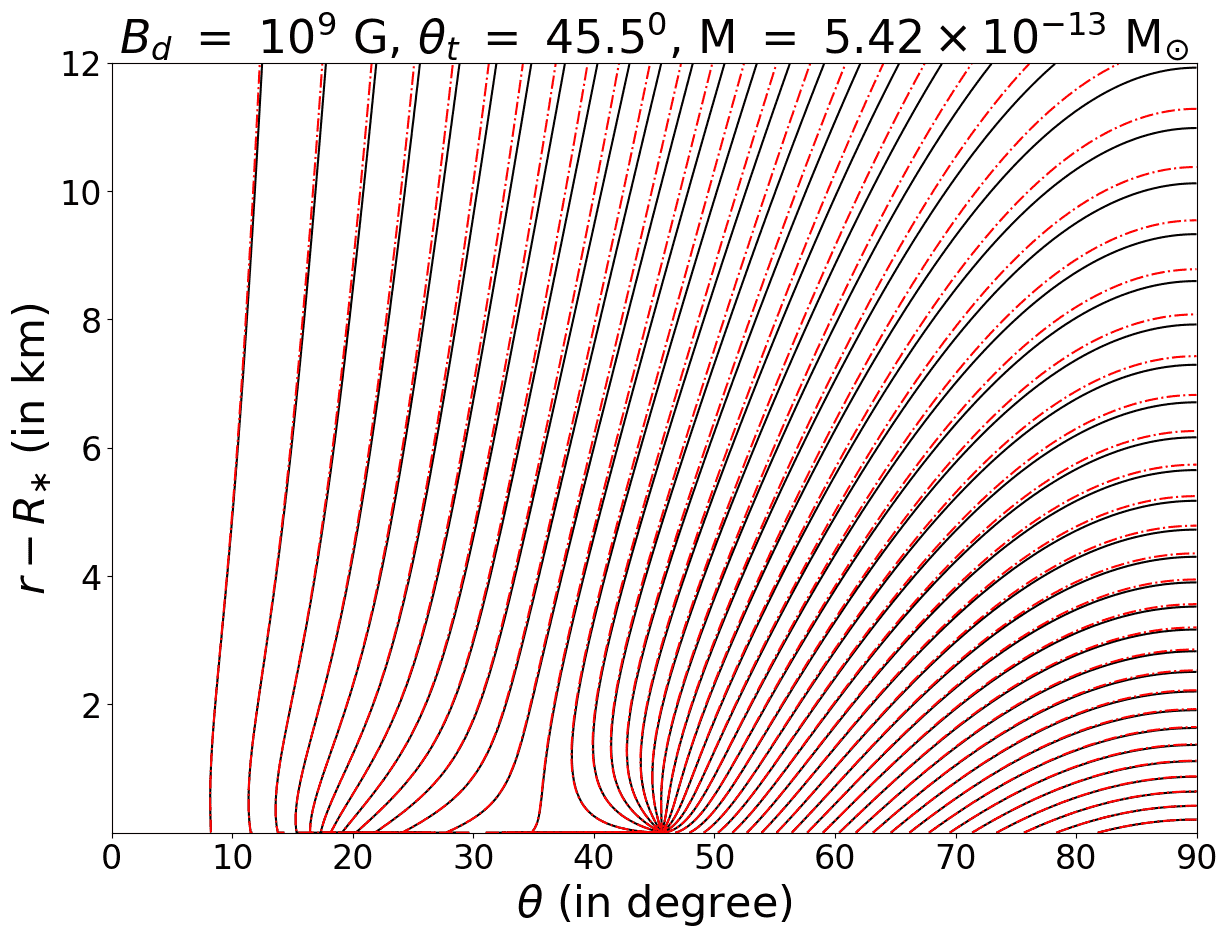}
\includegraphics[width=\columnwidth]{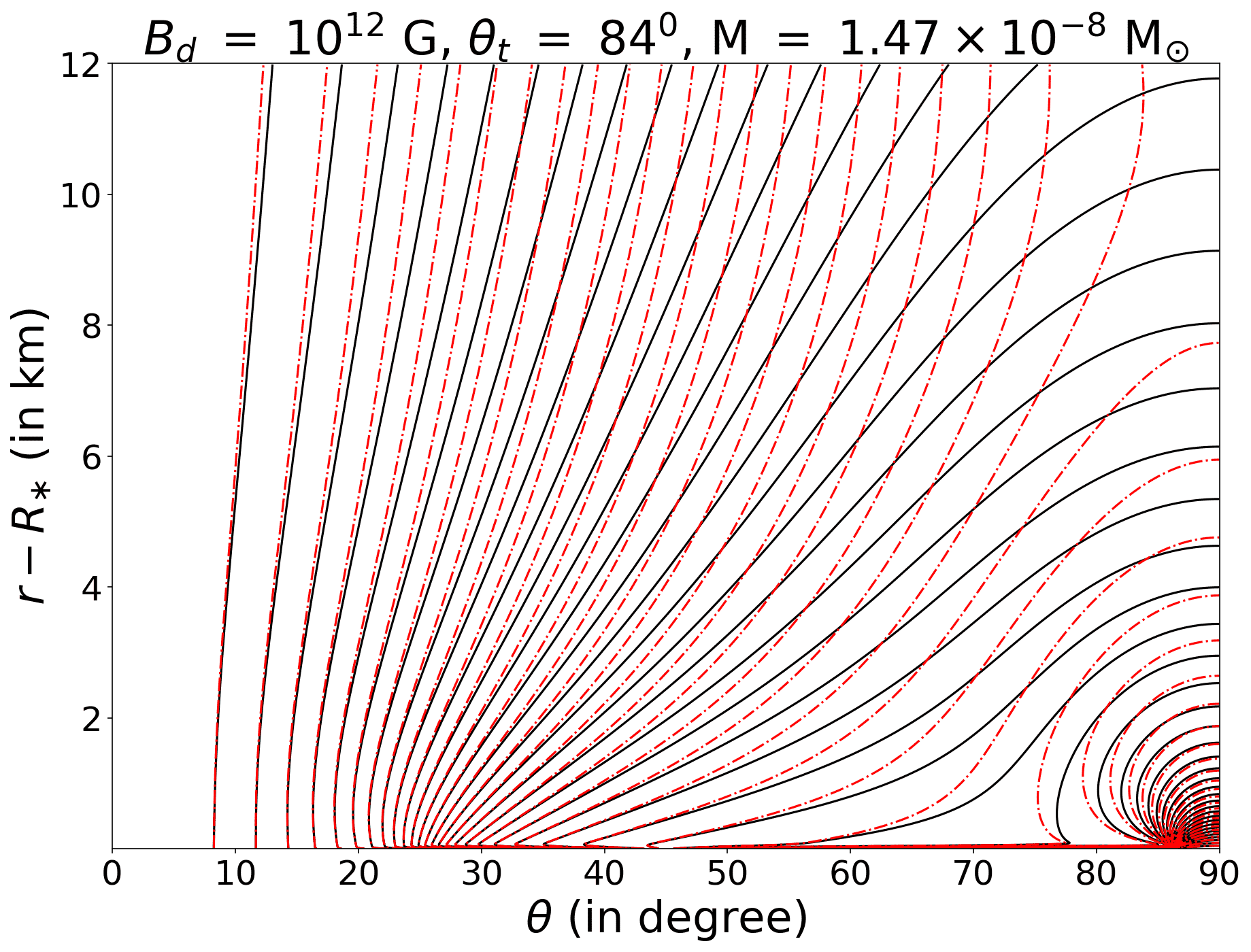}
}
\caption{Plots of magnetic field lines for fixed dipole condition (red dashdot lines) and CFB (black solid lines) at the outer radial boundary for Case 1 : $B_{d} = 10^{9}$ G, $\theta_{t} = 45.5^{0}$, $r_{c} =$ $ 5.5$ m, $M$ $= 5.42 \times 10^{-13}$ M$_{\odot}$ (top) and Case 2 : $B_{d} = 10^{12}$ G, $\theta_{t} = 84.0^{0}$, $r_{c} =$ $ 103.1$ m, $M$ $= 1.47 \times 10^{-8}$ M$_{\odot}$ (bottom). Magnetosphere gets compressed as dipole moment is lowered due to the accretion mound. Solutions for a larger mass (due to a high $B_{d}$ and high $\theta_{t}$) shows a larger deviation between the CFB and fixed dipole boundary condition.}
\label{fixedvsff}
\end{figure}

This subsection aims to highlight the change in the final solution due to the deployment of CFB at the outer radial boundary instead of a fixed dipole boundary. The dipolar boundary is fixed to the dipolar $\psi$ value for $B_{d}$ at the radius $R_{\rm out}$ i.e. $\psi^{\ast} \sin^{2}\theta (R_{\ast}/R_{\rm out})$. Figure \ref{fixedvsff} represents the difference between a solution with CFB (solid lines) and a solution with fixed boundary condition (red dot dashed lines) at the outer radial boundary ($5000\times5000$ simulation). We have plotted the solutions in Figure \ref{fixedvsff} for two cases :
\begin{itemize}
    \item Case $1$ (Figure \ref{fixedvsff}, Top) : \textit{Physically motivated} $\theta_{t}$ $=$ $\theta_{A}$ ; $B_{d} = 10^{9}$ G, $\theta_{t} = 45.5^{0}$, $r_{c} =$ $5.5$ m, $M$ $= 5.42 \times 10^{-13}$ M$_{\odot}$
    \item Case $2$ (Figure \ref{fixedvsff}, Bottom) : \textit{Arbitrarily large} $\theta_{t}$ \textit{for high} $B_{d}$ \textit{to set masses similar to solutions by}  \cite{priymak2011quadrupole} ; $B_{d} = 10^{12}$ G, $\theta_{t} = 84.0^{0}$, $r_{c} =$ $ 103.1$ m, $M$ $= 1.47 \times 10^{-8}$ M$_{\odot}$.  
\end{itemize}

From Figure \ref{fixedvsff}, we can conclude that CFB compresses or reduces the size of the magnetosphere and thus shows a reduction in the dipole moment at the outer radii. The change from a fixed boundary is greater for mounds with larger mass, as apparent in the lower part of Figure \ref{fixedvsff}. The fixed boundary condition restricts the movement of the magnetic field lines. The confined mass for the solution of Case 2 is qualitatively similar to the results of \cite{priymak2011quadrupole}, obtained using an outflow boundary condition. However, such a boundary results in radial magnetic fields, similar to a monopole and is not physically motivated. Thus the newly proposed current-free boundary yields results similar to maximal burial obtained earlier, with better physically motivated boundary constraints. In more recent works, bootstrapping methods have been applied to estimate the reduced dipole moment and calculate the outer boundary accordingly \citep{rossetto23}. However, such a boundary restricts the large scale fields to be dipolar in nature, preventing free evolution due to deformations at the base. The CFB gives magnetic field lines more freedom to move while also maintaining the value of multipole moments in vacuum.     

\subsubsection{Effect of outer radius of simulation domain}\label{rout_effect}
\begin{figure}
\vbox{
\includegraphics[width=\columnwidth]{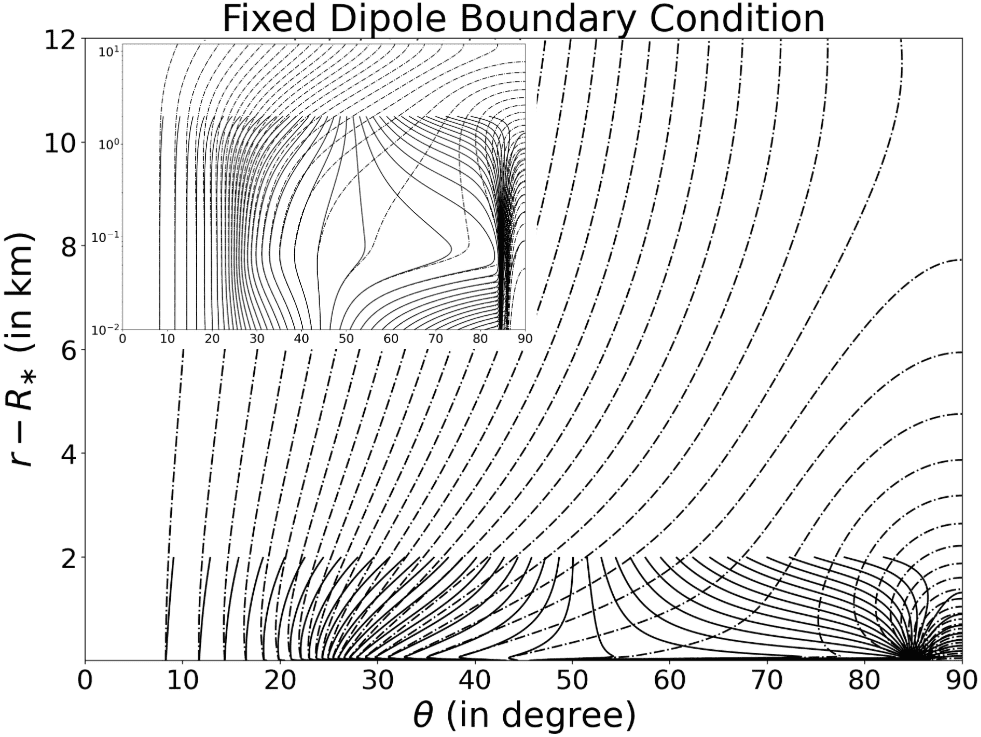}
\includegraphics[width=\columnwidth]{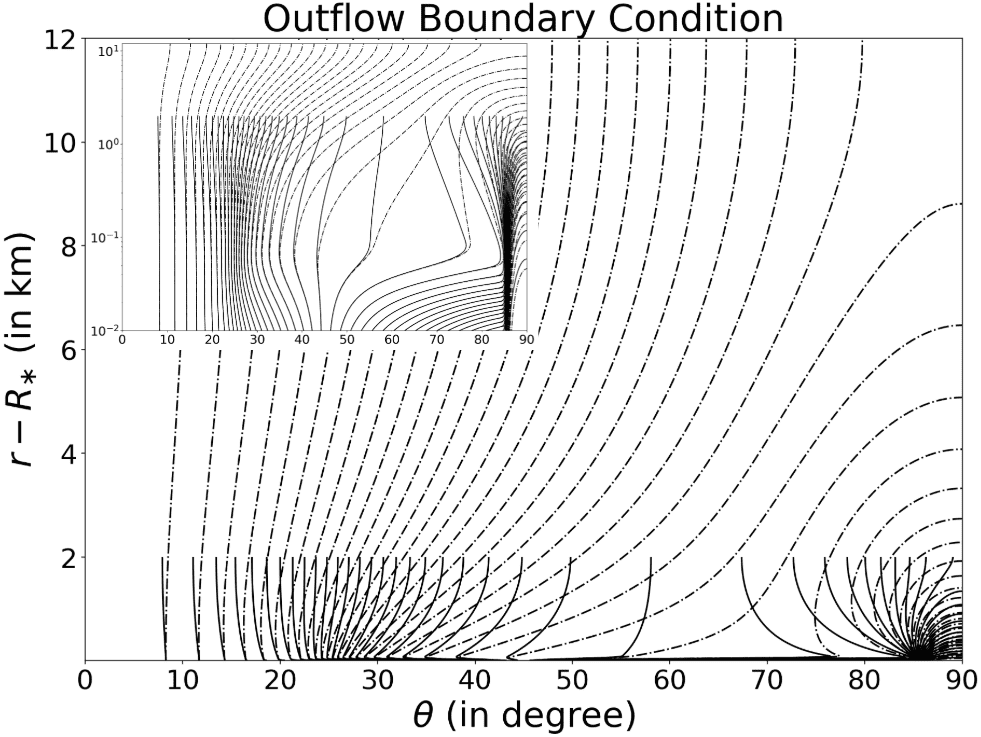}
\includegraphics[width=\columnwidth]{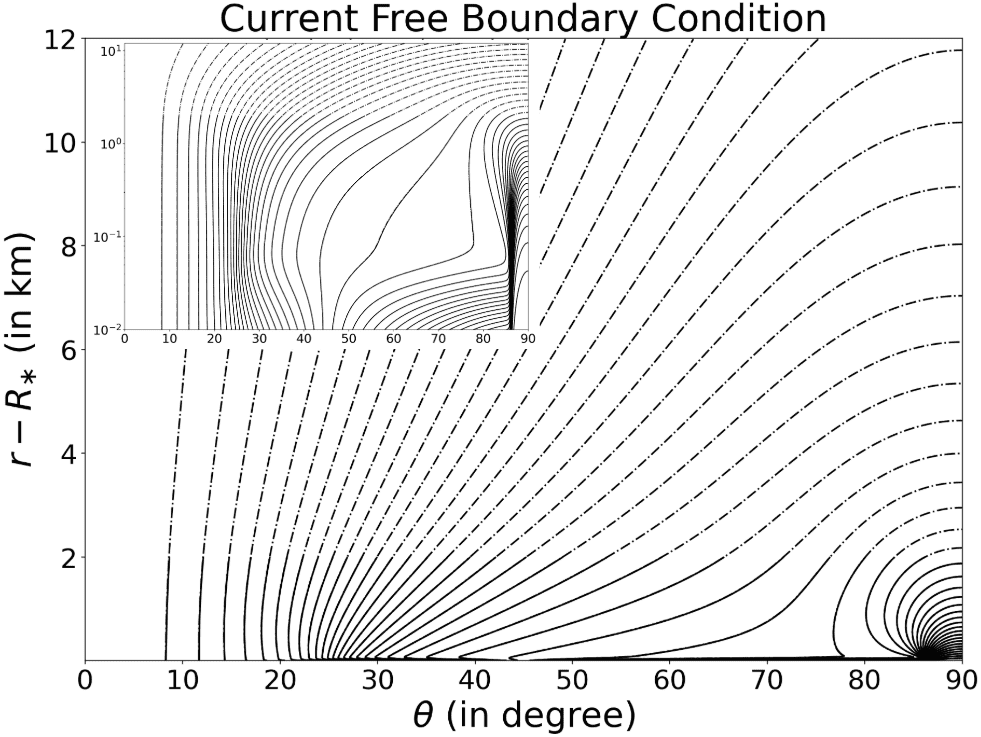}
}
\caption{Plots of magnetic field lines for a small radial domain $2$ km (solid lines) and a large radial domain $12$ km (dashdot lines) with a Fixed Dipole Boundary Condition (top), Outflow Boundary Condition (middle) and CFB (bottom) at the outer radial boundary for Case $2$ in subsection \ref{3.2.1}. The inset plots show the same plots in logarithmic scale. Magnetic field line solutions for CFB seems to be independent of the choice of outer radius of the domain unlike the solutions for outflow boundary condition and fixed dipole boundary condition.}
\label{smallrad}
\end{figure}

This subsection aims to highlight the effect of changing the value of the outer boundary of the solution domain on different boundary conditions. The accretion mound solution for Case $2$ from the previous subsection \ref{3.2.1} has been used to make these comparisons, as it denotes solutions with some of the largest mound masses and hence significant deviations from a dipolar magnetic field. In Fig.~\ref{smallrad} we present solutions with two different choices of the outer radius of the compute domain for the GS solver: a) $R_{\rm out} - R_{\ast} = 2$ km (solid lines) and b) $R_{\rm out} - R_{\ast} = 12$ km (dash dot lines). The panels denote solutions with three different boundary conditions for the outer boundary: a) fixed boundary in the upper panel, b) an outflow condition ($\frac{\partial \psi}{\partial y} = 0$) or a free boundary in the middle panel, and c) the lower panel shows a solution with the new CFB. All solution domains with two different outer boundaries have the same grid structure up to the $R_{\rm out} - R_{\ast} = 2$ km. 

For both the fixed and the outflow boundaries, the solutions for the smaller compute domains differ significantly from those with larger $R_{\rm out}$. This implies that the nature of the solutions depend on the choice of extent of the compute domain as larger deformations of field lines are supported in extended domains. Unlike the fixed and the outflow boundaries, solutions with a multipolar boundary for the two $R_{\rm out}$ are seen to overlap seamlessly. The change in the solutions with a multipolar boundary due the outer boundary are less than $0.01 \%$, unlike the solutions with fixed and outflow boundaries. However, to achieve the desired accuracy, solutions with smaller $R_{\rm out}$ require a larger number of multipole moments (Equation \ref{FFeq}). This inference has been established with a test in Appendix \ref{app-b1.3}.

Thus, a CFB can be used to find the solution with sufficient accuracy for a smaller outer radial domain, allowing for higher resolution with limited computational resources.  

\begin{figure}
\vbox{
\includegraphics[width=\columnwidth]{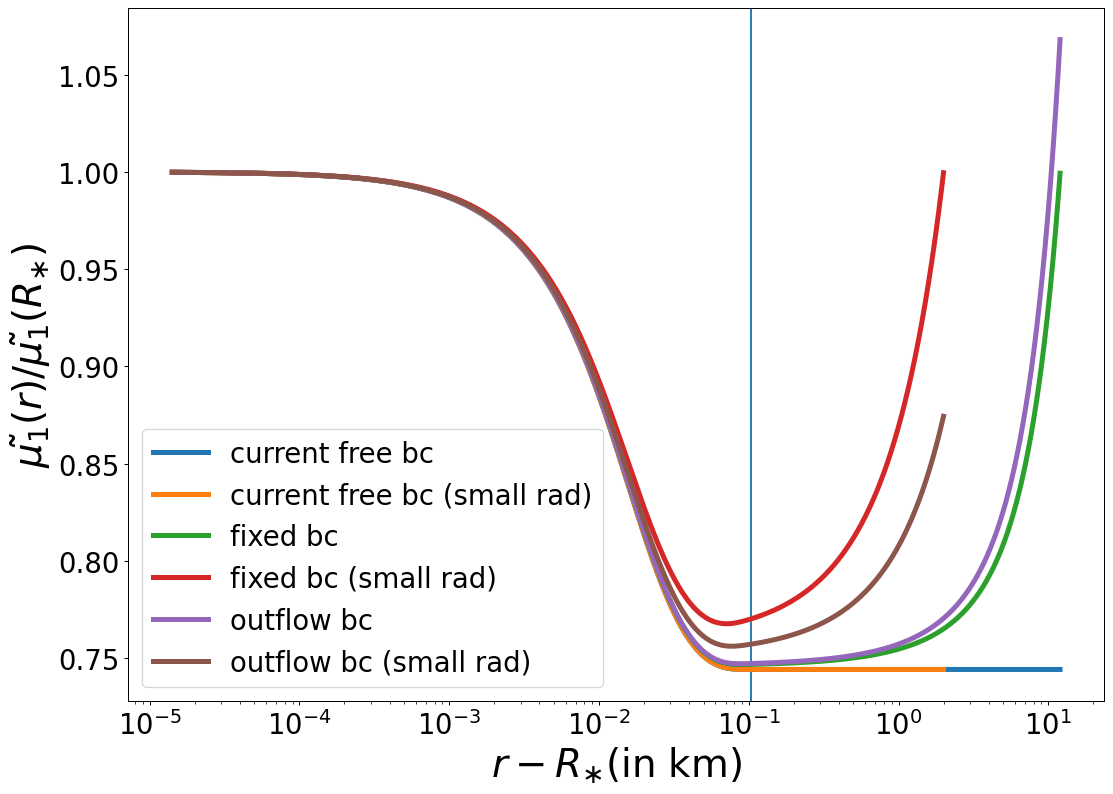}
}
\caption{Normalized Dipole Moment at a radius above the neutron star relative to the normalized dipole moment at the surface of the neutron star (Equation \ref{MUeq}) plotted for solutions with three boundary conditions and two different simulation domains i.e for six cases. Vertical line is the maximum height $r_{c}$ of the mound. Dipole moment for the CFB condition solutions are constant in vacuum irrespective of the extent of the radial domain unlike the dipole moments for the other two boundary conditions.}
\label{smallraddip}
\end{figure}

 The normalized $\ell-$th magnetic multipole moment of the magnetosphere can defined as \citep[][ and references therein]{suvorov2020multi,fujisawa2022magneticallymulti}
\begin{equation}\label{MUeq}
    \Tilde{\mu_{\ell}}(r) = r \int_{0}^{\pi} \psi(r,\theta) P_{\ell}^{1}(\cos \theta) d\theta
\end{equation}
The normalized dipole moment at any radius relative to the normalized dipole moment at the surface of the neutron star (Equation \ref{MUeq}, $\ell$ $=$ 1) has been plotted for the above solutions in Figure \ref{smallraddip} as a function of height above the surface of the neutron star. The vertical line in Figure \ref{smallraddip} is the $r_{c}$ (maximum height) of the mound. CFB solution for the smaller radial domain (orange curve) overlaps the CFB solution for the larger radial domain (blue curve) unlike the dipole moment ratios for the fixed boundary and outflow boundary condition. Dipole moment ratios of the larger domain solutions for the fixed (green curve) and outflow (purple) boundary conditions resemble the dipole moment ratios of the CFB solutions upto a particular height above $r_{c}$. However, the dipole moment ratios of the smaller domain solutions for the fixed (red curve) and outflow (brown) boundary conditions deviate from the CFB dipole moment ratios even below $r_{c}$. This shows that as the radial extent of the compute domain increases, the dipole moment ratio of a different boundary condition solution tends towards the dipole moment ratios of a CFB solution. This validates the chosen CFB condition, since as the radial extent of the compute domain increases, the effect of the outer radial boundary condition on the solution reduces.

The ratio of normalized dipole moments for a fixed boundary approach $1.0$ at the outer radius as expected, but the dipole moment ratios for the outflow boundary condition behave in a manner different from the solutions found by \cite{priymak2011quadrupole} and \cite{suvorov2020multi}. This could be due to the choice of the coordinates, as we assign an outflow boundary condition for our logarithmic radial coordinate while the above mentioned authors assign outflow boundary condition for the radial coordinate. 

\section{Solutions for large confined masses}\label{largeconfines}
One of the primary science goals in this domain is to understand how much mass can be contained by a mound for a given surface magnetic field ($B_{d}$) and angular extent ($\theta_{t}$). Large mound masses are expected to result in stronger field reduction at the outer radius (field burial) and higher mass ellipticities, which are of strong interest for searches of continuous gravitational waves \citep{bonazzola1996gravitational}. In previous studies \citep[such as][]{payne2004burial,priymak2011quadrupole,mukherjee2012phase} the maximum mass was limited by the appearance of closed field lines forming magnetic bubbles inside the solution domain, resulting in non-convergence of the GS iterations. In the first subsection, we show the results for the $1/cosh$ profile used by \cite{fujisawa2022magneticallymulti}. In the final two subsections, we investigate the nature of GS solutions for a physically motivated ring-shaped mound profile for different mound heights (and hence mass) to explore the appearance of a maximum mass, if any.

\subsection{Filled mound with exponential decay beyond $\psi_{a}$}\label{1cosh}
\begin{figure}
\vbox{
\includegraphics[width=\columnwidth]{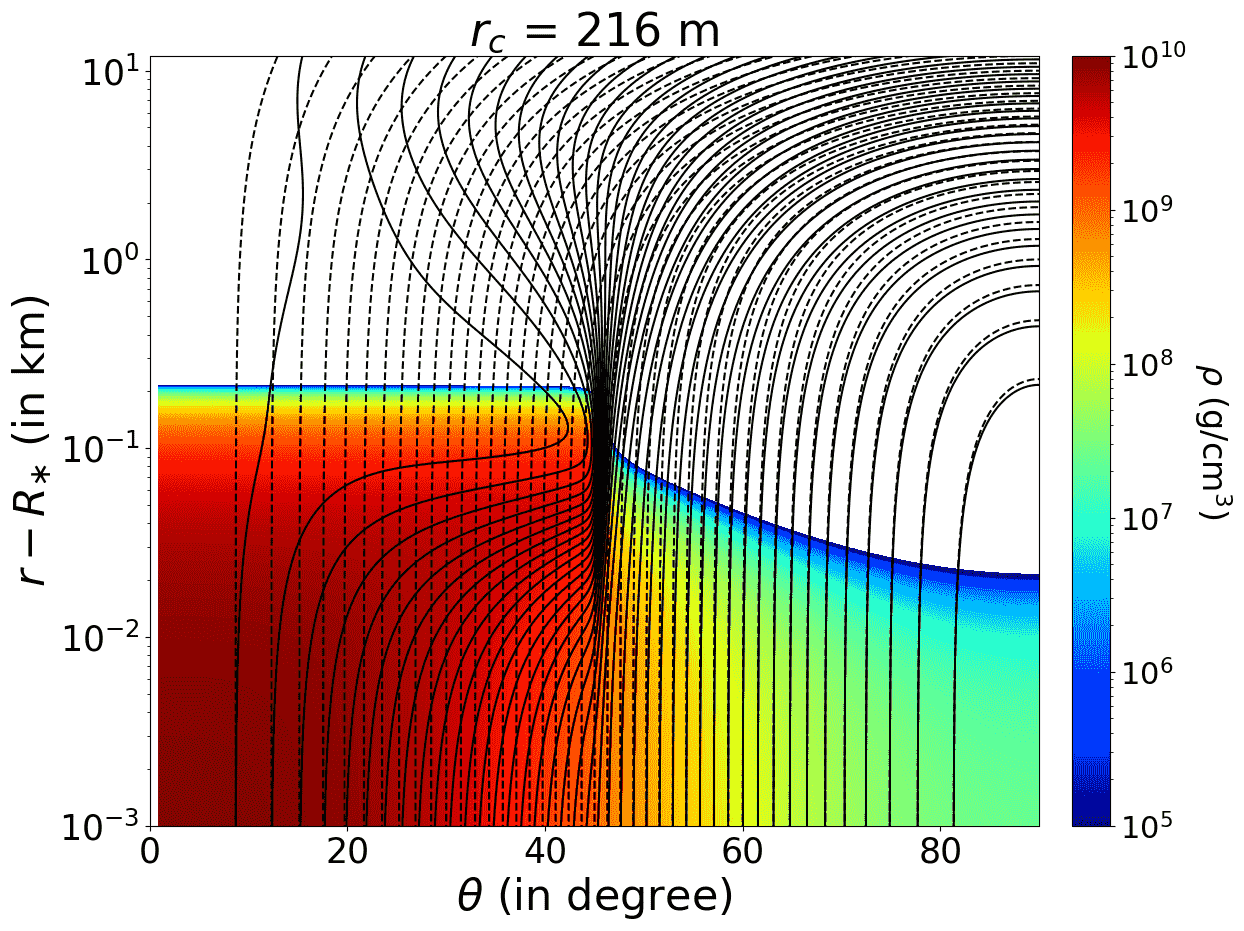}
\includegraphics[width=\columnwidth]{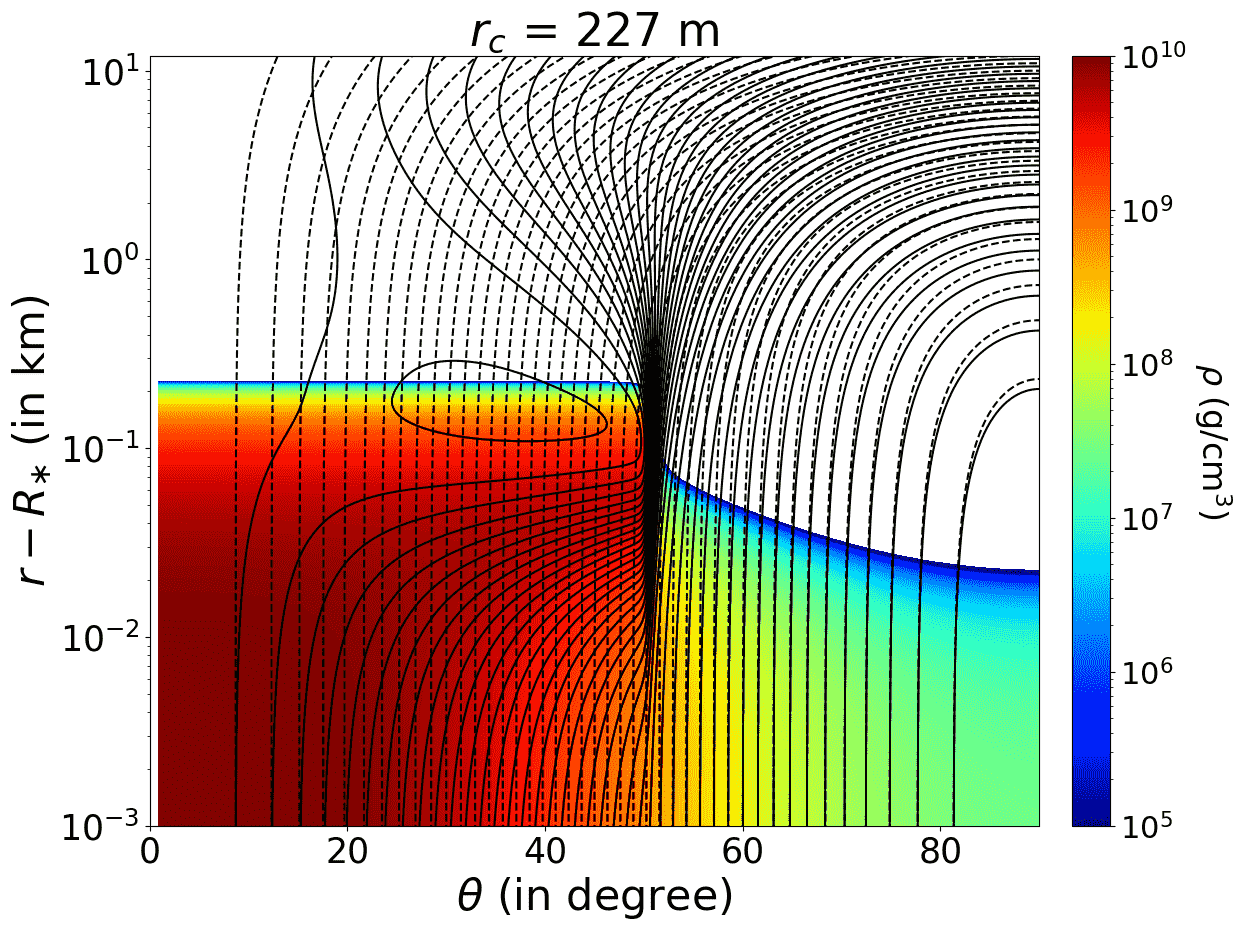}
}
\caption{Density profiles and magnetic field lines (solid) for the $1/$cosh profile for $r_{c} = 216$ m (top) and $r_{c} = 227$ m  (bottom). These solutions are calculated for a $B_{d}=10^{12.5}$ G. Dashed lines are the undistorted dipolar magnetic field lines. Masses with an order of $10^{-8}$ M$_{\odot}$ can be simulated with this profile. Closed Magnetic loops (bottom plot) are detected in the solutions beyond a certain mass.}
\label{cosh}
\end{figure}

In the past, several authors have modelled mounds to be not strictly confined to a chosen $\theta_t$, which would be expected in accreting systems with a well-defined inner radius \citep{mukherjee2012phase,muk2017revisit}. Instead, often an exponentially decaying profile function has been chosen \citep{payne2004burial,priymak2011quadrupole,priymak2014cyclotron,fujisawa2022magneticallymulti}. Such a profile results in significant mass loading to field lines beyond the $\theta(\psi_a)$ latitude, extending up to the equator \citep{muk2017revisit}. During a given accretion episode, such extended mass distribution is not possible for mass loaded field lines from an accretion disk with a finite inner edge, translating to a strict cut-off latitudinal angle ( $\theta_t$ as in Eq.~\ref{TTeq}). Nonetheless, an extended mass distribution may be used to motivate long-term accretion based confinement of matter. It is possible that during accretion, matter spreads out of confinement 
through reconfiguration of magnetic fields by ideal or resistive instabilities \citep{vigelius2009resistive,suvorov2019relaxation,Kulsrud_Sunyaev_2020}. Thus, the accretion mound might reconfigure to a larger size. To model such mounds and compare with earlier results, we perform GS solutions with a profile given by \citep{fujisawa2022magneticallymulti}: 
\begin{equation}
    r_{0}(\psi) = R_{\ast} + \frac{r_{c}}{\cosh{\frac{\psi}{\psi_{a}}}},
\end{equation}
Figure \ref{cosh} shows an accretion mound with this profile (resolution $5000\times5000$) for parameter $\theta_{t}=35.27^{0}$ and two values of $r_{c}$. The top plot in Figure \ref{cosh} has a single mound with a total mass of $4.23\times10^{-8}$ M$_{\odot}$, with a mass of $1.43\times10^{-9}$ M$_{\odot}$ beyond $\psi_{a}$. Accretion mounds of Mass $10^{-8}-10^{-7}$ M$_{\odot}$ can be set with this profile for a high magnetic field of $10^{12.5}$ G (assumed by \cite{priymak2011quadrupole}) before closed magnetic loops appear in the solution. Such closed magnetic loops have also been reported in several earlier works \citep{hameury1983magnetohydrostatics, payne2004burial, 
mukherjee2012phase, priymak2011quadrupole}. These closed loops have been identified to arise due to the inherent susceptibility of the bent field lines to MHD instabilities, confirmed in dynamic simulations \citep{payne2007burial,mukherjee2013mhd2d}.
 Closed magnetic loops do not appear in the ring-shaped mound profiles unless the resolution is low and the solution is found from a perturbed guess (Figure \ref{balloon}, Appendix subsection \ref{app3}). This likely results from the additional freedom of the field lines to bend towards the polar regions, which would otherwise have led to more severe distortions, as observed in filled mounds.

Thus, this profile is limited to a maximum mass above which closed loops related to magnetic field geometry appear in the solutions. Due to this limitation, this profile does not continue to spread without closed magnetic loops with increasing mass like the ring-shaped mound profile as discussed in the next subsection.

\subsection{Ring-shaped mound}
In this subsection, we outline the behaviour of GS solutions on increased mass loading for the physically motivated ring-shaped mound profile and describe different properties of the mound for this profile.

\subsubsection{Latitudinal spreading of matter}\label{3.4}
The mass of the mound for the ring-shaped mound profile could be increased by increasing the angular extent of the mound ($\theta_{t}$) and maximum mound height ($r_{c}$). For mound profiles modelled to be created from accretion episodes, $\theta_{t}$ is expected to be determined by the inner edge of the accretion disk, which is related to the Alfv\'en radius as $r_t=\zeta R_A$, as described in the subsection \ref{param}. However, in this subsection and some of the other subsections, solutions have also been calculated for an arbitrary $\theta_{t}$ to explore larger masses. Such solutions present hypothetical large mass loading, beyond the confines of a single accretion episode, as described in the earlier subsection (Sec.~\ref{1cosh}). Once $\theta_{t}$ is fixed, for a particular $B_{d}$ and $\theta_{t}$, an increase in the height ($r_{c}$) of the mound leads to a higher confined accreted mass and larger magnetic field deformation.

\begin{figure*}
\vbox{
\includegraphics[width=2\columnwidth]{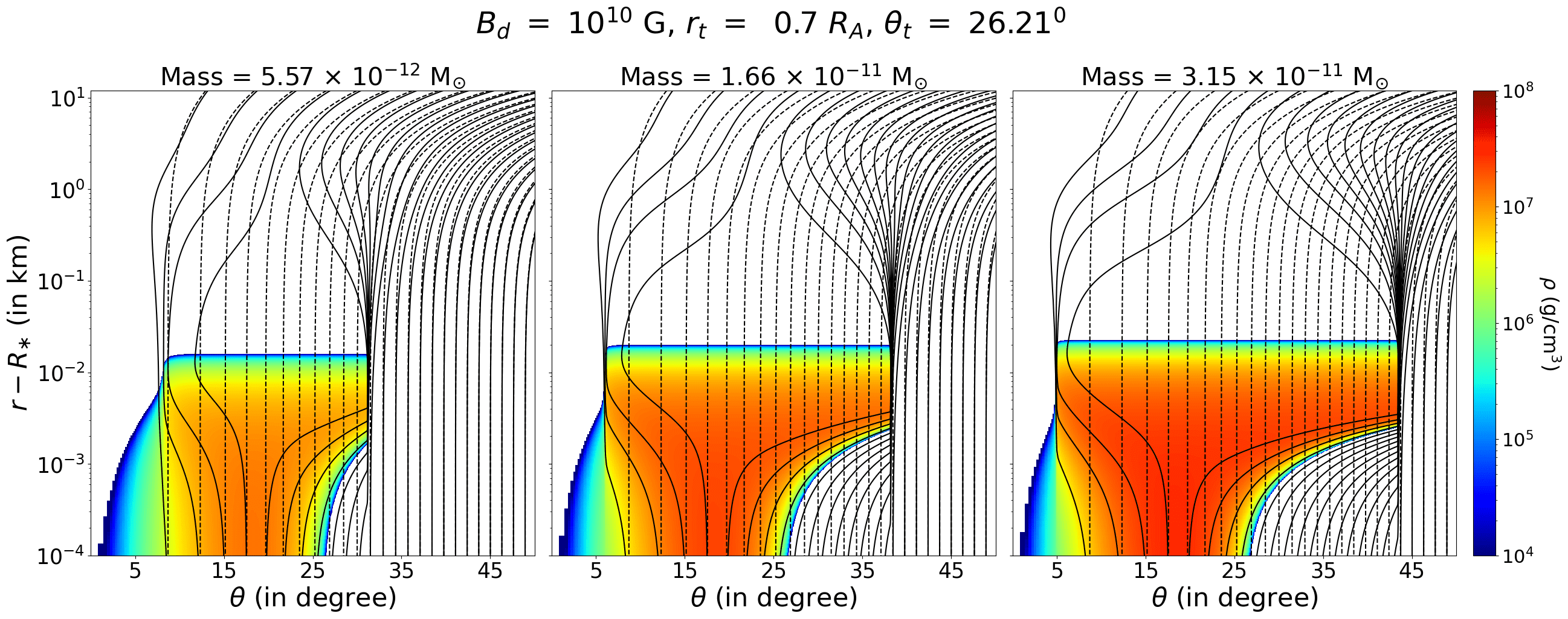}
\includegraphics[width=2\columnwidth]{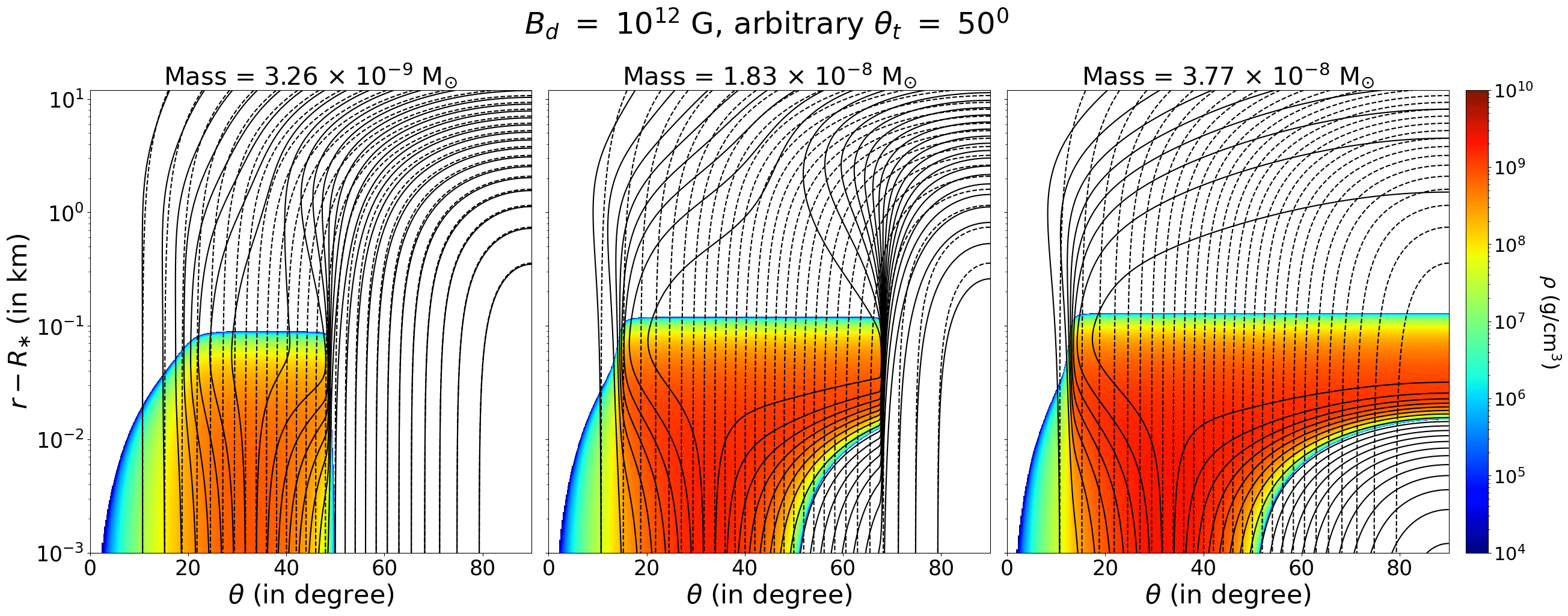}
}
\caption{Density profiles and magnetic field lines (solid) for the ring-shaped mound profile for parameters $B_{d}=10^{10}$ G, $r_{t}=0.7R_{A}$, $\theta_{t}=26.21^{0}$ (top plot) and parameters $B_{d}=10^{12}$ G, arbitrary $\theta_{t}=50^{0}$ (bottom plot). The top plot X-axis has a $\theta$ range of $0^{0}-50^{0}$, while the bottom plot X-axis has a $\theta$ range of $0^{0}-90^{0}$. Also, the colorbars have different ranges for the top and bottom plots. Both the top and bottom plots have resolution $5000\times5000$. $r_{c}$ is increased and thus Mass increases from top left to right and the same for the bottom plot. Dashed lines are the undistorted dipolar magnetic field lines. Mounds for the top plot with a physically motivated $\theta_{t}$ spread towards the equator, but the NMM is not evolved till the equator. Mounds for the bottom plot with an arbitrary $\theta_{t}$ spread towards the equator and the mounds are evolved till the equator. It must also be pointed out that the sequentially spreading solutions have a different $dM/d\psi$.}
\label{strange}
\end{figure*}

This study finds that with an increase in confined mass, the ring-shaped mound spreads latitudinally both towards the pole and the equator. Two examples of such spreading solutions are presented in Figure \ref{strange}, where we see that the density profile has extended up to an angle greater than $\theta_{t}$. A sequence of solutions with increasing $r_{c}$ has been calculated and plotted for two cases : a physically motivated $\theta_{t}$ ($r_{t}=\zeta R_{A}$) and an arbitrary value of $\theta_{t}$ (similar motivation to Case $2$ in subsection \ref{3.2.1}). We have calculated solutions till a maximum mass beyond which the uniqueness of solutions cannot be established and this maximum mass is limited by resolution. We call this the Numerical Maximum Mass (NMM). The procedure for finding the NMM for a particular $B_{d}$ and $\theta_{t}$ and the dependence of the NMM on resolution has been described for a case in Appendix \ref{numapp}. From Figure \ref{strange}, we can see that the magnetic field lines have a limited poleward spread and an extensive equator-ward spread that successively increases with mass. A higher equator-ward spread reduces the dipole moment above the mound as discussed later in subsection \ref{arbitrary_theta}. Such an equator-ward spread was also found by \cite{payne2004burial,priymak2011quadrupole,fujisawa2022magneticallymulti}. However, all such previous models populated the matter on all the magnetic field lines from the pole to the equator. Such spreading becomes more explicitly demonstrated in the current results, where the matter distribution is shown to have higher angular extent than the foot point of the last matter loaded field line, given the choice of the truncation angle ($\theta_t$). These results were found for all the mounds calculated in this paper. For physically motivated $\theta_{t}$ ranges, the latitudinal spread was limited to an angle less than $90^{0}$ (before the equator) due to NMM constraints as described in subsection \ref{physical_theta}. For an arbitrary $\theta_{t}$, the mound has a latitudinal spread till the equator as described in subsection \ref{arbitrary_theta}. 

One must note that the mass beyond $\theta_{t}$ with a height of $22.63$ meters in the upper part and rightmost plot of Figure \ref{strange} is supported against gravity by a vacuum region of height three meters (the Y-axis of Figure \ref{strange} is in log scale) with strong $B_{\theta}$ fields. Such solutions may not be stable and their dynamical stability should be verified. 

\subsubsection{Ellipticity and relative dipole moment behaviour for $r_{t}=R_{A}$}\label{ell_dip}

\begin{figure}
\vbox{
\includegraphics[width=0.45\textwidth]{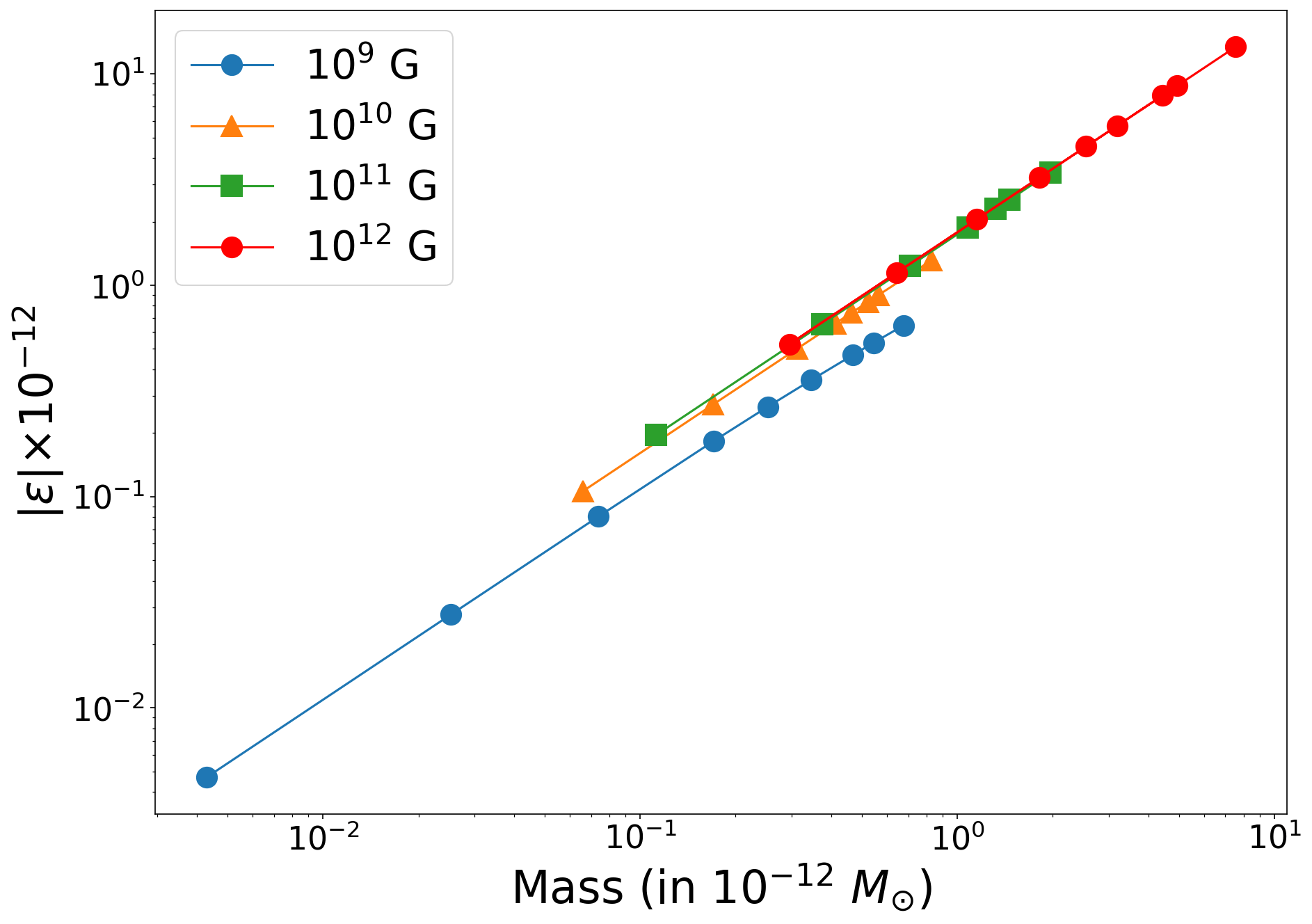}
\includegraphics[width=0.45\textwidth]{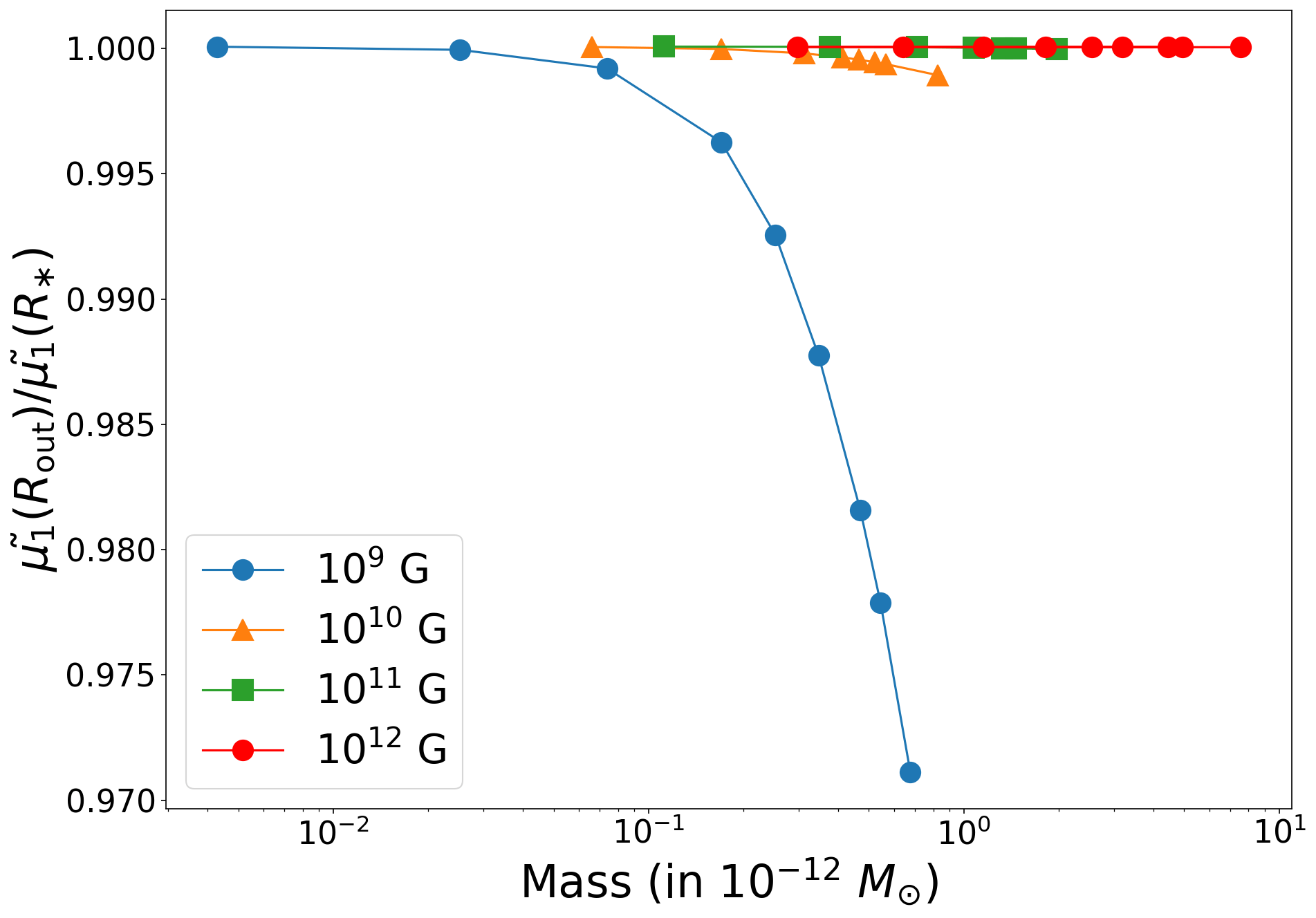}
}
\caption{Top plot has absolute mass ellipticity of the $2$ mounds on the Y-axis and mass of the two mounds on the X-axis for four values of $B_{d}$. Ellipticity is proportional to the accreted mass. Bottom plot has normalized dipole moment at the outer radius relative to its value at the neutron star surface varying with the accreted mass of the two mounds for four values of $B_{d}$. See subsection \ref{ell_dip} for further description.}
\label{ellmass}
\end{figure}

The results in this subsection are for simulations with grid size of $10000\times10000$, except for cases with $B_{d}= 10^{12}$ G, where low values of $\theta_{t} (\lesssim 5.7^{0})$, necessitates higher resolution ($12000\times12000$). $\theta_{A}$ has been calculated assuming the accretion disk to be truncated at $R_{A}$ (Equation \ref{RAeq}). Although we have run a suite of simulations with different mound masses (by varying $r_c$), the solutions have not been calculated until a maximum mass (NMM) due to the compute intensive nature of the exercise. However, the results depict the general trend of the variation of the derived parameters such as mass ellipticity and normalised relative outer dipole moment (eq.~\ref{MUeq}) with Mass. We define ellipticity $\epsilon$ \citep{priymak2011quadrupole,fujisawa2022magneticallymulti} as 
\begin{equation}
    \epsilon = \frac{I_{zz} - I_{xx}}{I_{0}},
\end{equation}
where $I_{0}=1.11\times10^{45}$ g$/$cm$^{2}$ is the moment of inertia of a spherically symmetric neutron star. $I_{jj}$ is the moment of inertia tensor about the j coordinate axis. Here $Z$ axis is the magnetic axis. The mass and absolute ellipticity for $4$ different $B_{d}$ are plotted in the top plot of Figure \ref{ellmass}. As found in previous works \citep{priymak2011quadrupole,suvorov2020multi,fujisawa2022magneticallymulti,Rossetto_2025}, the absolute ellipticity of our solutions is proportional to the accreted mass. Also, as shown in the bottom plot of Figure \ref{ellmass}, the value of $\tilde{\mu_{1}}(R_{\rm out})/\tilde{\mu_{1}}(R_{\ast})$ (Equation \ref{MUeq}) decreases with increasing accreted mass.

\subsubsection{Results for $r_{t}=\zeta R_{A}$}\label{physical_theta}

\begin{figure*}
\begin{multicols}{2}
    \includegraphics[width=0.9\linewidth]{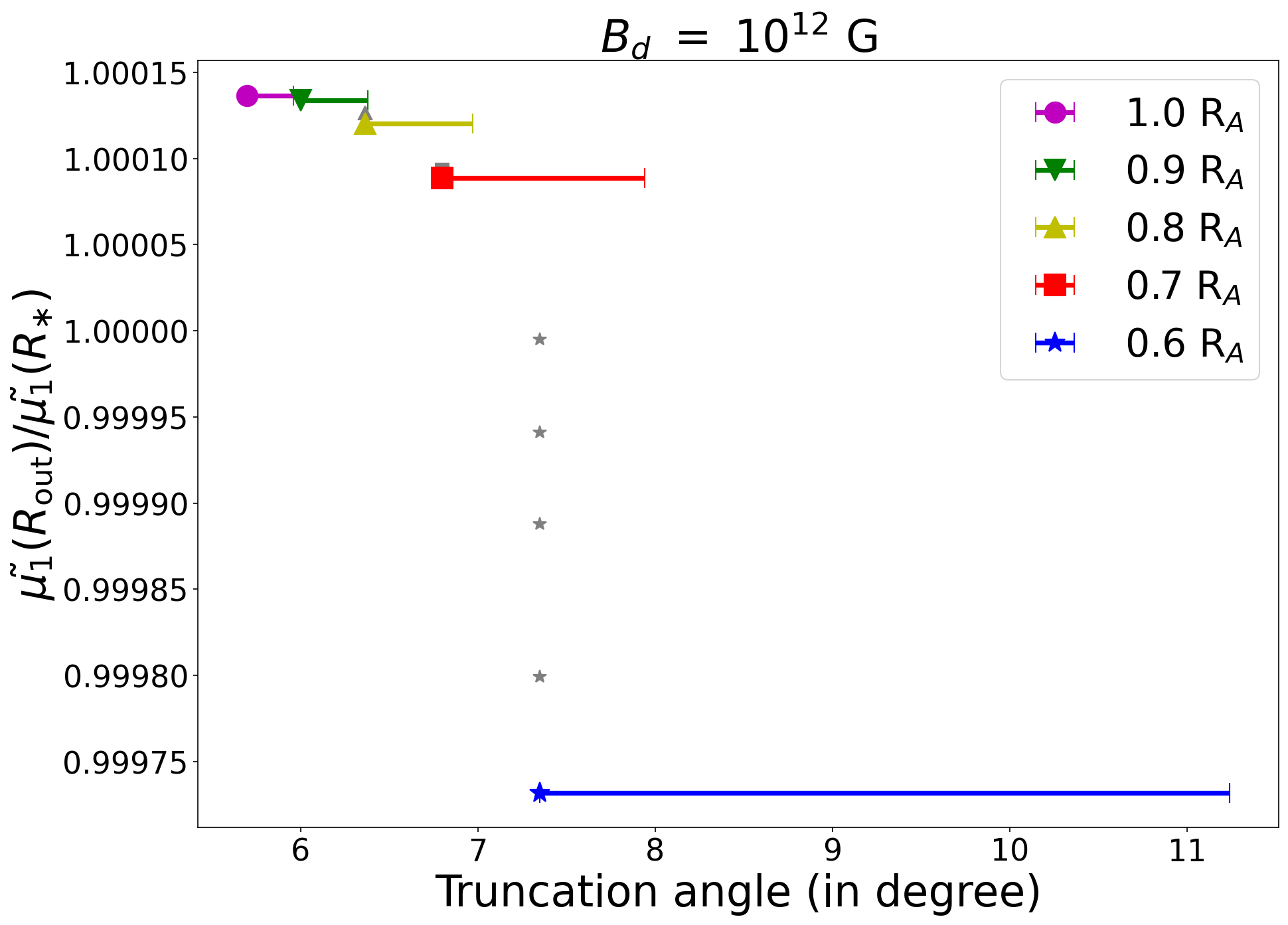}\par
    \includegraphics[width=0.9\linewidth]{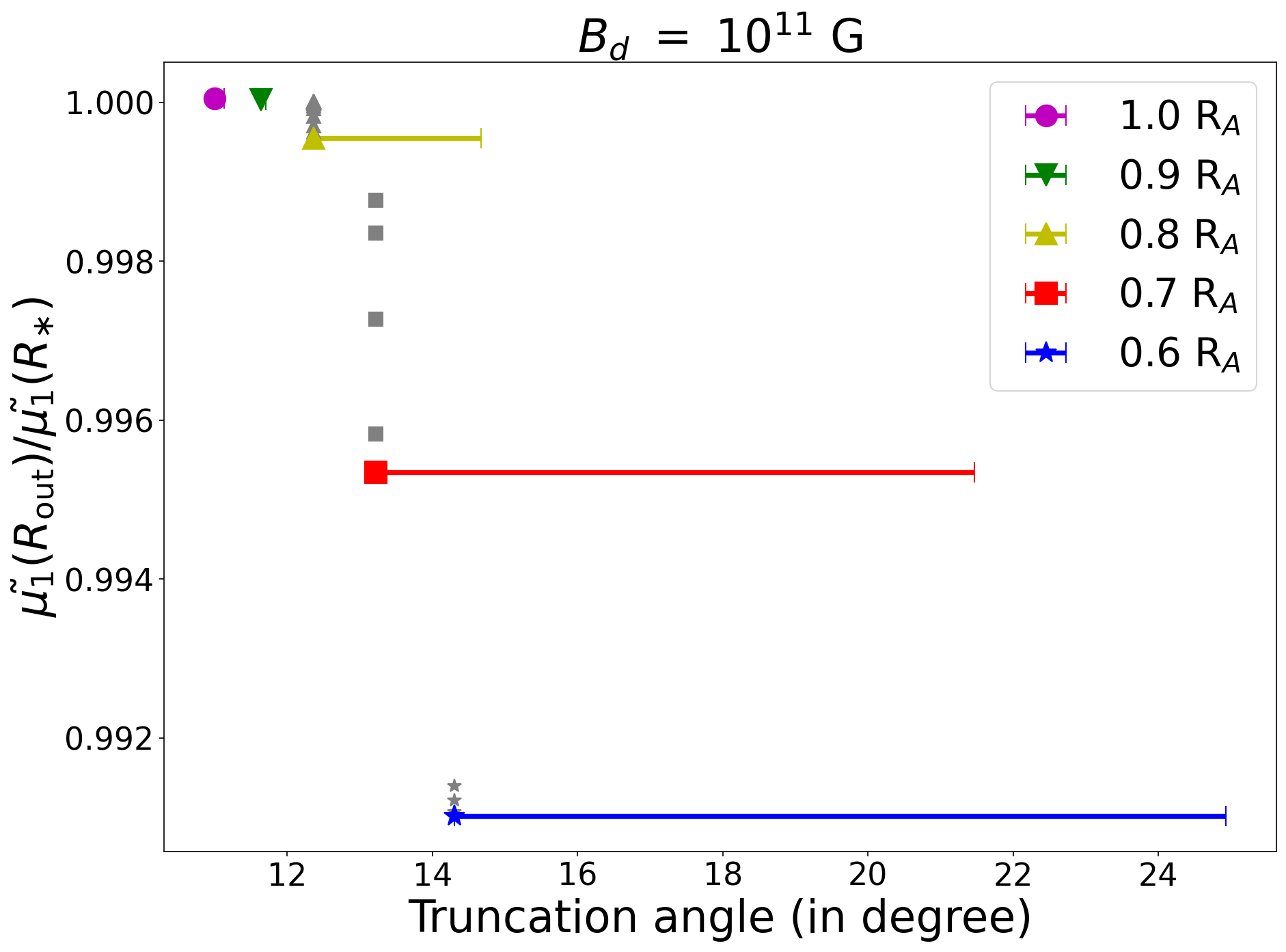}\par 
    \end{multicols}
\begin{multicols}{2}
    \includegraphics[width=0.9\linewidth]{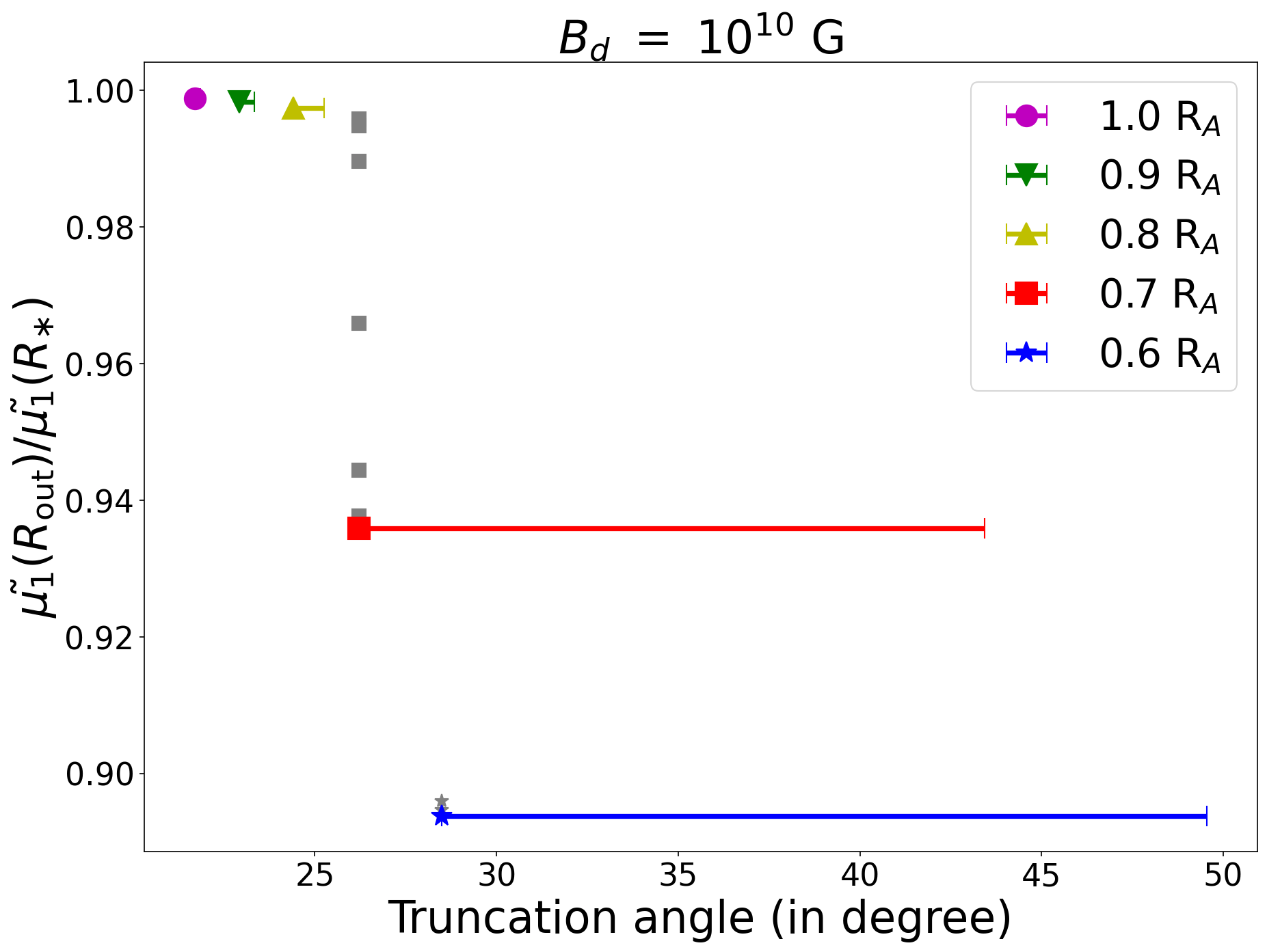}\par
    \includegraphics[width=0.9\linewidth]{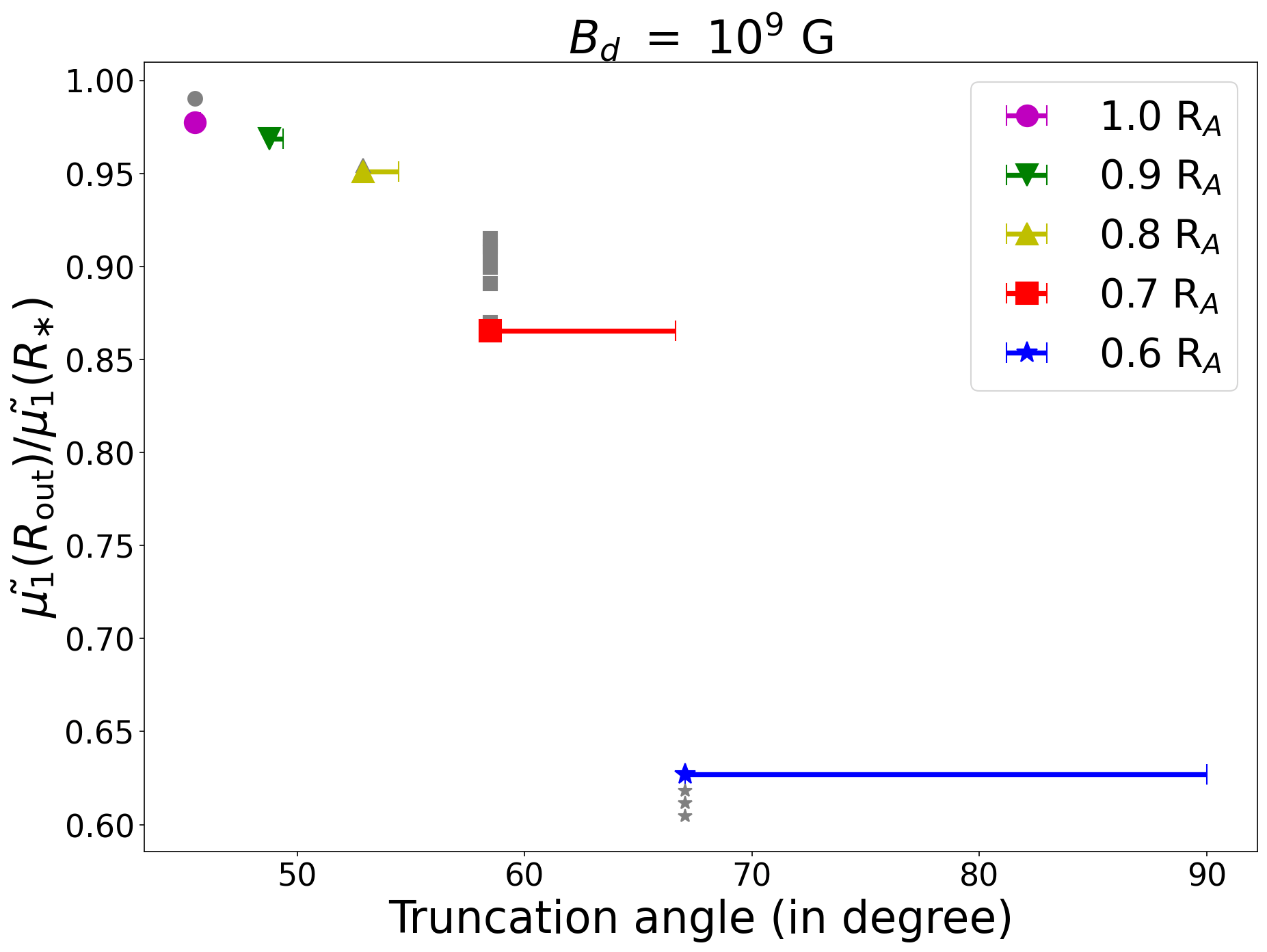}\par
\end{multicols}
\caption{Plot of normalized dipole moment at the outer radius relative to its value at the NS surface for $5$ different truncation angles ($\theta_{t}$) made for $4$ surface magnetic field strengths ($B_{d}$). The colored symbols indicate the Maximum Mass (NMM) set by this simulation for a respective $\theta_{t}$, while the gray symbols are the solutions for a lower mass. Errorbars show the equator-ward angular spread of matter beyond $\theta_{t}$. See subsection \ref{physical_theta} for further description.}
\label{truncdip}
\end{figure*}

To find the impact of mound masses on different properties of such systems such as mass ellipticities, efficiency of field burial and truncation angles $\theta_{t}$, we have performed a suite of GS solutions. Numerical simulations of magnetized accretion onto neutron stars \citep[such as][]{Long_2005,Romanova_2008,bessolaz2008,zanni08,kulkarni2013,parfrey,parfrey17b} predict the truncation radius of the accretion disk to be smaller than the traditional Alfv\'en radius ($r_{t} \sim 0.5 - 1.0 R_{A}$). We have thus varied $r_{t}$ in the range $0.6 - 1.0$ $R_{A}$, to obtain different values of $\theta_{t}$ for a given $B_{d}$ and repeated the activity for four different magnetic fields ($B_{d} \in( 10^{9}, 10^{10}, 10^{11}, 10^{12})$ G), for a total of $20$ solutions. For each of the $20$ solutions, $r_{c}$ is increased to get a higher confined mass. Although we are limited by numerical diffusion, we chose to perform all simulations with the same resolution $5000\times5000$ to draw a generic set of conclusions for the physical parameters of interest. An example of the density profiles and magnetic field lines of one of these results are plotted in the upper part of Figure \ref{strange}. The mass and dipole moments for all the results are depicted in Figures \ref{truncmass} and \ref{truncdip}.

 In Figures \ref{truncmass} and \ref{truncdip}, the coloured points correspond to the maximum mass allowed by our simulations (NMM) (Appendix \ref{numapp}). The gray points in these figures are the other solutions with a lower $r_{c}$. The four figures plotted are for four values of $B_{d}$. For each $B_{d}$ figure, accreted mass and $\tilde{\mu_{1}}(R_{\rm out})/\tilde{\mu_{1}}(R_{\ast})$ are plotted against the five values of $\theta_{t}$ in the figures \ref{truncmass} and \ref{truncdip} respectively. Most of these solutions spread towards the equator with matter going beyond $\theta_{t}$ as in Figure \ref{strange}. The horizontal bar in Figures \ref{truncmass} and \ref{truncdip} shows the extent of the angular spread of the mound beyond $\theta_{t}$. Here are the primary results :

\begin{itemize} 
\item Higher mass of the mounds could be accreted and confined for a higher $B_{d}$. For a particular $B_{d}$, figure \ref{truncmass} clearly shows that if $\theta_{t}$ is higher, the angular extent of matter is greater (both $\theta_{t}$ and the equator-ward spread is high), and thus the amount of accreted mass is higher.    

\item Figure \ref{truncmass} also shows the maximum absolute mass ellipticity obtained by our calculations for the four magnetic fields. For the four magnetic fields ($10^{9},10^{10},10^{11},10^{12}$) G, the maximum absolute ellipticities are $9.52\times10^{-13},4.22\times10^{-11},2.1\times10^{-10},3.75\times10^{-10}$ respectively. The value of these ellipticities serves as a lower limit (since most of the mounds have a further chance of spreading). However, the impact of MHD instabilities has not been considered. There could also be a contribution to the ellipticity purely due to the presence of a magnetic field \citep{nazari2020,Rossetto_2025}. \cite{Rossetto_2025} found the absolute magnetic ellipticity to be six orders lower than the absolute mass ellipticity. In this paper, the maximum absolute magnetic ellipticity for the four magnetic fields ($10^{9},10^{10},10^{11},10^{12}$) G were found to be $2.43\times10^{-17},9.32\times10^{-16},10^{-14},5.6\times10^{-14}$ respectively, which is approximately four orders lower than the maximum absolute mass ellipticity.

\begin{figure*}
\begin{multicols}{2}
    \includegraphics[width=0.9\linewidth]{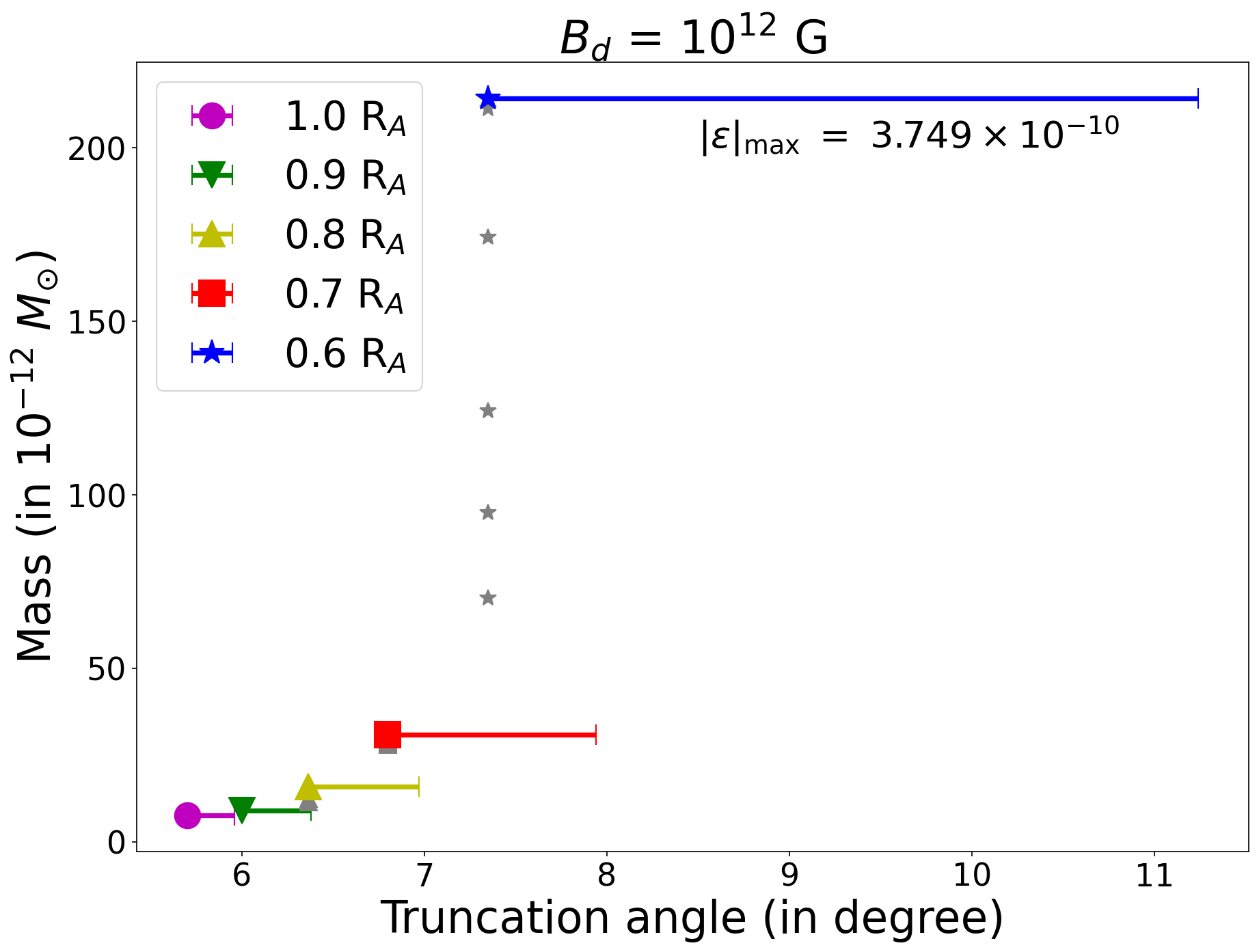}\par 
    \includegraphics[width=0.9\linewidth]{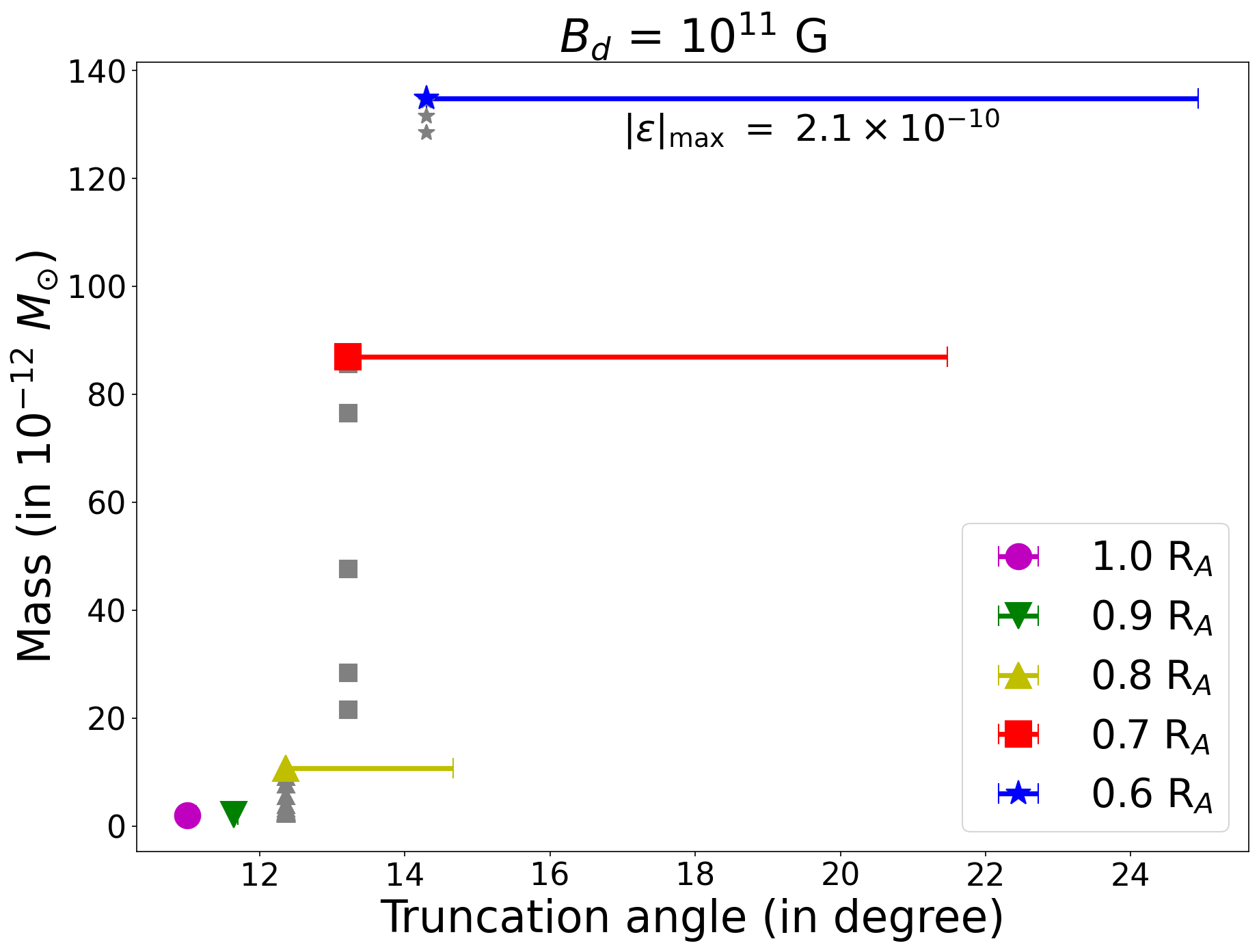}\par 
    \end{multicols}
\begin{multicols}{2}
    \includegraphics[width=0.9\linewidth]{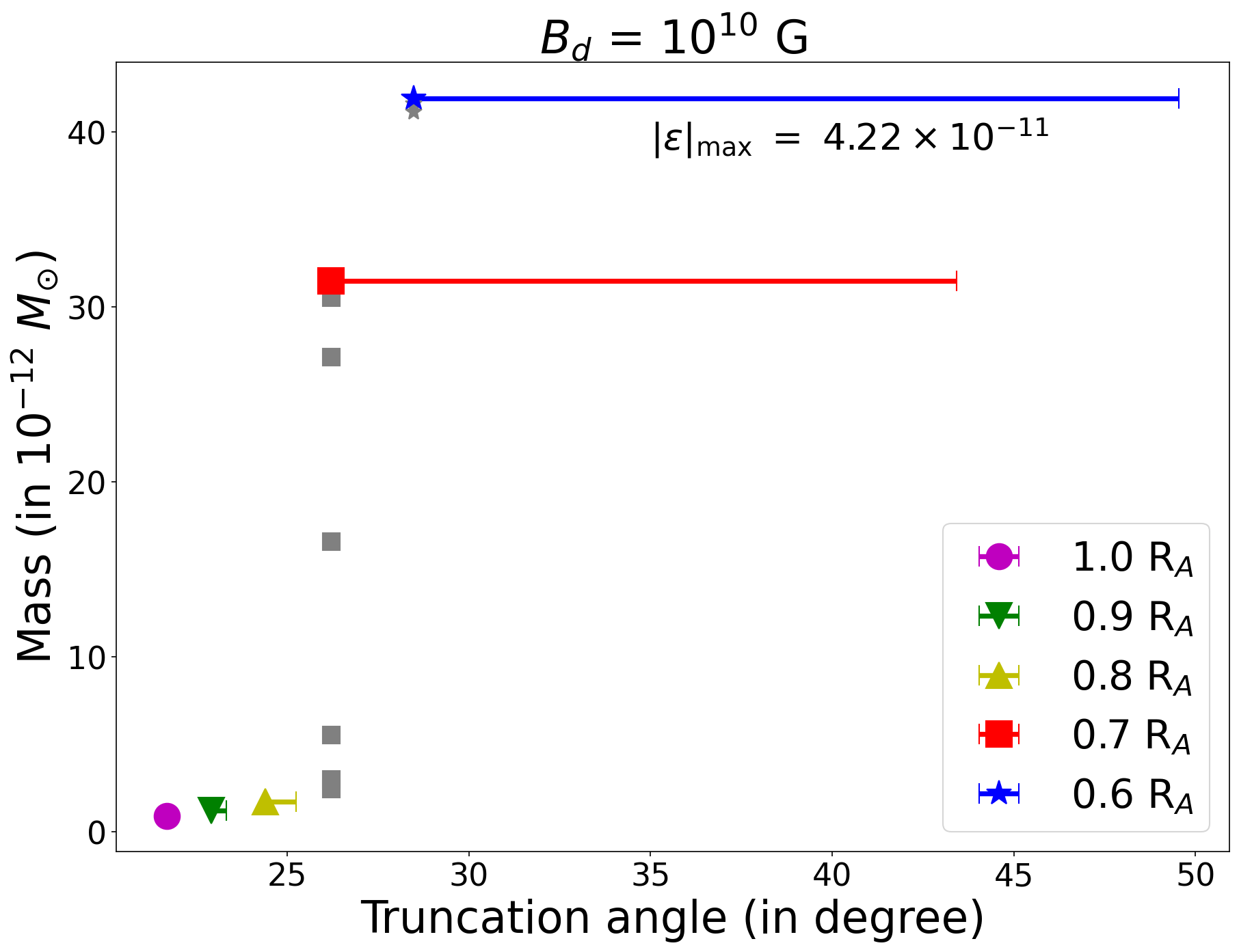}\par
    \includegraphics[width=0.9\linewidth]{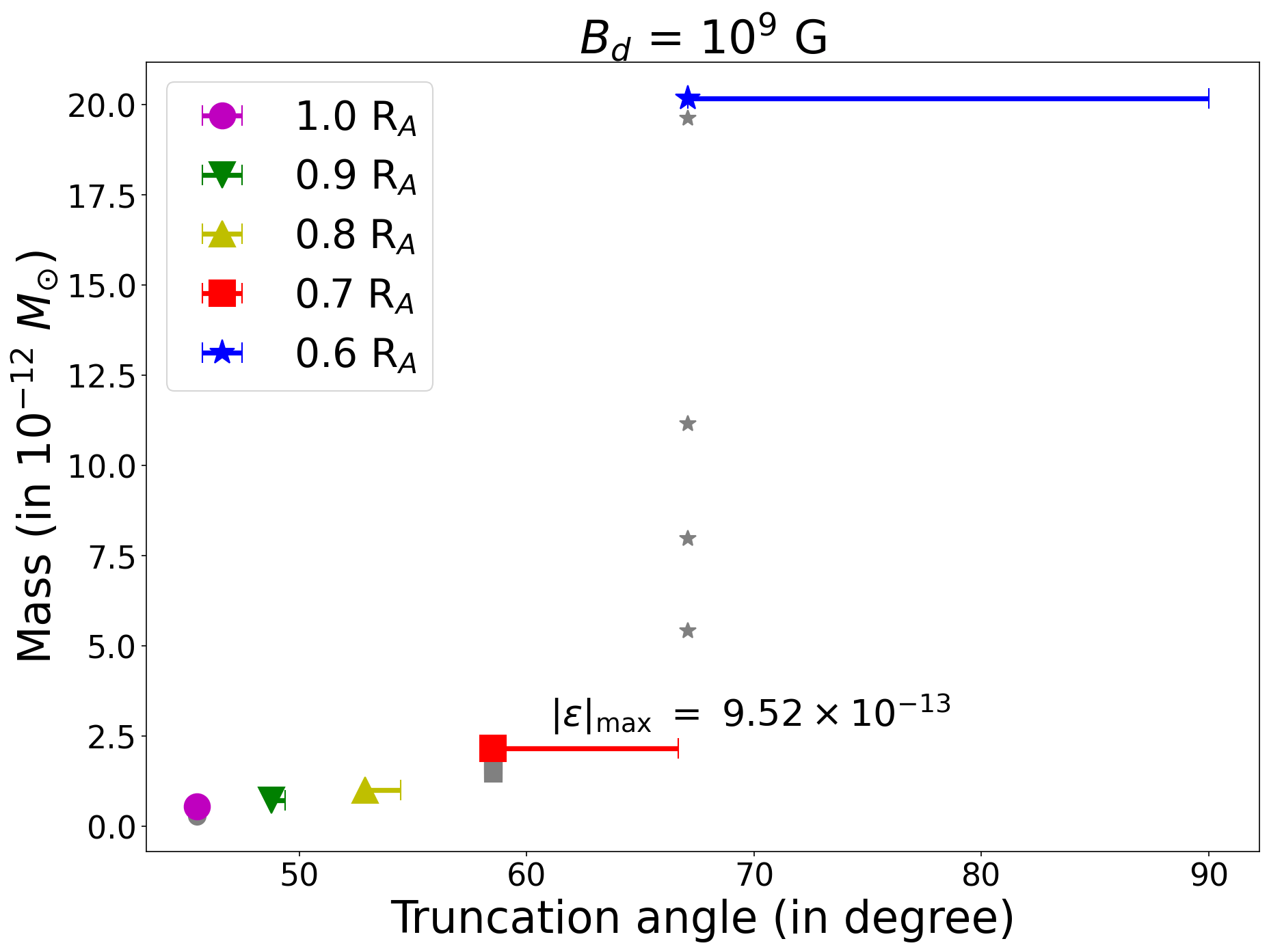}\par
\end{multicols}
\caption{Plot of Mass (in $10^{-12}$ M$_{\odot}$) for $5$ different possible truncation angles ($\theta_{t}$) made for $4$ surface magnetic field strengths ($B_{d}$). The colored symbols indicate the maximum mass (NMM) set by this simulation for a respective $\theta_{t}$, while the gray symbols are the solutions for a lower height. The errorbars show the latitudinal angular spread of matter beyond $\theta_{t}$. See subsection \ref{physical_theta} for further description.}
\label{truncmass}
\end{figure*}

\item Figure \ref{truncdip} shows the value of $\tilde{\mu_{1}}(R_{\rm out})/\tilde{\mu_{1}}(R_{\ast})$ for all solutions. For a particular $B_{d}$, since larger masses lead to lower dipole moments at the outer radius, an increase in the $\theta_{t}$ values decreases the values of $\tilde{\mu_{1}}(R_{\rm out})/\tilde{\mu_{1}}(R_{\ast})$. The solution with $B_{d}=10^{9}$ G, $\zeta = 0.6$ has its matter extended till the equator, and we find the highest field burial for this solution with a value of $\tilde{\mu_{1}}(R_{\rm out})/\tilde{\mu_{1}}(R_{\ast})=0.627$.  

\item Change in higher moments ($\ell > 1$) is insignificant for our solutions, except for the low field $10^{9}$ G, $\zeta = 0.6$ solution, where the relative octupole moment is about $0.12$.

\end{itemize}

 The latitudinal spreading by the ring-shaped mound profile is the true depiction of field burial predicted before through semi-analytic studies \citep{hameury1983magnetohydrostatics,melatos2001hydromagnetic}. We have a solution with a latitudinal spread till the equator only for the case $B_{d} = 10^{9}$ G, $\zeta = 0.6$. For all the other cases, we have not calculated solutions that have a latitudinal spread till the equator due to computational constraints. In a future work, we will attempt to find the latitudinal spreading of mounds with $\theta_{t}$ near the poles, spreading from the pole to the equator, to explore the highest field burial achievable. 

\subsubsection{Results for arbitrary $\theta_{t}$}\label{arbitrary_theta}

Though $\theta_{t}$ for a mound with $B_{d}=10^{12}$ G is as low as $5.7 - 7.3$ degree if constrained to $\theta_t = \theta_A$, for the purpose of demonstration and to have a high $\theta$ resolution near $\theta_{t}$ we simulate accretion mounds for an arbitrary $\theta_{t}=50^{0}$. As shown in the lower part of Figure \ref{strange}, the mound starts to spread towards the equator, but the mass beyond $\theta > \theta_{t}$ is completely supported by the vacuum magnetic field against gravity. Before the mound spreads towards the equator as shown in the lower part of Figure \ref{strange}, some resistive instabilities \citep{vigelius2009resistive,suvorov2019relaxation,Kulsrud_Sunyaev_2020} could result in diffusion of matter near $\theta_{t}$ where the magnetic field lines bunch together. This might change $\theta_{t}$ and the magnetic geometry, thus automatically invalidating the solutions given in Figure \ref{strange}. Even cumulative accretion episodes could change $\theta_{t}$. In addition, the solutions shown in the lower part of Figure \ref{strange} increase in mass sequentially, but we should understand that they do not come from sequential accreting episodes, since they have a different $dM/d\psi$.

\begin{figure}
\vbox{
\includegraphics[width=\columnwidth]{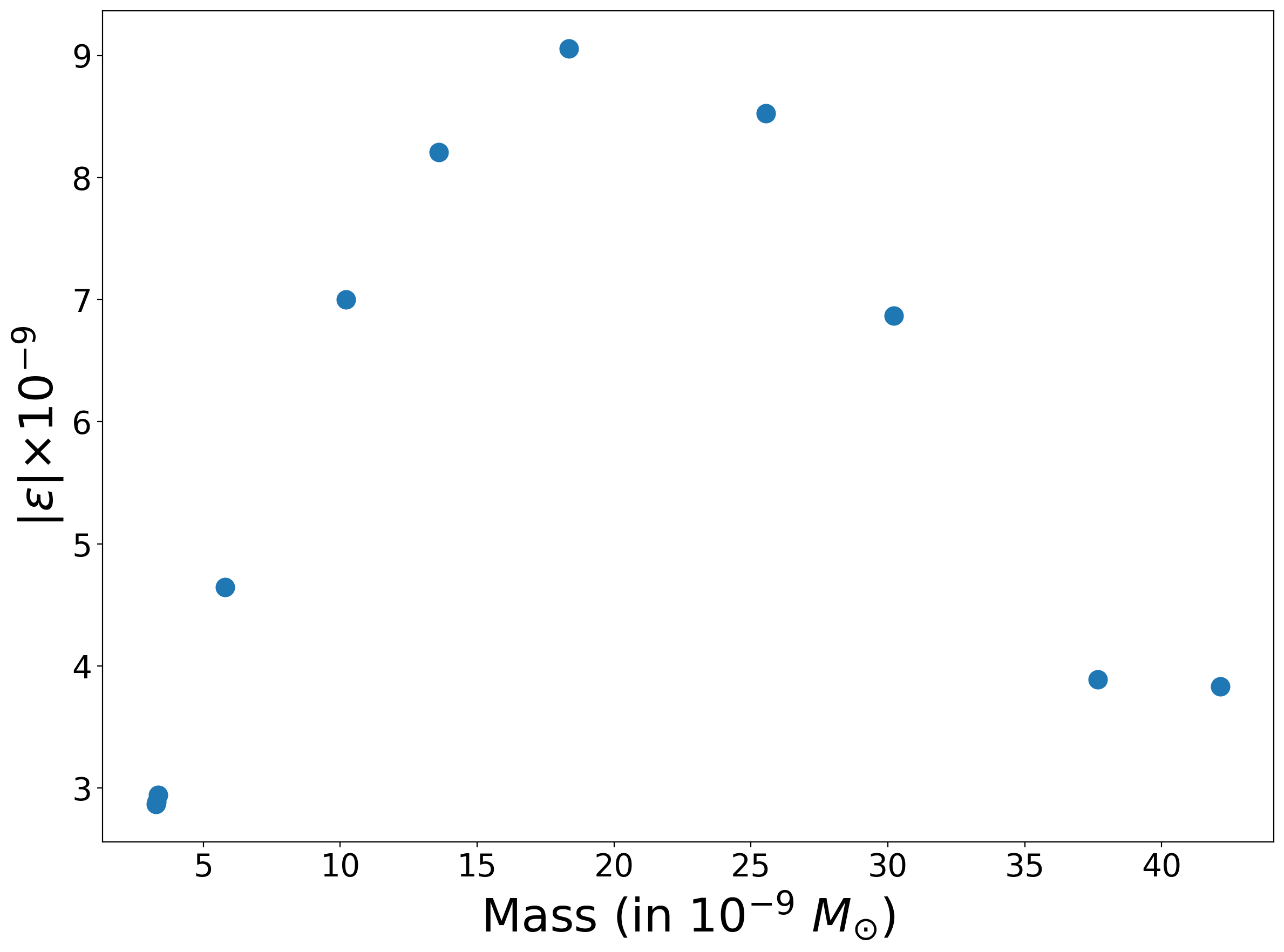}
\includegraphics[width=\columnwidth]{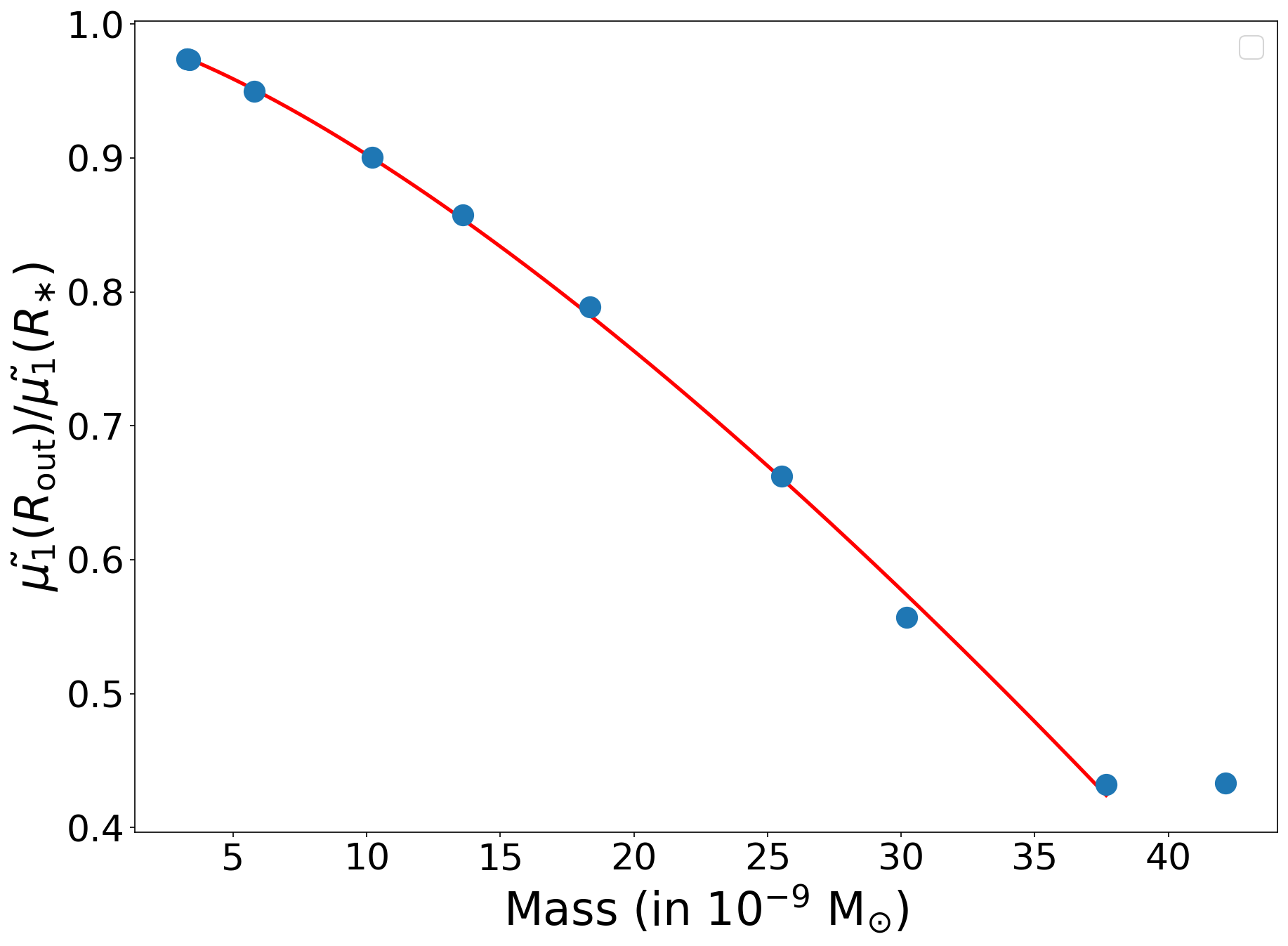}
\includegraphics[width=\columnwidth]{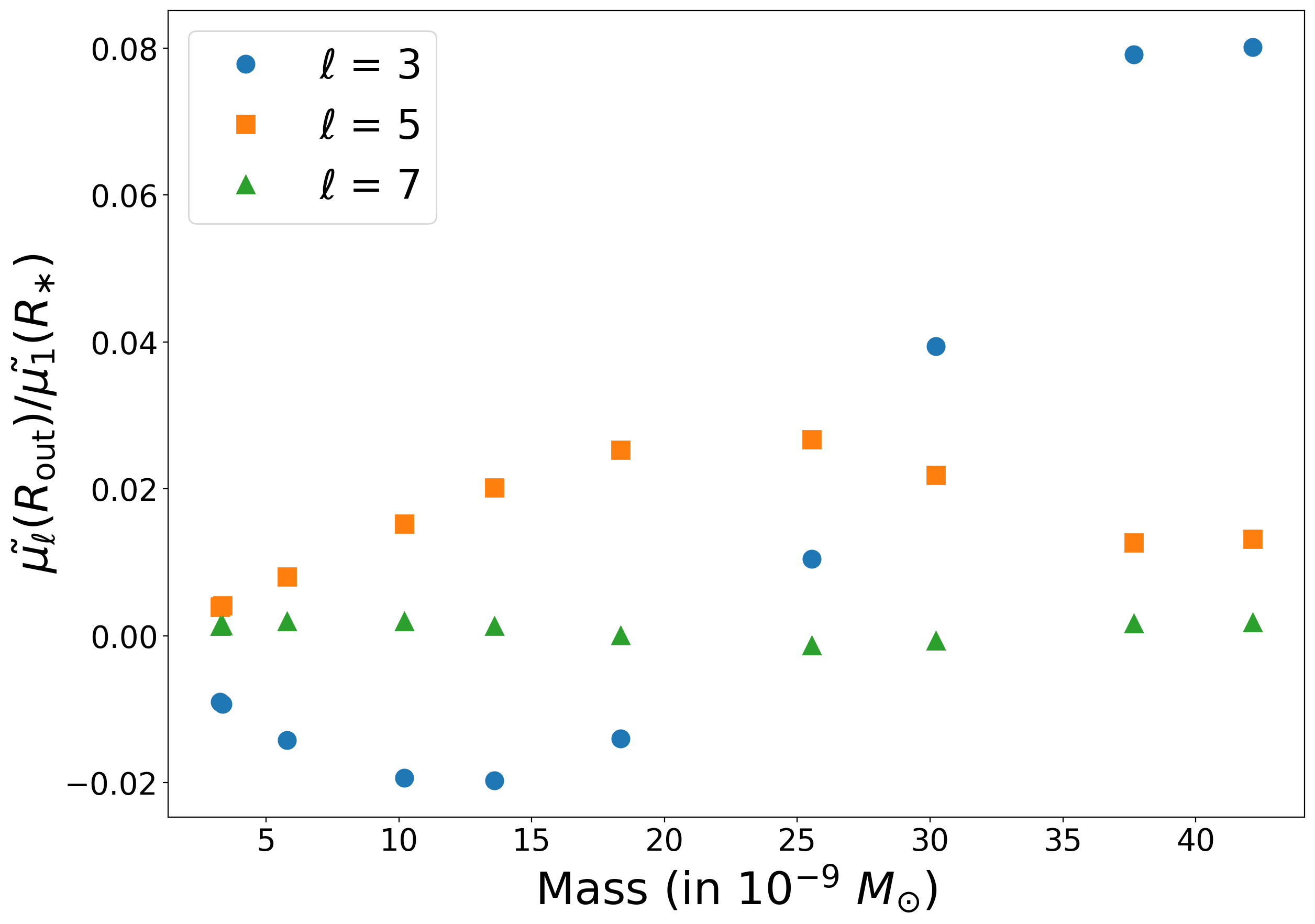}
}
\caption{The parameters for all the three plots are $B_{d}=10^{12}$ G and an arbitrary $\theta_{t}=50^{0}$. The topmost plot has shown absolute ellipticity as a function of Mass. The middle plot has shown normalized dipole moment at the outer radius relative to its value at the neutron star surface as a function of Mass. Blue points in this plot (except the last point) is fitted to the function $y=ax^{b} + k$ with a red line. The bottom plot has shown higher normalized multipole moments at the outer radius relative to dipole moment value at the neutron star surface as a function of Mass. See subsection \ref{arbitrary_theta} for further description.}
\label{ell_l1_l357}
\end{figure}

Ellipticity and change in multipole moments of these solutions are plotted in Figure \ref{ell_l1_l357}. The topmost plot of Figure \ref{ell_l1_l357} shows that there is a turnover in ellipticity after a certain mass, as the mound spreads towards the equator, unlike the ellipticity in Figure \ref{ellmass} which is not evolved till equator. Finally, as the mound reaches the equator, the ellipticity does not change appreciably. The middle plot of Figure \ref{ell_l1_l357} shows that the dipole moment keeps reducing as the mound mass increases, and the mound stretches the magnetic field lines and matter towards the equator. As the mound reaches the equator, the dipole moment remains more or less constant. The ratio of normalized dipole moments (except the last point) is found to have the behaviour
\begin{equation}\label{50degl1}
    \frac{\tilde{\mu_{1}}(R_{\rm out})}{\tilde{\mu_{1}}(R_{\ast})} = 0.994 - (0.00386\times \left(\frac{Mass}{10^{-9} M_{\odot}}\right)^{1.376})
\end{equation} 
The bottom plot of Figure \ref{ell_l1_l357} shows that the octupole moment initially decreases and then increases as the mass keeps on increasing. When the mound reaches the equator, octupole moment remains the same. $\ell=5$ and $\ell=7$ moments do not show a significant change as the mass of the mound increases.

\subsection{Ring-shaped mound on a neutron star atmosphere and ocean} \label{sunkmound}

 The neutron star surface is expected to have a liquid ocean and gas layer over the solid crust with a height ranging from $\sim 3 - 1700$ cm \citep{Nättilä_2024}. Current models of accretion mounds obtained using the GS solver do not take into consideration a pre-existing ocean and atmosphere. Here, we try to construct a profile by considering a mound mixed with and sunk into the neutron star ocean and atmosphere. This is similar to the profile suggested by \cite{melatos2001hydromagnetic} but with significant changes.

\begin{figure}
\includegraphics[width=\columnwidth]{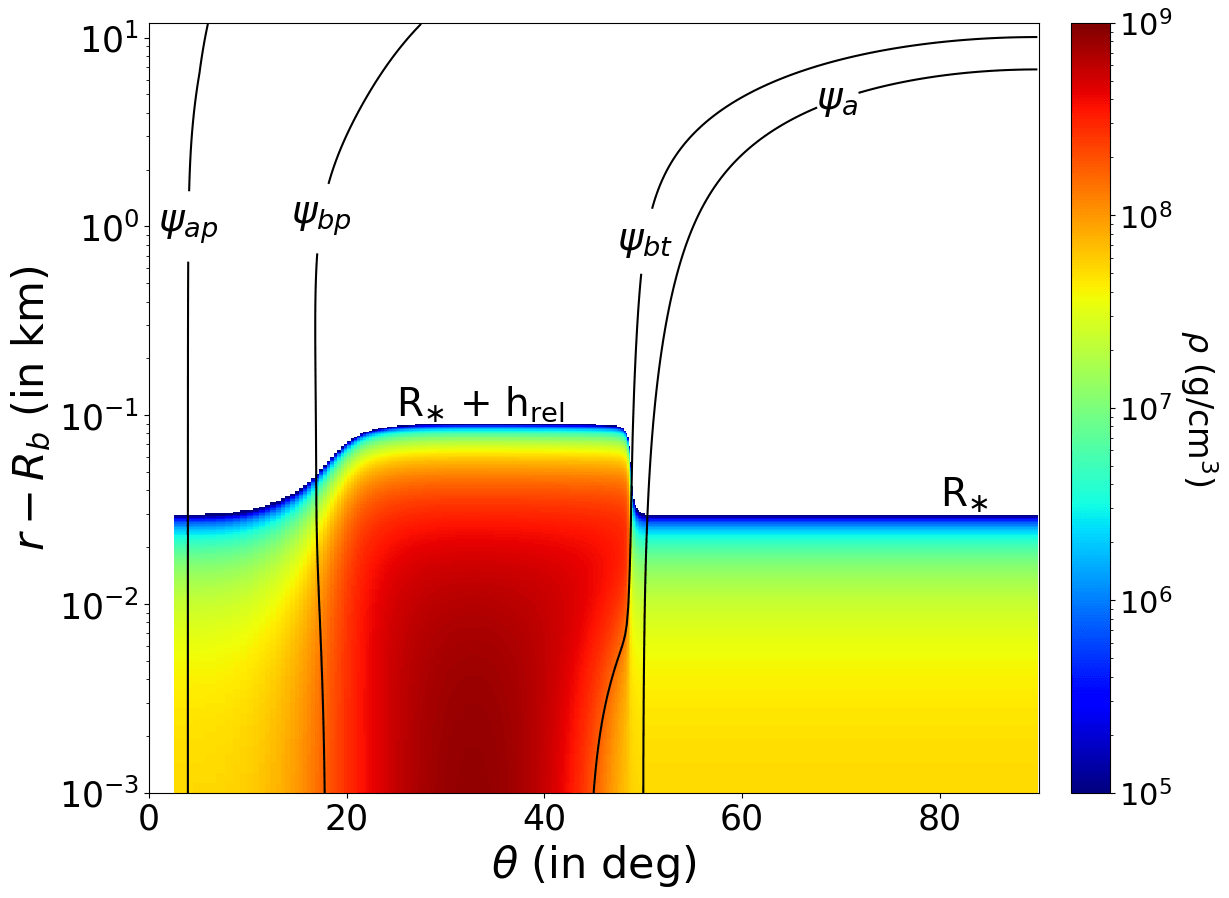}
\caption{Density profile of the mound from Equation \ref{eqntransitn} for parameters $B_{d}=10^{12}$ G, arbitrary $\theta_{t}=50^{0}$, arbitrary $\psi_{\rm ap} = \sin^{2}(4.0^{0})$, $r_{c}=$ $90$ m, $h_{\rm env}=$ $30$ m, accreted mass = $3.15\times10^{-9}$ M$_{\odot}$. Magnetic field lines for $\psi_{\rm ap}$,$\psi_{\rm bp}$,$\psi_{\rm bt}$ and $\psi_{a}$ are also plotted here. The five regions separated by the field lines from $\theta=0^{0}$ to $\theta=90^{0}$ are the current free envelope, the transition region, the ring-shaped mound profile region, the transition region and the neutron star current free envelope respectively.}
\label{transitn}
\end{figure}

A pre-existing current free neutron star ocean and atmosphere threaded with an initial dipolar magnetic field which is fixed at a radius $R_{b}$ is assumed here. The matter falls on the neutron star ocean through a surface angular range of $\theta_{\ell}\leq\theta\leq\theta_{t}$ confined by the magnetic field lines with $\theta_{\ell}$ constrained by the light cylinder and $\theta_{t}$ by the accretion disk inner radius. Matter mixes with the ocean and atmosphere, modifies the density profile and magnetic geometry above $R_{b}$ in the range $\psi_{\rm ap}\leq\psi\leq\psi_{a}$ and thus forms the accretion mound. It is assumed that beyond this range, the envelope is current free even after accretion and the height of the envelope above the inner boundary of the domain ($R_{b}$) is fixed. Thus, for this profile, the inner boundary ($R_{b}$) is not the surface of the neutron star, but it is a radial limit above which the mass loading profile per magnetic field line ($dM/d\psi$) of the neutron star has been modified by accretion, and the dipolar magnetic field is fixed at this radius. This is modelled by the following profile :
\begin{alignat}{2}\label{eqntransitn}
    r_{0}(\psi) & = R_{b} +  h_{\rm env} = R_{\ast} \quad \quad \quad \quad \quad \quad \quad \quad \quad \quad \quad & 0.0\leq\psi<\psi_{\rm ap} \nonumber\\
       & = R_{\ast} + h_{p}\left(\frac{\psi}{\psi_{a}(\theta_{t})} - \frac{\psi_{\rm ap}}{\psi_{a}(\theta_{t})}\right)^{2} & \psi_{\rm ap}\leq\psi<\psi_{\rm bp} \nonumber\\
       & = R_{b} + \frac{r_{c}}{0.25}\left(0.25-\left(\frac{\psi}{\psi_{a}(\theta_{t})}-0.5\right)^{2}\right) & \psi_{\rm bp}\leq\psi<\psi_{\rm bt} \nonumber\\
       & = R_{\ast} + h_{t}\left(1.0 - \frac{\psi}{\psi_{a}(\theta_{t})}\right)^{2} & \psi_{\rm bt}\leq\psi<\psi_{a} \nonumber\\
       & = R_{b} +  h_{\rm env} = R_{\ast} & \psi_{a}\leq\psi\leq\psi^{\ast}.
\end{alignat}
Here, $R_{\ast}$ is the radius of the neutron star and is fixed at $10$ km. $R_{b}$ is the radius of the inner boundary. $\psi^{\ast}$ and $\psi_{a} (\theta_{t})$ are the same parameters used in the ring-shaped mound profile, and they vary with $B_{d}$ and $\zeta$, but now $r_{c}$ is the maximum height of the mound above $R_{b}$. The new free parameters for this profile are 
\begin{itemize}
    \item $h_{\rm env}$ - This free parameter decides the depth of the current free neutron star envelope below $R_{\ast}$. The value of the inner radial boundary $R_{b} = R_{\ast} - h_{\rm env}$. Hence, the true height of the accreted mound above $R_{\ast}$ will be given by $h_{\rm rel}=r_{c}-h_{\rm env}$.
    \item $\psi_{\rm ap}$ - This parameter decides the extent of the ring-shaped mound near the pole, below this $\psi$ is the envelope of the neutron star. This could be arbitrarily decided (since its value is dependent on the accretion profile) or it could be calculated from the light cylinder radius. When calculated from light cylinder constraints, for pulsars with periods in the range of $0.001-1$ seconds, the angular extent of the open field lines on the neutron star surface from the pole varies from $27^{0}-0.8^{0}$. We take a conservative estimate of $4^{0}$ for the $10^{12}$ G pulsar assumed in this section.
\end{itemize}
The other new parameters that are dependent are calculated as follows:
\begin{itemize}
    \item $h_{p}$ and $\psi_{\rm bp}$ are calculated by ensuring continuity of $r_{0}(\psi)$ and $\frac{dr_{0}}{d\psi}$ at $\psi_{\rm bp}$.
    \item $h_{t}$ and $\psi_{\rm bt}$ are calculated by ensuring continuity of $r_{0}(\psi)$ and $\frac{dr_{0}}{d\psi}$ at $\psi_{\rm bt}$.
\end{itemize}
The $\psi_{\rm ap}\leq\psi<\psi_{\rm bp}$ and $\psi_{\rm bt}\leq\psi<\psi_{a}$ ranges define smooth transitions from the magnetostatic mound to the current free neutron star envelope. Since we need to have a particular value of initial $B_{d}$ on the pole at $R_{\ast}$, we fix the value at $R_{b}$ to $B_{d}\times(R_{\ast}/R_{b})^{3}$. The gravitational acceleration term $g$ is calculated using $g=\frac{GM_{\ast}}{R_{b}^{2}}$ (the mass of the neutron star is assumed to be unchanged as the envelope mass is relatively negligible). The calculation for $\theta_{t}$ was performed using the new inner boundary $R_{b}$. Figure \ref{transitn} describes the form of the new ring-shaped mound profile with an envelope for a specific set of parameters. 

To demonstrate the effects of this profile on solutions, we choose a $B_{d}=10^{12}$ G and an arbitrary $\theta_{t}=50^{0}$ (similar to the solutions in the lower part of Figure \ref{strange}). Although the value of $\theta_{t}$ is not physically motivated, it helps to demonstrate the nature of the solutions with better $\theta$ resolution and larger mass, since $\theta$ resolution is higher near a higher $\theta_{t}$. All solutions in this subsection have been calculated for a resolution of $5000\times5000$. The accreted mass is calculated after finding the total mass in the simulation domain and subtracting it from the pre-existing envelope mass. This likely underestimates the accreted mass, since in a realistic scenario some of the pre-existing mass will be gravitationally compressed within the neutron star while the remaining mass will be mixed with the incoming accreted mass. We do not model such gravitational compression here.

\subsubsection{Approximating sinking}\label{approx_sink}
\begin{figure}
\vbox{
\includegraphics[width=\columnwidth]{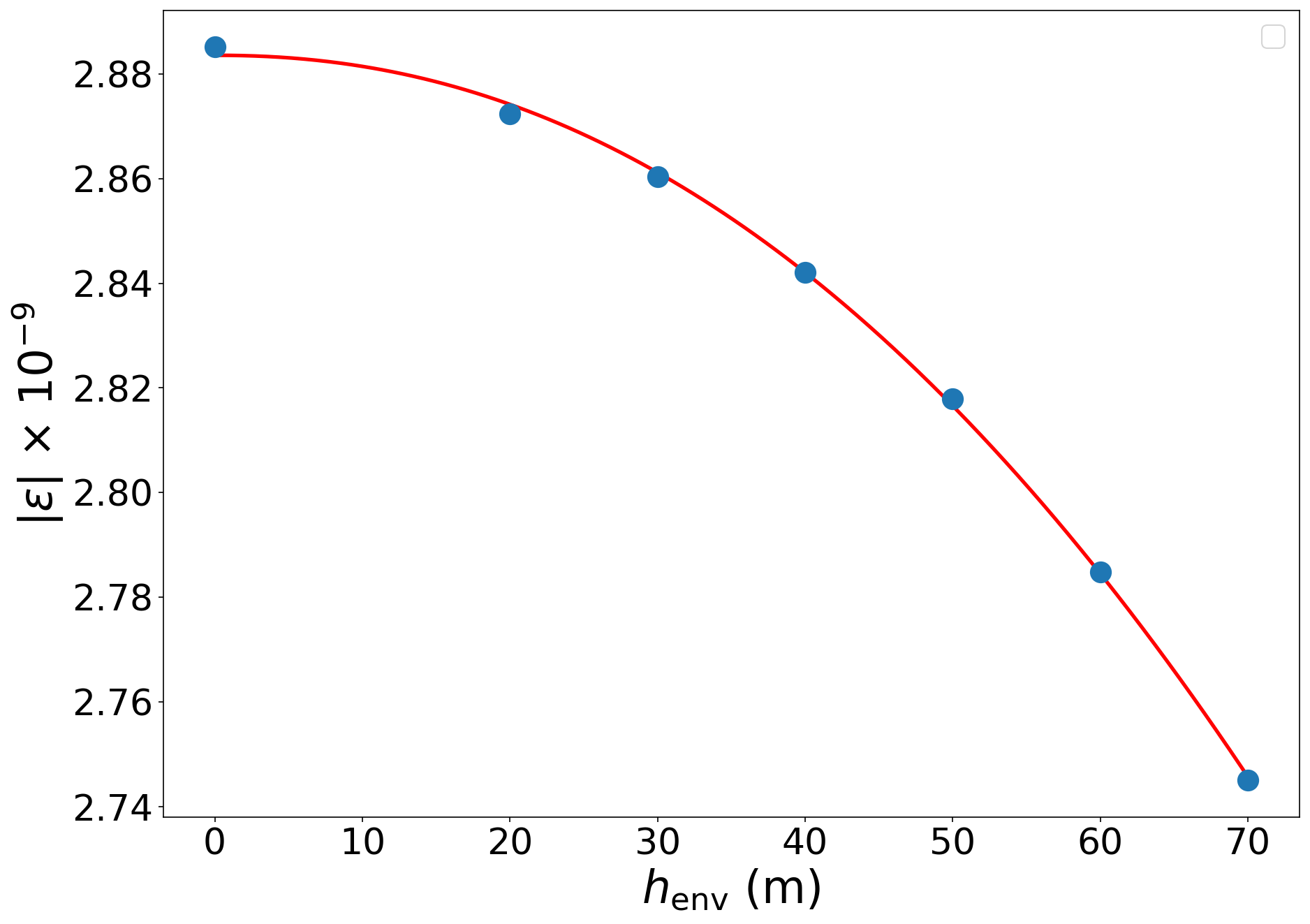}
\includegraphics[width=\columnwidth]{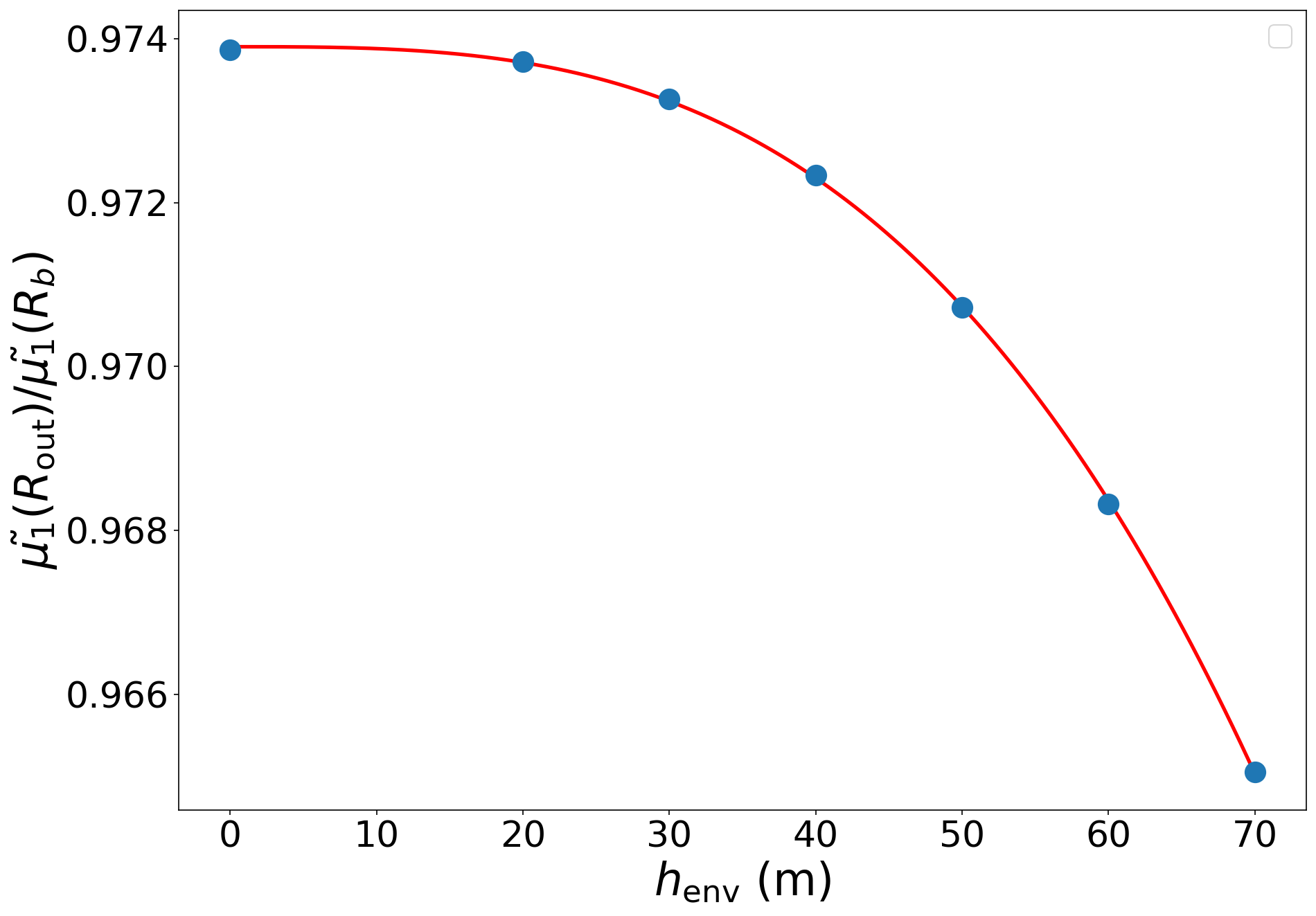}
}
\caption{Absolute Ellipticity versus Envelope height (top plot) and Normalized dipole moment at the outer radius with respect to the surface versus Envelope height (bottom plot). Absolute ellipticity reduces and relative dipole moment decreases with increase in the envelope depth below the surface. Blue points in the plots is fitted to the function $y=ax^{b} + k$ with a red line. Fitting parameters are given in equations \ref{ellfit} and \ref{l1fit}.}
\label{sinkell}
\end{figure}

To understand the effect of sinking the same amount of accreted mass to different depths in the neutron star, we change the values of $h_{\rm env}$ or $h_{\rm rel}=r_{c}-h_{\rm env}$ (maximum height of the mound above the surface $R_{\ast}$). We compute GS solutions for seven different values of $h_{\rm env}$ (seven different $R_{b}$) such that the amount of accreted mass $3.28\times10^{-9}$ M$_{\odot}$ is the same for all runs. The first solution with the same mass and $h_{\rm env}=0.0$ is the pure ring-shaped mound profile without the pre-existing ocean, since for the ring-shaped mound profile $R_{b} = R_{\ast}$. The result for these simulations is noted in Table \ref{sink} and plotted in Figure \ref{sinkell}. Since $h_{\rm env}$ is the envelope depth below the surface $R_{\ast}$, an increase in $h_{\rm env}$ would imply a successive increase in the sinking of the mound. When the mound sinks, ellipticity and dipole moments both are seen to reduce in Figure \ref{sinkell}. Dipole moments are calculated with respect to the bottom radial boundary $R_{b}$, which is different for the seven examples. Considering gravitational compression will further reduce the ellipticity and will increase the value of dipole moments. Both absolute ellipticity and relative dipole moments are fitted with respect to the envelope depth as a red line in Figure \ref{sinkell}. The normalized dipole moment ratios are found to have the behaviour
\begin{equation}\label{l1fit}
    \frac{\tilde{\mu_{1}}(R_{\rm out})}{\tilde{\mu_{1}}(R_{b})} = 0.9739 - (2.02\times10^{-8}\times h^{3.058}_{\rm env})
\end{equation}
where $h_{\rm env}$ is in m, while absolute ellipticity is found to have the behaviour
\begin{equation}\label{ellfit}
    \frac{|\epsilon|}{10^{-9}} = 2.8835-(1.5\times10^{-5}\times h^{2.145}_{\rm env})
\end{equation} 
where $h_{\rm env}$ is in m.

\cite{vigelius2010sinking} performed ideal MHD axisymmetric simulations on a neutron star by growing an accreted mound with an isothermal EOS either on a hard surface or inside a fluid base. For the same accreted mass and the same $dM/d\psi$, they found that the accretion mound grown inside a fluid base reduces the ellipticity by $25-60\%$ relative to the accretion mound grown on a hard surface. They calculated $h_{\rm env}$ by assuming the mass of the isothermal fluid base to be at least ten times the accreted mass. For the Paczynski EOS and the ring-shaped mound on ocean $r_{0}(\psi)$ (not same $dM/d\psi$), if a similar assumption of envelope mass equals ten times accreted mass ($M_{\rm env}=10M_{\rm acc}=3.28\times10^{-8}$ M$_{\odot}$) is made in this work, it is found that $h_{\rm env}=112.6$ m and the absolute mass ellipticity is reduced by $13 \%$ for the solution with this $h_{\rm env}$ and accreted mass relative to the pure ring-shaped mound profile.

\begin{table}
\caption{Radius of the neutron star is fixed at $R_{\ast}=10$ km. For the same accreted mass $3.28\times10^{-9}$ M$_{\odot}$, the values of $h_{\rm env}$ (depth below $R_{\ast}$), $R_{b}$ (radius of inner boundary), $r_{c}$ (mound height above $R_{b}$), absolute ellipticity and dipole moment at the outer radius relative to the inner boundary is tabulated here.}
\label{sink}
\begin{center}
\begin{tabular}{|c|c|c|c|c|}
\hline
 $h_{\rm env}$ (m) & $r_{c}$ (m) & $R_{b}$ (km) & $|\epsilon|\times10^{-9}$ & $\tilde{\mu_{1}}(R_{\rm out})/\tilde{\mu_{1}}(R_{b})$ \\ 
 \hline\hline
$0$ & $90.1$ & $10$ & $2.885$ & $0.9739$ \\
 \hline
$20$ & $90.09$ & $9.98$ & $2.872$ & $0.9737$ \\
 \hline 
$30$ & $90.43$ & $9.97$ & $2.86$ & $0.9733$ \\
  \hline
$40$ & $91.19$ & $9.96$ & $2.842$ & $0.9723$ \\
  \hline  
$50$ & $92.58$ & $9.95$ & $2.818$ & $0.9707$ \\
  \hline
$60$ & $94.78$ & $9.94$ & $2.785$ & $0.968$ \\
  \hline  
$70$ & $97.95$ & $9.93$ & $2.745$ & $0.965$ \\
 \hline\hline
\end{tabular}
\end{center}
\end{table}

\subsubsection{Fixing $R_{b}$}
In a realistic system, it is better to fix the depth at which magnetic fields are frozen from the crystallization properties of the crust. For an OCP (one component plasma), ion lattice melts at a Coulomb coupling parameter (i.e. electrostatic to thermal energy) $\Gamma_{m}$ $\leq$ $175$ \citep{potekhin2000}. The density at which crystallization occurs depends on the temperature and composition of the crust. For an isolated cold neutron star, temperatures are low and matter is composed of iron (depending on the initial formation scenario and accretion process during the proto-neutron star stage) and thus ions crystallize at low densities such as $10^{6}$ g cm$^{-3}$ \citep{Nättilä_2024}. However, an increase in temperature during accretion leads to higher crystallization densities. Also, accreted hydrogen helium matter undergoes burning at densities close to $10^{5}$ g cm$^{-3}$ \citep{brown1998ocean}, adding hydrogen, helium, carbon and certain heavy elements till $Z=26$ to the neutron star matter with certain mass fractions at different column depths which are dependent on the mass accretion rates \citep{brown1998ocean}. This also changes the density at which matter is crystallized. For an accreting neutron star, we assume that crystallization occurs at a density of $10^{9}$ g cm$^{-3}$ \citep{brown1998ocean} and thus fix the magnetic fields at this density. From our chosen EOS and gravity, we can find the density at any prescribed depth assuming a constant height envelope. Thus, we decide to simulate for inner boundary density of $10^{9}$ g$/$cm$^{3}$. 

To understand the change in the solution with composition, we compare a solution with a composition of pure ionized helium matter ($\mu_{e}=2.0$) and a composition of pure ionized iron ($\mu_{e}=2.148$). For a neutron star with radius $R_{\ast}=10$ km, the helium mound has a density of $10^{9}$ g cm$^{-3}$ at the inner radial boundary $R_{b}=9.9074$ km while an iron mound has a density of $10^{9}$ g cm$^{-3}$ at the inner radial boundary $R_{b}=9.9158$ km i.e. at a lower depth. Now, we calculate the solutions for the two mounds with a different composition but with the same accreted mass of $1.85\times10^{-8}$ M$_{\odot}$. We find the absolute ellipticity for the iron mound to be $7.75\times10^{-9}$ which is higher than that of the helium mound : $6.67\times10^{-9}$. Dipole moments are also higher for the iron mound ($0.746$) relative to the helium mound ($0.7$). 

\begin{figure}
\vbox{
\includegraphics[width=\columnwidth]{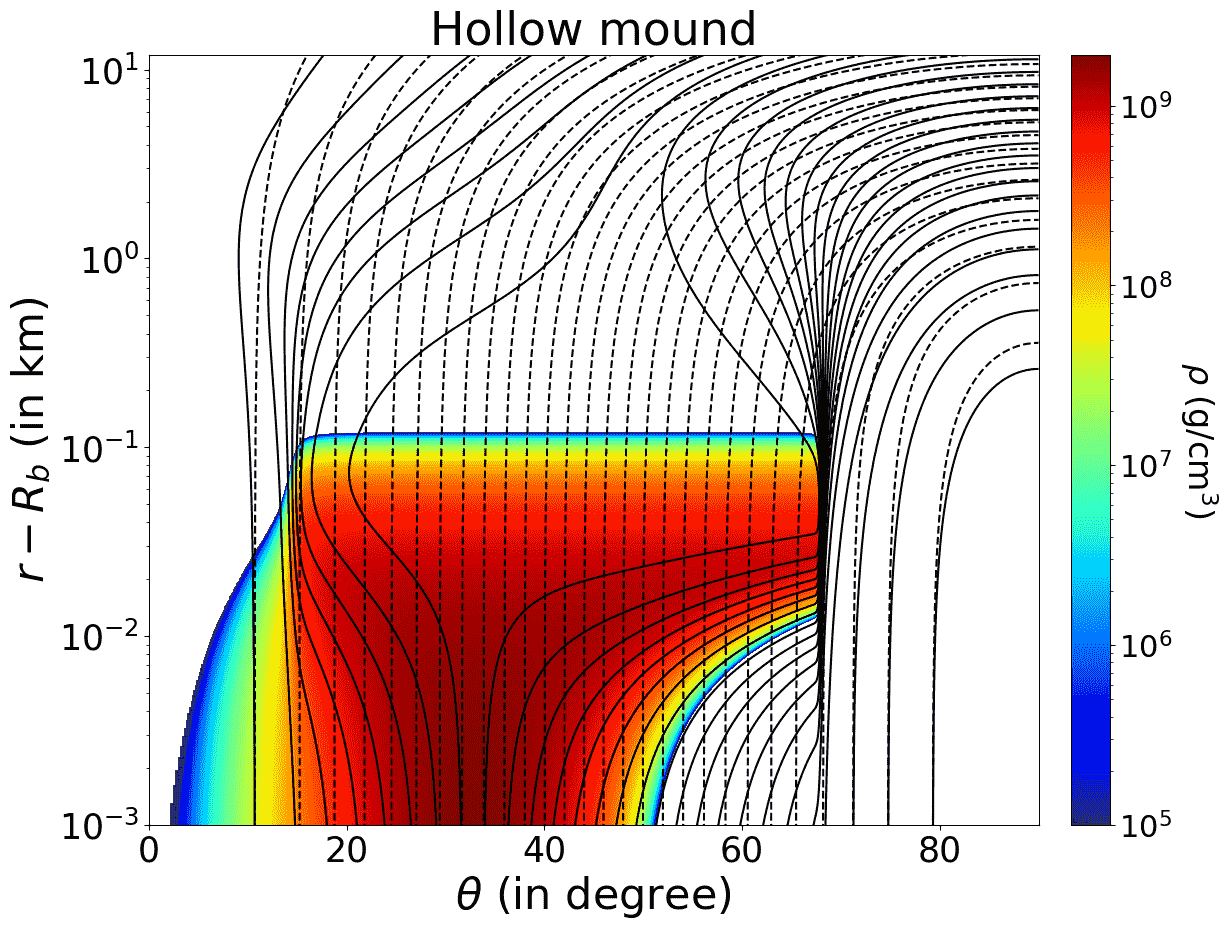}
\includegraphics[width=\columnwidth]{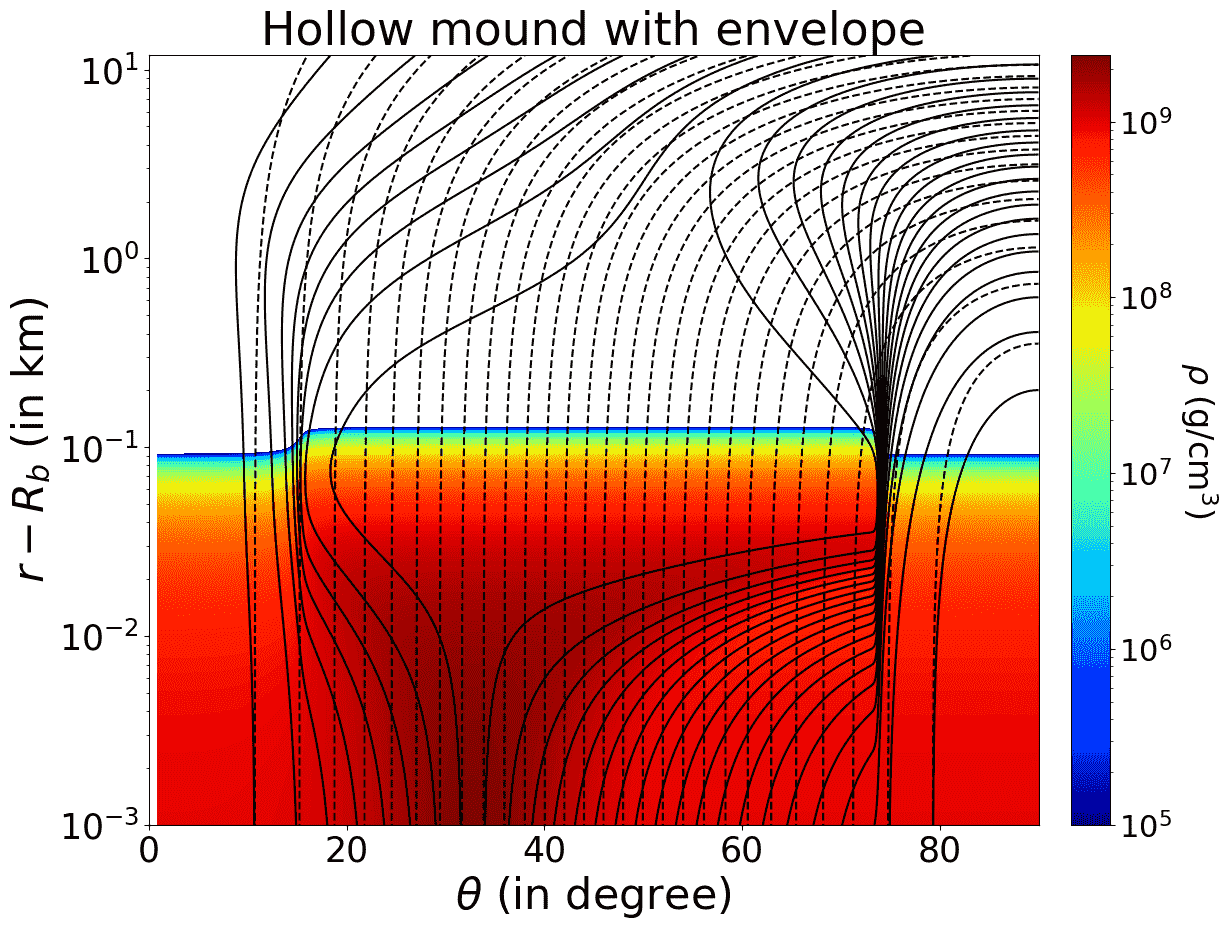}
}
\caption{Density profiles and magnetic field lines (solid) for ring-shaped mound profile with ionized helium composition without envelope (top) and with envelope (middle). Dashed lines are the undistorted dipolar magnetic field lines. Density profile for the ring-shaped mound profile with ionized iron composition and with envelope is plotted at the bottom. Both the solutions have the same $R_{\ast}$, but a different domain inner boundary $R_{b}$. They also have the same amount of accreted mass.}
\label{extrasupp}
\end{figure}

Assuming a helium mound and fixing $R_{b}$ at a density of $10^{9}$ g cm$^{-3}$ for an accreted mass $1.85\times10^{-8}$ M$_{\odot}$, we find a latitudinally spreading mound profile with a pre-existing envelope as plotted in the bottom plot of Figure \ref{extrasupp}. A ring-shaped mound profile with the same accreted mass has been plotted in the top plot of Figure \ref{extrasupp} for comparison. We have already deduced in the previous subsection \ref{approx_sink} that the sinking of the mound from the ring-shaped mound profile to the ring-shaped mound profile inside the neutron star envelope reduces the absolute ellipticities and the dipole moments. We can also observe from Figure \ref{extrasupp} that for the pure ring-shaped mound profile, the mound matter beyond $\theta_{t}$ is supported against gravity by a vacuum region and strong magnetic fields, while for the ring-shaped mound profile with the envelope, the mound matter beyond $\theta_{t}$ is supported against gravity by relatively low density matter and strong magnetic fields. Both the solutions are prone to MHD instabilities \citep{newcomb1961,2004goed}. But, a line-tying boundary condition provides stability to the system \citep{payne2007burial,vigelius2008three}. However, it is difficult to determine the combined effects on the solutions without a non linear evolution. Thus, the dynamical stability should be verified for both the mounds in a future work.

\section{Quadru-dipolar inner boundary for ring-shaped mound profile}\label{SEC4}

Multipolar magnetic fields were invoked to explain significant pair multiplication above the polar cap \citep{ruderman1975}. Recently, there has been strong observational evidence of relatively smaller X-ray thermal emission areas than expected, which have been explained due to the presence of multipolar magnetic fields \citep{Geppert2017,refId0}. However, \cite{prakash2019} has found that this estimation of the polarcap areas could be unreliable. Analysis of observations from NICER also suggests the presence of multipolar magnetic fields \citep{Bilous_2019,Chen_2020,Kalapotharakos_2021,Riley_2021}. Thus, neutron stars may have an intrinsic multipolar magnetic field before accretion from the binary companion begins. To understand the qualitative difference in the solutions relative to an initial dipolar field, we computed the GS solutions for an initial quadru-dipolar field. 

\subsection{Description of the method}\label{quad_method}
As noted in \cite{fujisawa2022magneticallymulti}, quadru-dipolar solution outside neutron star with radius $R_{\ast}$ is
\begin{align}\label{quadip}
    \psi(r,\theta) = \frac{\psi^{\ast} \sin^{2}\theta R_{\ast}}{r} + f_{\mbox{qd}} \frac{\psi^{\ast} R_{\ast}^{2} \sin^{2}\theta \cos{\theta} }{r^{2}}.
\end{align}
Here $\psi^{\ast} = \frac{1}{2}B_{d}R_{\ast}^{2}$ and $f_{\mbox{qd}} = \frac{B_{q}}{B_{d}}$ (surface quadrupolar to dipolar strength fraction). The first term corresponds to a dipole and the second term is the quadrupolar term. For a neutron star with some quadrupolar contribution to the magnetic field, the angular extent of the simulations is from $\theta=0^{0}$ (north magnetic pole) to $\theta=180^{0}$(south magnetic pole) as the magnetic fields are not equatorially symmetric. At both $\theta=0^{0}$ and $\theta=180^{0}$, we assume $\psi=0$. At $r=R_{\rm in}$, a fixed quadru-dipolar field (Equation \ref{quadip}) is assumed. For some quadrupolar contribution to the neutron star magnetic field, current free boundary (CFB) condition equation for the outer radial boundary changes to (as both odd and even components contribute)
\begin{align}
    \psi(R_{\rm out},\theta^{\prime}) = \sum_{\ell=1}^{\ell_{\rm max}} & \left( \frac{(2\ell+1) P_{\ell}^{1}(\cos \theta^{\prime})sin \theta^{\prime}}{2\ell(\ell+1)} \left(\frac{R_{\rm in}}{R_{\rm out}}\right)^{\ell} \right.\nonumber \\
    & \left. \quad \times \int_{-1}^{1} \frac{\psi(R_{\rm in},\cos\theta) P_{\ell}^{1}(\cos \theta)}{\sqrt{1-\cos^{2}\theta}} d(\cos\theta) \right).
\end{align}

The axisymmetric nature of the Grad Shafranov equation only allows us to model the dipolar and quadrupolar components of the field aligned to the magnetic axis with the accretion disk approaching the neutron star at the equator. We assume that the quadrupolar component is aligned parallel to the dipolar component. 

The truncation angle ($\theta_{t}$) calculation for a dipolar field is invalid for a quadru-dipolar field as the Alfv\'en radius differs due to an extra radial field (from the quadrupolar component) at the equator. \cite{cikintoglu} has made an estimation of the value of the Alfv\'en radius ($R_{A^{\ast}}$) for a nonrotating aligned quadru-dipolar neutron star (axisymmetric assumption) by using conservation of angular momentum in steady state. For a particular $B_{q}$,$B_{d}$,$\Dot{M}$ and certain accretion disk parameters $\alpha$,$\varepsilon$, $h/r$, Alfv\'en radius ($R_{A^{\ast}}$) can be calculated from Equation (25) in \cite{cikintoglu} as:
\begin{equation}\label{CKeq}
    \left(\frac{R_{A^{\ast}}}{R_{A}}\right)^{3.5} + \beta \left(\frac{R_{A^{\ast}}}{R_{A}}\right)^{-2} = 1
\end{equation}
where 
\begin{equation*}
    \beta = \frac{27}{\alpha \sqrt{2} \varepsilon^{3.5}} \frac{h}{r} \left(f_{\mbox{qd}}\right)^{2} \frac{R_{\ast}^{2}}{R_{A}^{2}}.
\end{equation*}

Here $R_{A}$ is the Alfv\'en radius for dipolar solution (Equation \ref{RAeq}). We have assumed $\Dot{M}$ here to have the same value as the earlier dipolar simulations above, and $\alpha=0.1$, $h/r=0.1$, $\varepsilon=1.0$ as assumed by \cite{cikintoglu}.

$\theta_{t}$ can be found from the roots of equation below, which is obtained by equating $\psi$ at $\theta=90^{0}$, $r=R_{A^{\ast}}$ (truncated accretion disk) to the $\psi$ at $\theta=\theta_{t}$, $r=R_{\ast}$ (neutron star surface) as follows
\begin{equation}
    -f_{\mbox{qd}}\cos^{3}\theta_{t} - \cos^{2}\theta_{t} + f_{\mbox{qd}}\cos\theta_{t} + 1.0 - \frac{R_{\ast}}{R_{A^{\ast}}} = 0.0.
\end{equation}
The two roots of the equation above provide the values of $\theta_{t}$ in both the hemispheres.

\subsection{Results}\label{quad_results}
\begin{figure*}
\begin{multicols}{3}
    \includegraphics[width=0.9\linewidth]{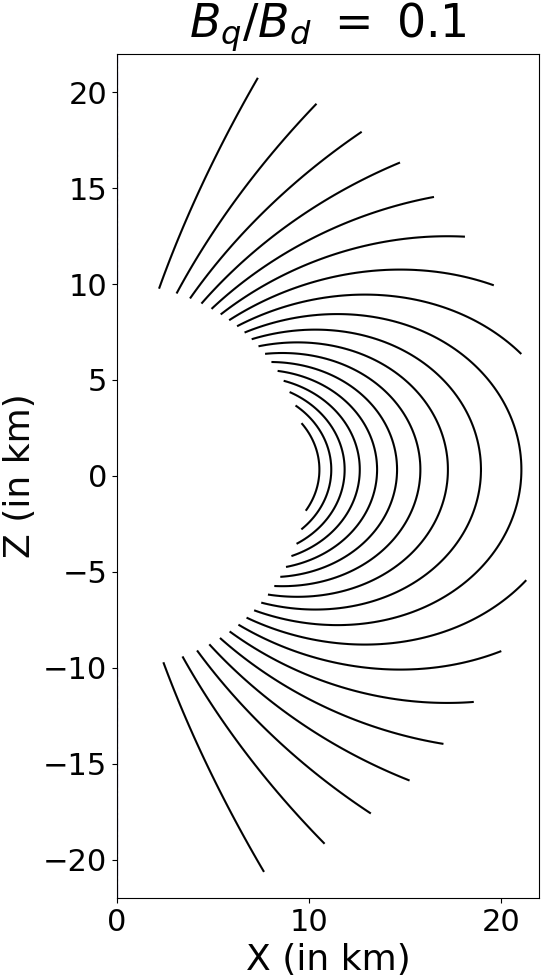}\par 
    \includegraphics[width=0.9\linewidth]{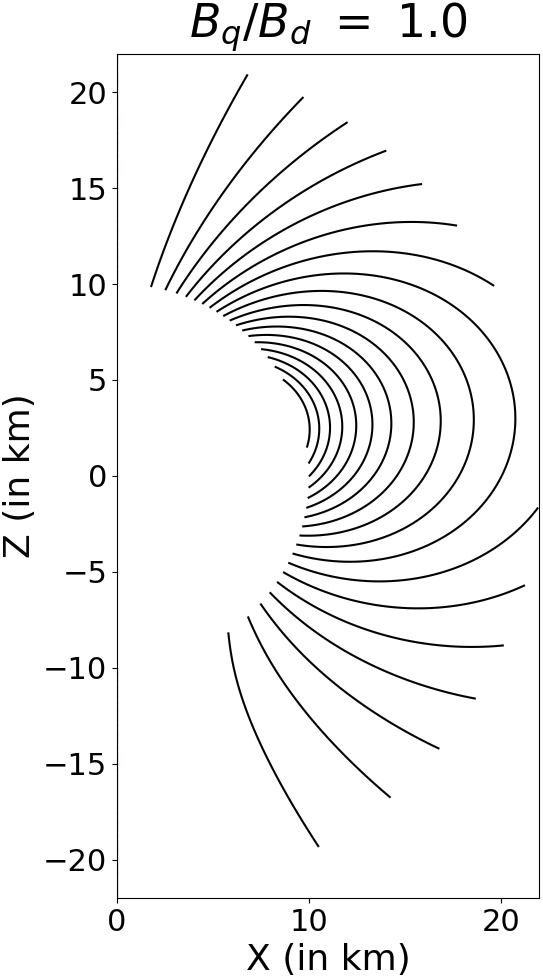}\par 
    \includegraphics[width=0.9\linewidth]{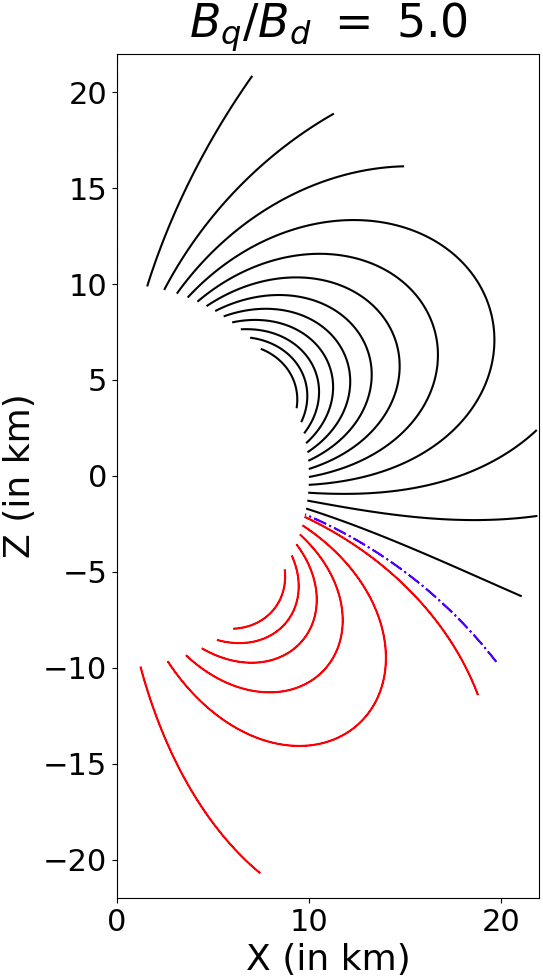}\par
\end{multicols}
\caption{Polar Plot of magnetic field lines for three different surface quadrupolar field to surface dipolar field ratios. Black lines are the magnetic field lines with positive $\psi$, while the red magnetic field lines are opposite to the black lines with negative $\psi$. Blue dotted line is the magnetic field line with $\psi=0$. As $f_{\mbox{qd}}$ increases from left to right, the equatorial asymmetry of the magnetic field geometry increases, while in the rightmost plot the quadrupole component dominates near the surface and thus the radial field is strong near the equator considerably helping the funneling of matter.}
\label{diff_bqbd}
\end{figure*}

For any particular value of $B_{d}$ and $f_{\mbox{qd}}$, we calculate the Alfv\'en radius ($R_{A^{\ast}}$) from eq.~\ref{CKeq} and then subsequently $\theta_{t}$. Beyond a certain $f_{\mbox{qd}}$ (for a particular $B_{d}$ and $\Dot{M}$), no real solutions for $R_{A^{\ast}}$ are found, as the strong quadrupolar field results in matter falling directly on the neutron star surface through the quadrupolar funnel (i.e close to the blue dashdot line in the rightmost plot of Figure \ref{diff_bqbd}). Figure \ref{diff_bqbd} shows the magnetic field lines in polar coordinates for three values of $f_{\mbox{qd}}$. In this work, we choose the orientation of the quadrupolar field such that it lies in the direction of the dipolar field near the Magnetic North pole ($\theta=0^{0}$) (MNP), while it is in the opposite direction to the dipolar field near the Magnetic South pole ($\theta=180^{0}$) (MSP). Thus, the magnetic field is strengthened near the MNP while it gets weakened near the MSP making the field geometry asymmetric about the equator. As $f_{\mbox{qd}}$ increases, the field geometry becomes more asymmetric and eventually the quadrupolar field dominates (as shown in Figure \ref{diff_bqbd}). In the rightmost plot of Figure \ref{diff_bqbd} with $f_{\mbox{qd}}=5$, the black lines are the magnetic field lines with positive $\psi$ i.e. in the direction of the dipolar magnetic field while the red lines are the magnetic field lines with negative $\psi$ i.e. opposite direction to the dipolar magnetic field and in the direction of the quadrupolar field. Since for this case, the accretion disk at the equator is only threaded by the black field lines and not the red lines, matter is accreted near one of the poles and the equator. But after a certain higher $f_{\mbox{qd}}$, the accreted matter will be channeled towards the neutron star surface through the quadrupolar funnel, following the radial field lines connecting to the equatorial region (see the blue dotted line in the rightmost plot of Figure ~\ref{diff_bqbd}). We have used the ring-shaped mound profile for all cases here. We investigated for two values of $B_{d}$, as discussed below.

\subsubsection{$B_{d}=10^{10}$ G}
GS solutions discussed in this section are for grids of size $10000\times5000$. Solutions can only be calculated using this method until $f_{\mbox{qd}}$ of $1.0$ since for $B_{d}=10^{10}$ G and $f_{\mbox{qd}}>1.0$, matter channels directly through the quadrupolar funnel as discussed in the first paragraph of subsection \ref{quad_results}. The parameters and results for the different GS solutions have been noted in Table \ref{tab1}. The solutions have been evolved till the NMM for the current choice of grid size.

\begin{figure*}
\begin{multicols}{2}
    \includegraphics[width=0.95\columnwidth]{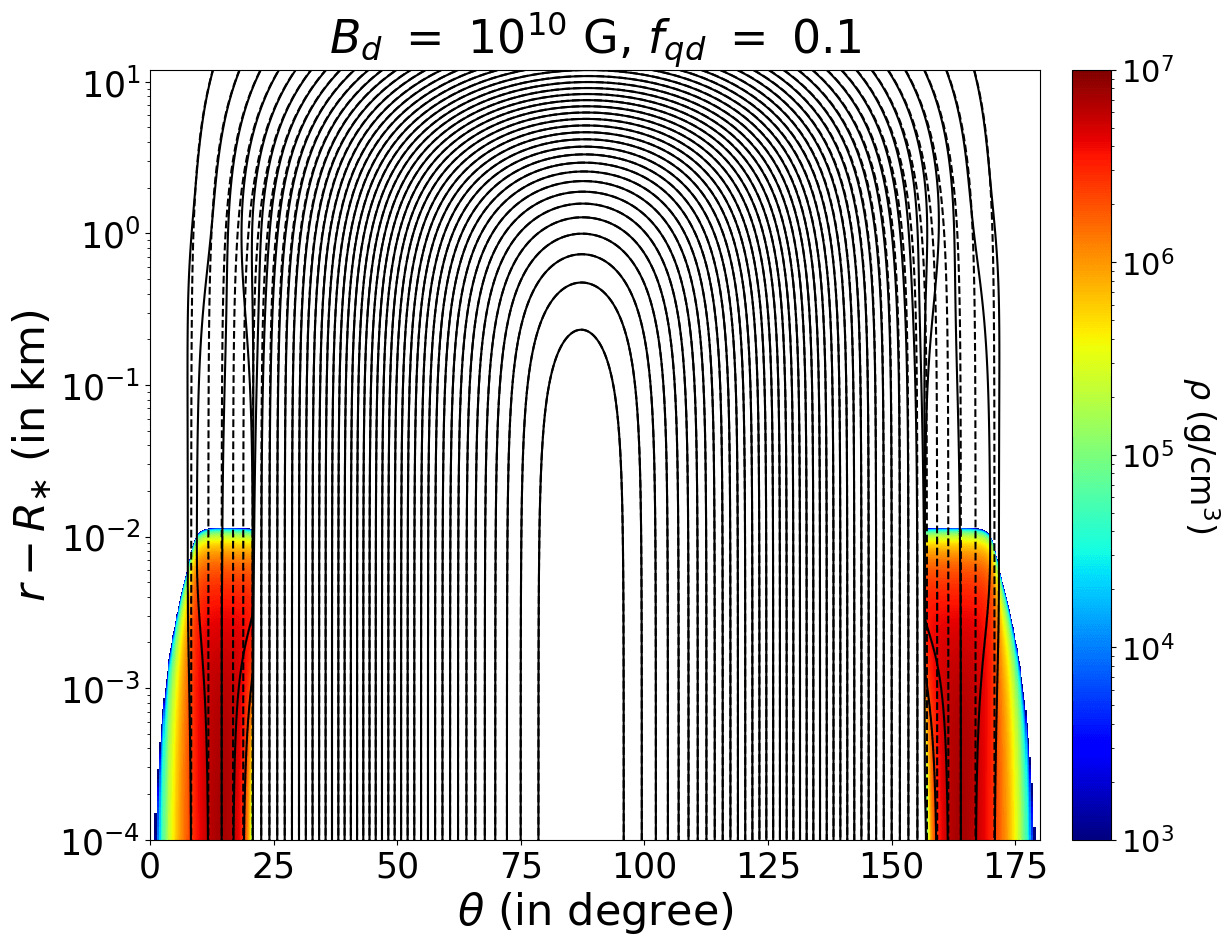}\par
    \includegraphics[width=0.95\columnwidth]{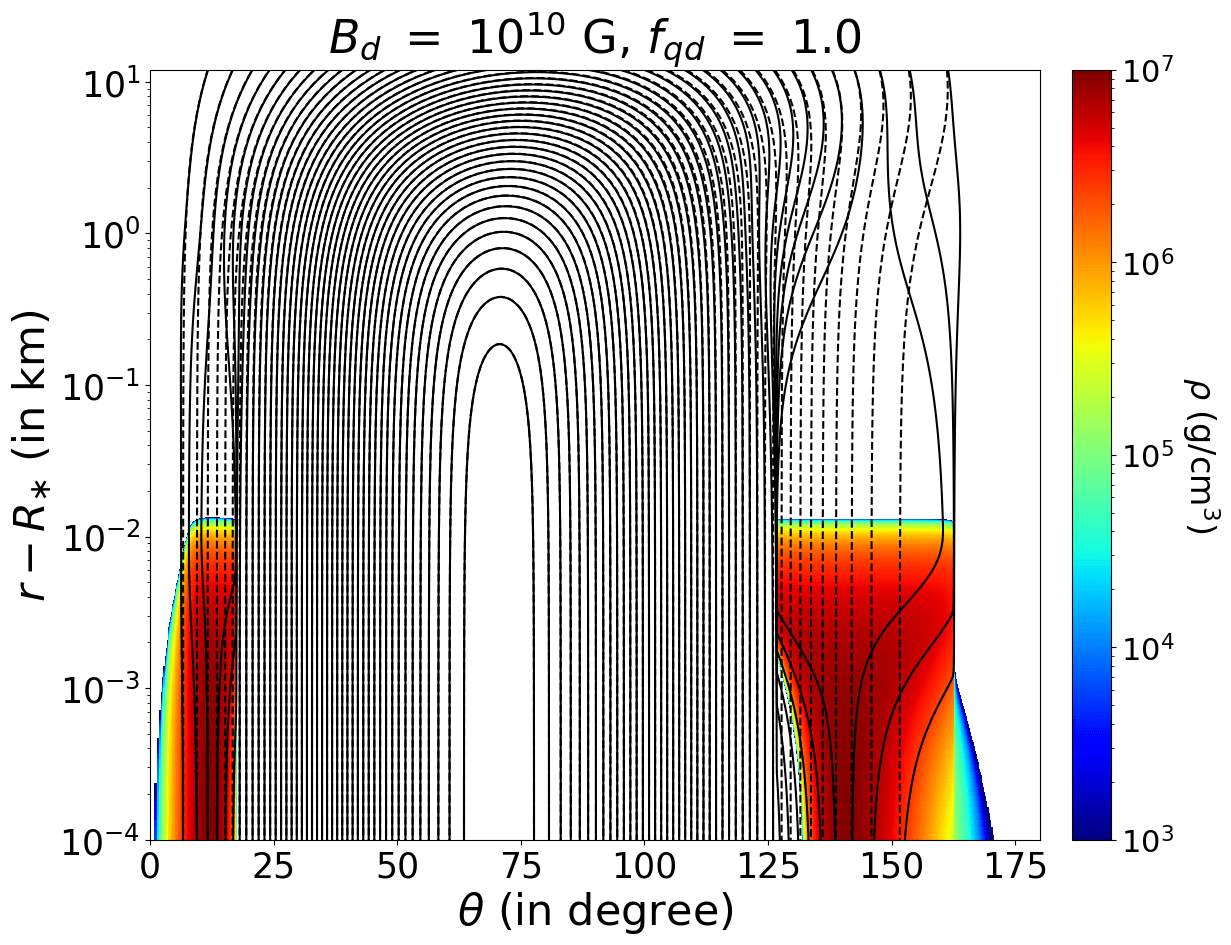}\par 
    \end{multicols}
\caption{ Plot of Density Profile and Magnetic field lines (solid) for parameters $B_{d}$ $=$ $10^{10}$ G and an initial quadrudipolar magnetic field of the neutron star. Plots have been made for Case B (left) and Case D (right). The quadrupole to dipole fractions ($f_{\mbox{qd}}$) have been indicated in the plots. Dashed lines are the initial undistorted quadrudipolar magnetic field lines. Increase in asymmetry due to a larger quadrupolar fraction can lead to a larger angular extent of the mass near one of the magnetic poles relative to the other pole (right plot).}
\label{bd10}
\end{figure*}

For a non zero $f_{\mbox{qd}}$, the angular extent of matter for the mound near the MSP is relatively higher than the angular extent of matter for the mound near MNP (Figure \ref{bd10}). Some of the conclusions from the four cases (Table \ref{tab1}) are as follows:

\begin{itemize}
    \item Case A ($f_{\mbox{qd}}=0.0$) - Two mounds at the two poles are symmetric and have the same mass, height and angular extent.
    \item Case B ($f_{\mbox{qd}}=0.1$) - As a small quadrupolar field is introduced, there is a lower mass mound near the MNP than the MSP, for the same maximum mound heights ($r_c$) at both the poles. Introduction of a quadrupolar field also increases the absolute mass ellipticity and the final heights of the mounds relative to the previous case. 
    \item Case C ($f_{\mbox{qd}}=1.0$, $r_{c}$(MNP)=$r_{c}$(MSP)) - As the surface quadrupolar field is further strengthened, mass near the MNP does not change significantly, but mass near the MSP increases by one order of magnitude relative to case B. Height of the two mounds near MNP and MSP is considered to be the same here (evolved till NMM for only the mound near MSP) as well but it is higher than the mounds in case B. Mass ellipticity for case C is $2.5$ times higher relative to case B.
    \item Case D ($f_{\mbox{qd}}=1.0$, $r_{c}$(MNP)>$r_{c}$(MSP)) - The heights of the mounds could differ near the two poles due to variations in the local mass accretion rates. Assuming a different height for case D, we evolve the solutions till NMM for the mounds near both MNP and MSP. The mass and height of the mound near MNP increase while the mass and height of the mound near MSP are unchanged relative to the previous case. 
\end{itemize}

Plots have been made for Case B and Case D on the left and right respectively in Figure \ref{bd10}. The ratio of normalized dipole moments for cases A, B, C, and D is $0.9989$, $0.9986$, $0.9954$ and $0.9953$ respectively. Changes in the dipole moment ratios are insignificant for these solutions. From these results, we can conclude that increasing the quadrupolar field (to a certain $f_{\mbox{qd}}$) considerably increases the allowed mass of the mounds and also the mass ellipticity. In addition, it is possible to have a different height of the mound near both poles. Considering this may also add to the mass of the mounds and the ellipticity as a result of it. 

\subsubsection{$B_{d}=10^{12}$ G} 

\begin{figure*}
\vbox{
\includegraphics[width=2\columnwidth]{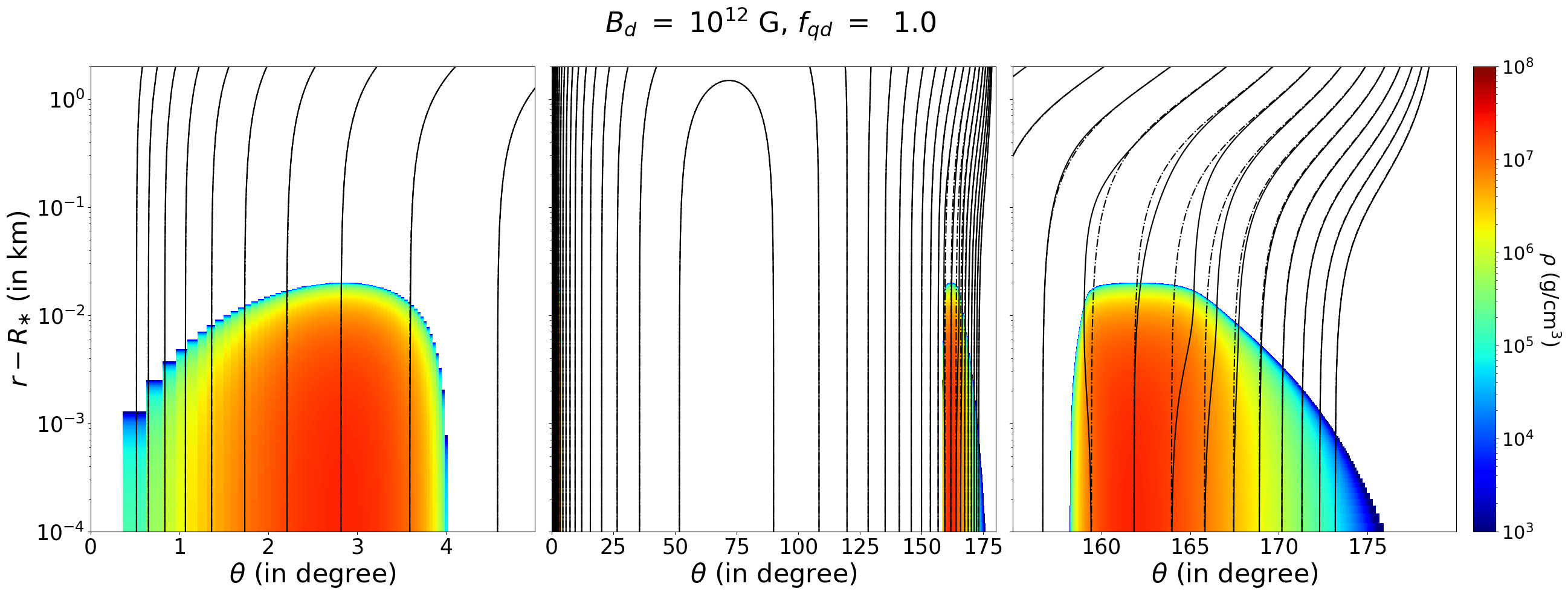} 
\includegraphics[width=2\columnwidth]{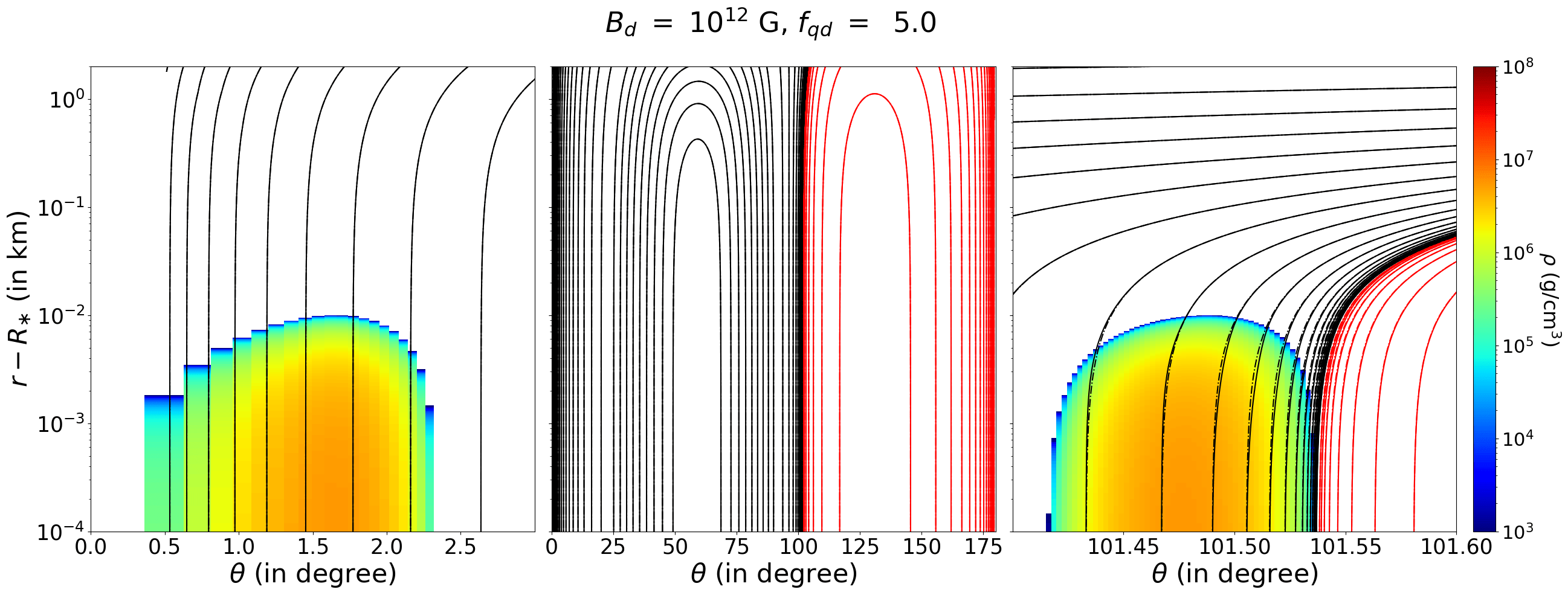}
}
\caption{Plot of Density Profile and Magnetic field lines (solid) for parameters $B_{d}$ $=$ $10^{12}$ G and an initial quadrudipolar magnetic field of the neutron star. Plots have been made for Case E (top middle) and Case F (bottom middle). Red lines are magnetic field lines with negative $\psi$ (opposite magnetic field direction (in the sense of clockwise or anticlockwise) to the black lines). The plots at the left and right of the middle plots are magnified plots of the mounds. Beyond a quadrupolar fraction, the angular extent of the mounds near MNP and equator becomes very narrow (bottom plot).}
\label{bd12}
\end{figure*}

All results described in this section are for simulations with grids: $50000\times2000$ for an $R_{\rm out}-R_{\ast}=2$ km. Solutions have been calculated using the method described in the previous subsection \ref{quad_method} up to a $f_{\mbox{qd}}$ of $5.0$ since for $B_{d}=10^{12}$ G and $f_{\mbox{qd}}>5.0$, matter channels directly through the quadrupolar funnel as discussed in the first paragraph of subsection \ref{quad_results}. Masses have not been calculated here upto NMM, since no significant change is expected in the qualitative output and such small angular extents require high angular resolution and thus high computational costs. The results of two cases E and F for representation have been presented here in Table \ref{tab1} and Figure \ref{bd12} (top and bottom respectively). Following are the inferences from the two cases
\begin{itemize}
    \item Case E ($f_{\mbox{qd}}=1.0$) - The angular extent of the mounds near both the poles are different, similar to Case C, but since the magnetic field is higher here, the angular extent of the mounds is lower for this case relative to Case C. 
    \item Case F ($f_{\mbox{qd}}=5.0$) - The mounds have a very low angular extent here due to the large surface magnetic field relative to the previous case E. The mounds are present near MNP and near the equator (close to $100^{0}$) as the surface quadrupolar component dominates over the surface dipolar component. The mound near the equator has an extremely narrow angular extent (Figure \ref{bd12}), but it has a mass greater than the mound near the pole.
\end{itemize}

\begin{table*}
\caption{Accretion mound results for quadrudipolar inner boundary are tabulated here for two magnetic fields. $f_{\mbox{qd}}$ is the surface quadrupole to dipole fraction. The truncation angle of the mound ($\theta_{t}$), mass of the mound and maximum height of the mound ($r_{c}$) are noted down here for the two mounds near Magnetic North Pole (MNP) and Magnetic South Pole (MSP). $|\epsilon|$ is the absolute ellipticity due to both the mounds.}
\label{tab1}
\begin{tabular}{|c|c|c|c|c|c|c|c|c|c|}
\hline
 \multirow{2}{*}{Sr.No.} & \multirow{2}{*}{$B_{d}$ (G)} & \multirow{2}{*}{$f_{\mbox{qd}}$} & \multirow{2}{*}{$|\epsilon|$ ($10^{-12}$)} & \multicolumn{3}{c}{MNP} & \multicolumn{3}{c}{MSP}  
 \\\cmidrule(lr){5-7} \cmidrule(lr){8-10}
   &  &  &   & $\theta_{t}$ & M($10^{-13}$ M$_{\odot}$) & $r_{c}$ (m) 
           &  $\theta_{t}$ & M($10^{-13}$ M$_{\odot}$) & $r_{c}$ (m) \\
 \cmidrule(lr){1-1} \cmidrule(lr){2-2} \cmidrule(lr){3-3} \cmidrule(lr){4-4} \cmidrule(lr){5-7} \cmidrule(lr){8-10}
  A & $10^{10}$ & $0.0$ & $1.47$ & $21.7^{0}$ & $4.659$ & $11.33$ & $158.3^{0}$ & $4.659$ & $11.33$ \\
  B & $10^{10}$ & $0.1$ & $1.55$ & $20.7^{0}$ & $4.337$ & $11.45$ & $157.17^{0}$ & $5.53$ & $11.45$ \\
  C & $10^{10}$ & $1.0$ & $3.91$ & $17.575^{0}$ & $4.39$ & $13.14$ & $132.36^{0}$ & $42.09$ & $13.14$ \\
  D & $10^{10}$ & $1.0$ & $3.96$ & $17.575^{0}$ & $4.73$ & $13.38$ & $132.36^{0}$ & $42.09$ & $13.14$ \\
  \hline
  E & $10^{12}$ & $1.0$ & $2.3$ & $4.0^{0}$ & $0.55$ & $20$ & $158.22^{0}$ & $14.9$ & $20$ \\
  F & $10^{12}$ & $5.0$ & $8.6\times10^{-4}$ & $2.336^{0}$ & $2.5\times10^{-2}$ & $10$ & $101.42^{0}$ & $6.4\times10^{-2}$ & $10$ \\
 \hline
\end{tabular}
\end{table*}

\section{Summary and Discussion}

In this work, we have solved for the magnetic field geometry of accreted matter on a neutron star with a zero temperature degenerate electron gas EOS (Paczynski EOS) by the non linear partial differential Grad Shafranov equation with zero toroidal field using the numerical relaxation method red black SOR. The numerical method is solved using an MPI framework and a new stretched radial coordinate that allows us to present solutions with improved resolution and an extended numerical domain. We have incorporated a new current free boundary condition at the outer radial boundary. Solutions for three profile functions ($r_{0}(\psi)$) have been discussed in this work. One of the profiles is a mound filled from the pole to the equator, with the profile decaying exponentially towards the equator. The ring-shaped mound profile first introduced in \cite{muk2017revisit} has been considerably used in this work. This profile accretes matter till an angle on the neutron star surface from the pole called the truncation angle $\theta_{t}$. A physically motivated $\theta_{t}$ is the angle at which the magnetic field line from the NS surface connects the truncation radius $r_{t}$ or the inner radius of the accretion disk. $r_{t}$ is calculated here to be $\zeta$ times the Alfv\'en radius $R_{A}$. 

We have explored a wide range of parameter space for the ring-shaped mound profile solutions by varying the maximum height $r_{c}$, magnetic field $B_{d}$ ($10^{9}-10^{12}$ G) and multiple values of $\zeta$ \citep[$\zeta=0.6-1$ motivated from dynamic simulations e.g.][]{parfrey,parfrey17b}. A physically motivated $\theta_{t}$ is used for a single accretion event. However, either cumulative accretion episodes or diffusion of matter over time could lead to matter accumulation till an arbitrarily large $\theta_{t}$. Thus, in certain sections, we have used an arbitrary $\theta_{t}$ motivated by these possibilities. The pure ring-shaped mound profile assumes no pre-existing neutron star atmosphere and ocean. The third profile in this work is called the ring-shaped mound on ocean profile assuming accreted matter mixed into the pre-existing ocean and assuming no gravitational compression. We have attempted to calculate solutions by incorporating the effect of sinking using this profile. For all the profiles, we have calculated the accreted mass, absolute ellipticity and dipole moment at the outer radius relative to the NS surface. The solutions for the above three profiles have been calculated by using a dipolar magnetic field at the inner boundary. We have also calculated solutions for a quadru-dipolar magnetic field at the inner boundary by using only the ring-shaped mound profile and by varying the surface quadrupole to dipole fractions. 

\begin{table}
\caption{For the four magnetic fields and for the ring-shaped mound profile, Numerical Maximum Mass (NMM), time taken to accrete the mound masses ($T_{\mbox{acc}}$), the relative dipole moment $\tilde{\mu_{1}}(R_{\rm out})/\tilde{\mu_{1}}(R_{\ast})$ has been noted down in the table for the case $\zeta=0.6$ (i.e the case with the maximum masses and lowest dipole moments).}
\label{times}
\begin{center}
\begin{tabular}{|c|c|c|c|}
\hline
 $B_{d}$ (G) & NMM ($10^{-12}$ M$_{\odot}$) & $T_{\mbox{acc}}$ (hours) & $\tilde{\mu_{1}}(R_{\rm out})/\tilde{\mu_{1}}(R_{\ast})$ \\ 
 \hline\hline
$10^{9}$ & $20.18$ & $9.11$ & $0.627$ \\
 \hline
$10^{10}$ & $41.92$ & $18.93$ & $0.894$ \\
  \hline
$10^{11}$ & $134.91$ & $60.92$ & $0.991$ \\
  \hline
$10^{12}$ & $214.2$ & $96.72$ & $0.999$ \\
 \hline\hline
\end{tabular}
\end{center}
\end{table}

Some of the important inferences from this work are as follows :

\begin{enumerate}
    \item \textbf{Current Free Boundary (CFB)} : The CFB condition is found to be more efficient than other boundary conditions. We have demonstrated in Figures \ref{smallrad} and \ref{smallraddip} that the solution using the CFB is independent of the chosen domain size unlike the fixed and free boundary.
    \item \textbf{Spreading ring-shaped mound profile solutions} : The ring-shaped mound profile is found to latitudinally spread with an increase in mass. The ring-shaped mound profile does not show closed magnetic loops in the solution beyond a certain mass like the filled mound profiles because magnetic field lines are stretched in both directions i.e. direction of the equator and direction of the poles. The ring-shaped mound profile is more physically motivated than the filled mound since the ring-shaped mound profile populates relatively lower mass on open magnetic field lines near the magnetic pole. Such spreading solutions have significant implications as outlined in the next point.  
    \item \textbf{Implications for field burial} : From the spreading solutions, we have demonstrated the onset of field burial. The efficiency of field burial depends on the extent of the angular spread of the accreted matter, beyond the confinement from $\theta_{t}$ at the base. For the assumed ring-shaped mound profile and a physically motivated $\theta_{t}$, we find the solutions to be non-unique beyond a particular mass we call the Numerical Maximum Mass (NMM) (Appendix \ref{numapp}) and this NMM is limited by resolution. We have evolved all solutions till NMM by considering a single resolution, to evaluate qualitative trends of various physical quantities. Lowest dipole moment values (or maximum field burial) calculated for the four $B_{d}$ values are noted in Table \ref{times}. We show field burial to be effective for the low field pulsars. This is evident for the solution with $B_{d}=10^{9}$ G and $\zeta=0.6$ which has a dipole moment reduction of $37\%$ and has a mound latitudinally spread till the equator. We could not show significant field burial for the high magnetic field pulsars since these solutions have not been explored to a maximum field burial possible i.e they have not been evolved till they latitudinally spread to the equator. Such an endeavour will require larger resolution and grid sizes, to go beyond the current limitations posed by the maximum numerical mass. This will be explored in a future work. 

    Table \ref{times} has noted down the maximum mass found by our solutions for the four magnetic fields and the time taken to accrete these masses. We can observe that we have modelled mounds which have been accreted onto the surface of the neutron star in a few hours to few days i.e the very initial accretion phase of a single outburst event. \cite{Patruno_2012} interpreted the exponential decay of spin frequency derivative from the X-ray data of a pulsar as evidence of magnetic field burial. The timescales required to accrete the masses and the magnetic screening found by our ring-shaped mound solutions provide support to this argument of short term field burial.

    Accretion mound solutions can also be calculated for an arbitrary choice of $\theta_{t}$ which may be the result of cumulative accretion episodes. For example, in subsection \ref{3.4}, we find the lowest dipole moment to be $0.43$ for a solution with $B_{d}=10^{12}$ G and an arbitrary $\theta_{t}=50^{0}$ for a mound latitudinally spread till the equator.
    \item \textbf{Ring-shaped mound on ocean} : For the ring-shaped mound profile, matter at an angle greater than $\theta_{t}$ is supported by vacuum and strong magnetic fields, but for the ring-shaped mound with the envelope, matter at an angle greater than $\theta_{t}$ is supported by a relatively less dense envelope and strong magnetic fields, thus increasing the chance of stability. By approximating sinking using this profile, we find that sinking of the same amount of accreted mass ignoring gravitational compression reduces the absolute mass ellipticity and dipole moments. A change in composition from helium to iron does not cause a significant change in the final solutions. Using this profile, we suggest choosing the inner boundary inside the neutron star to fix the magnetic fields based on the current crystallization properties of the crust in any future work to model a realistic system. The results for the pure ring-shaped mound profile provide upper limits to the mass and absolute ellipticity of the ring-shaped mound on the ocean.
    \item \textbf{Quadru-dipolar magnetic fields} : Due to a quadrupolar field in addition to the dipolar field, the magnetic field gets stronger near one pole and weaker near the other pole (middle plot of Figure \ref{diff_bqbd}). As the quadrupole to dipole fraction further increases, the dipolar magnetic field is strengthened by the quadrupolar field near one pole while the quadrupolar field dominates near another pole (rightmost plot of Figure \ref{diff_bqbd}). From the calculation of angular extents from Alfv\'en radius constraints for the quadru-dipolar magnetic fields, it is found that the angular extent of the mound near one magnetic pole decreases relative to the other magnetic pole. Due to this, asymmetric mounds are accreted with a different mass and different heights near the two magnetic poles. Eventually, beyond a certain large $f_{\mbox{qd}}$ limit ($1.0$ for $B_{d}=10^{10}$ G and $5.0$ for $B_{d}=10^{12}$ G), mass should flow through the quadrupolar funnel (blue dotted line in the rightmost plot of Figure \ref{diff_bqbd}) and these solutions cannot be modelled by the method used in this work. The mass flow through the quadrupolar funnel is depicted in GRMHD simulations of accreting neutron stars in \cite{pushpitadas}. Through preliminary analysis, we find that as the quadrupole field to dipole field fraction increases, the equatorial asymmetry of the magnetic field geometry, the mass and absolute ellipticity of the accretion mounds increase. Instead of using Alfv\'en radius constraints, \cite{fujisawa2022magneticallymulti} populates matter on all the magnetic field lines in both hemispheres. This is unlikely to happen from a single accretion episode, but it is possible for the case of fallback accretion on central compact objects (CCOs). 
\end{enumerate}

Limitations of this work are noted down below
\begin{enumerate}
    \item \textbf{Numerical Maximum Mass (NMM)} : NMM is limited by resolution and thus limits the mass of the solutions presented here (Appendix \ref{numapp}). In a future work, we will attempt to find the maximum mass of the solutions limited by physical restraints by modelling GS solutions on very large grid sizes to accommodate equator-ward spread from a polar cap based accretion profile. Such solutions will utilise the numerical framework established in this paper and the preliminary results presented here, to compute large scale solutions on extended grids.  \
    \item \textbf{No fixed mass-loading per flux tube between different solutions} : Following \cite{mukherjee2012phase,muk2017revisit,fujisawa2022magneticallymulti}, we assume an analytical profile function ($r_{0}(\psi)$) without explicit constraints on the form of mass distribution per flux tube ($dM/d\psi$). Magnetostatic solutions are obtained by a numerical relaxation method for an initially unknown particular $dM/d\psi$. One of the limitations in using such an analytical profile function $r_{0}(\psi)$ is that two different GS solutions, such as with different maximum height ($r_{c}$), will have different mass loading profile per field line ($dM/d\psi$). But, it does not mean that the solutions found here are incorrect. Each solution gives a possible equilibrium configuration for a given mass, although with a different $dM/d\psi$. In future, if a long-term burial is to be modelled, the solution framework needs to be adapted to conserve the mass-loading at different iteration steps \citep[e.g.][etc.]{payne2004burial,priymak2011quadrupole}. 
    \item \textbf{Artificial treatment of mixing and sinking} : Through axisymmetric MHD simulations, \cite{vigelius2010sinking} found that sinking reduces the ellipticity by $25-60 \%$ for an isothermal EOS. For the Paczynski EOS, we find that sinking reduces ellipticity, but not as significantly as the above work. But, we ignore gravitational compression and we have not treated sinking and mixing of matter dynamically. This problem must be explored dynamically. 
    \item \textbf{No GR effects} : \cite{rossetto23} numerically solved the general relativistic formulation of the Grad Shafranov equation (GRGS) for an isothermal EOS. After comparing the dimensionless dipole moment, they found that the GRGS solutions show approximately three times less screening than the Newtonian GS solutions. In addition, the characteristic scale height of the mound is reduced by $40\%$ for GRGS solutions for an isothermal EOS. The current work explores the efficacy of the CFB for a Newtonian framework for ease of applicability. For any future work, GS formalism and the CFB condition should be extended to a GR framework for a better self-consistent treatment of the problem.
    \item \textbf{Stability of the mounds} : The stability of accretion mounds has been probed by various authors \citep{payne2007burial,vigelius2008three,vigelius2009resistive} through MHD simulations, and they find axisymmetric stability and non axisymmetric modes making the mounds unstable. \cite{mukherjee2013mhd2d} performed axisymmetric 2d mhd simulations of a section of the mound and found that the mounds are stable to interchange instabilities when perturbed but increasing the mass of the mound destabilizes the mound. \cite{mukherjee2013mhd3d} performed 3D MHD simulations of the same in PLUTO and found multiple radially
    elongated streams in the azimuthal direction due to pressure-driven toroidal instabilities. \cite{Kulsrud_Sunyaev_2020}, through an analytical analysis, suggests that an efficient mechanism based on a strong ideal Schwarzschild instability will be responsible for mass flow across flux surfaces or magnetic field lines (instability creates a cascade of eddies down to resistive scales at which the mass flow occurs). Although the above results predict and demonstrate the onset of MHD instabilities, the nature of the solutions in the non-linear saturation regime is not well known. \cite{vigelius2009a} through MHD analysis predict that a non-axisymmetric equilibria may be achieved starting from the unstable GS solutions. The stability of the mound profiles described in this work needs to be explored using dynamical simulations. Non-ideal effects such as resistivity \citep{vigelius2009resistive} and thermal conduction \citep{suvorov2019relaxation} should also be included in such simulations. 
    \item \textbf{Mass accretion rate} : The mass accretion rate has been assigned to an arbitrarily high value here, which is the same for all $B_{d}$. This may not be true, and the values of the mass accretion rate should be motivated in future simulations by observations.
    \item \textbf{No toroidal field} : \cite{fujisawa2022magneticallymulti} found that a large toroidal field can be present inside the mounds, although this does not significantly affect the GS solutions. There will be an attempt in the future to check the effects of toroidal fields on our solutions.
\end{enumerate}

\section*{Acknowledgements}

The authors are thankful to the anonymous referee for the thorough review and detailed suggestions that helped improve the clarity and presentation of the article. We gratefully acknowledge the use
of high performance computing facilities at IUCAA, Pune\footnote{\href{http://hpc.iucaa.in}{http://hpc.iucaa.in}}. DM thanks Yuri Levin for a discussion on the idea of a current free boundary condition. The authors thank Ashwin Devaraj for his useful input on our plots. The authors also thank Prathamesh Ratnaparkhi for useful discussions on finding NMM.

\section*{Data availability}

No new data was generated in support of this research. The simulations used in this work are available from the corresponding authors upon reasonable request.



\def\apj{ApJ}%
\def\mnras{MNRAS}%
\def\aap{A\&A}%
\def\apjl{ApJL}
\def\physrep{PhR}
\def\apjs{ApJS}
\def\pasa{PASA}
\def\pasj{PASJ}
\def\nat{Nature}
\def\memsai{MmSAI}
\def\aj{AJ}%
\def\aaps{A\&AS}%
\def\iaucirc{IAU~Circ.}%
\def\sovast{Soviet~Ast.}%
\def\apss{Ap\&SS}
\def\aplett{ApJL}
\def\aapr{Astron Astrophys Rev}
\def\pasp{PASP}
\def\araa{ARAA}%
\def\nar{New Astronomy Reviews}

\bibliographystyle{mnras}
\bibliography{mnras_main} 

\begin{thebibliography}{}
\makeatletter
\relax
\def\mn@urlcharsother{\let\do\@makeother \do\$\do\&\do\#\do\^\do\_\do\%\do\~}
\def\mn@doi{\begingroup\mn@urlcharsother \@ifnextchar [ {\mn@doi@} {\mn@doi@[]}}
\def\mn@doi@[#1]#2{\def\@tempa{#1}\ifx\@tempa\@empty \href {http://dx.doi.org/#2} {doi:#2}\else \href {http://dx.doi.org/#2} {#1}\fi \endgroup}
\def\mn@eprint#1#2{\mn@eprint@#1:#2::\@nil}
\def\mn@eprint@arXiv#1{\href {http://arxiv.org/abs/#1} {{\tt arXiv:#1}}}
\def\mn@eprint@dblp#1{\href {http://dblp.uni-trier.de/rec/bibtex/#1.xml} {dblp:#1}}
\def\mn@eprint@#1:#2:#3:#4\@nil{\def\@tempa {#1}\def\@tempb {#2}\def\@tempc {#3}\ifx \@tempc \@empty \let \@tempc \@tempb \let \@tempb \@tempa \fi \ifx \@tempb \@empty \def\@tempb {arXiv}\fi \@ifundefined {mn@eprint@\@tempb}{\@tempb:\@tempc}{\expandafter \expandafter \csname mn@eprint@\@tempb\endcsname \expandafter{\@tempc}}}

\bibitem[\protect\citeauthoryear{{Alford} \& {Schwenzer}}{{Alford} \& {Schwenzer}}{2014}]{alford2014}
{Alford} M.~G.,  {Schwenzer} K.,  2014, \mn@doi [\apj] {10.1088/0004-637X/781/1/26}, \href {https://ui.adsabs.harvard.edu/abs/2014ApJ...781...26A} {781, 26}

\bibitem[\protect\citeauthoryear{{Alpar}, {Cheng}, {Ruderman}  \& {Shaham}}{{Alpar} et~al.}{1982}]{Alpar1982}
{Alpar} M.~A.,  {Cheng} A.~F.,  {Ruderman} M.~A.,   {Shaham} J.,  1982, \mn@doi [\nat] {10.1038/300728a0}, \href {https://ui.adsabs.harvard.edu/abs/1982Natur.300..728A} {300, 728}

\bibitem[\protect\citeauthoryear{{Arons} \& {Lea}}{{Arons} \& {Lea}}{1976a}]{aronslea1976a}
{Arons} J.,  {Lea} S.~M.,  1976a, \mn@doi [\apj] {10.1086/154562}, \href {https://ui.adsabs.harvard.edu/abs/1976ApJ...207..914A} {207, 914}

\bibitem[\protect\citeauthoryear{{Arons} \& {Lea}}{{Arons} \& {Lea}}{1976b}]{aronslea1976b}
{Arons} J.,  {Lea} S.~M.,  1976b, \mn@doi [\apj] {10.1086/154888}, \href {https://ui.adsabs.harvard.edu/abs/1976ApJ...210..792A} {210, 792}

\bibitem[\protect\citeauthoryear{{Arons} \& {Lea}}{{Arons} \& {Lea}}{1980}]{aronslea1980}
{Arons} J.,  {Lea} S.~M.,  1980, \mn@doi [\apj] {10.1086/157706}, \href {https://ui.adsabs.harvard.edu/abs/1980ApJ...235.1016A} {235, 1016}

\bibitem[\protect\citeauthoryear{{Arumugasamy} \& {Mitra}}{{Arumugasamy} \& {Mitra}}{2019}]{prakash2019}
{Arumugasamy} P.,  {Mitra} D.,  2019, \mn@doi [\mnras] {10.1093/mnras/stz2299}, \href {https://ui.adsabs.harvard.edu/abs/2019MNRAS.489.4589A} {489, 4589}

\bibitem[\protect\citeauthoryear{{Baym}, {Pethick}, {Pines}  \& {Ruderman}}{{Baym} et~al.}{1969}]{baym1969}
{Baym} G.,  {Pethick} C.,  {Pines} D.,   {Ruderman} M.,  1969, \mn@doi [\nat] {10.1038/224872a0}, \href {https://ui.adsabs.harvard.edu/abs/1969Natur.224..872B} {224, 872}

\bibitem[\protect\citeauthoryear{{Becker} \& {Wolff}}{{Becker} \& {Wolff}}{2007}]{Becker_2007}
{Becker} P.~A.,  {Wolff} M.~T.,  2007, \mn@doi [\apj] {10.1086/509108}, \href {https://ui.adsabs.harvard.edu/abs/2007ApJ...654..435B} {654, 435}

\bibitem[\protect\citeauthoryear{{Bessolaz}, {Zanni}, {Ferreira}, {Keppens}  \& {Bouvier}}{{Bessolaz} et~al.}{2008}]{bessolaz2008}
{Bessolaz} N.,  {Zanni} C.,  {Ferreira} J.,  {Keppens} R.,   {Bouvier} J.,  2008, \mn@doi [\aap] {10.1051/0004-6361:20078328}, \href {https://ui.adsabs.harvard.edu/abs/2008A&A...478..155B} {478, 155}

\bibitem[\protect\citeauthoryear{{Bildsten}}{{Bildsten}}{1998}]{bildsten1998}
{Bildsten} L.,  1998, \mn@doi [\apjl] {10.1086/311440}, \href {https://ui.adsabs.harvard.edu/abs/1998ApJ...501L..89B} {501, L89}

\bibitem[\protect\citeauthoryear{{Bilous} et~al.,}{{Bilous} et~al.}{2019}]{Bilous_2019}
{Bilous} A.~V.,  et~al., 2019, \mn@doi [\apjl] {10.3847/2041-8213/ab53e7}, \href {https://ui.adsabs.harvard.edu/abs/2019ApJ...887L..23B} {887, L23}

\bibitem[\protect\citeauthoryear{{Bisnovatyi-Kogan} \& {Komberg}}{{Bisnovatyi-Kogan} \& {Komberg}}{1974}]{1974bisnov}
{Bisnovatyi-Kogan} G.~S.,  {Komberg} B.~V.,  1974, \sovast, \href {https://ui.adsabs.harvard.edu/abs/1974SvA....18..217B} {18, 217}

\bibitem[\protect\citeauthoryear{{Blandford}, {Decampli}  \& {Konigl}}{{Blandford} et~al.}{1979}]{1979blandford}
{Blandford} R.~D.,  {Decampli} W.~M.,   {Konigl} A.,  1979, in Bulletin of the American Astronomical Society. p.~703

\bibitem[\protect\citeauthoryear{{Blondin} \& {Freese}}{{Blondin} \& {Freese}}{1986}]{Blondin1986}
{Blondin} J.~M.,  {Freese} K.,  1986, \mn@doi [\nat] {10.1038/323786a0}, \href {https://ui.adsabs.harvard.edu/abs/1986Natur.323..786B} {323, 786}

\bibitem[\protect\citeauthoryear{{Bonazzola} \& {Gourgoulhon}}{{Bonazzola} \& {Gourgoulhon}}{1996}]{bonazzola1996gravitational}
{Bonazzola} S.,  {Gourgoulhon} E.,  1996, \mn@doi [\aap] {10.48550/arXiv.astro-ph/9602107}, \href {https://ui.adsabs.harvard.edu/abs/1996A&A...312..675B} {312, 675}

\bibitem[\protect\citeauthoryear{{Brown} \& {Bildsten}}{{Brown} \& {Bildsten}}{1998}]{brown1998ocean}
{Brown} E.~F.,  {Bildsten} L.,  1998, \mn@doi [\apj] {10.1086/305419}, \href {https://ui.adsabs.harvard.edu/abs/1998ApJ...496..915B} {496, 915}

\bibitem[\protect\citeauthoryear{{Caiazzo} \& {Heyl}}{{Caiazzo} \& {Heyl}}{2021}]{caiazzo_2021}
{Caiazzo} I.,  {Heyl} J.,  2021, \mn@doi [\mnras] {10.1093/mnras/staa3428}, \href {https://ui.adsabs.harvard.edu/abs/2021MNRAS.501..109C} {501, 109}

\bibitem[\protect\citeauthoryear{{Chandrasekhar}}{{Chandrasekhar}}{1970}]{chandrashekhar1970}
{Chandrasekhar} S.,  1970, \mn@doi [\prl] {10.1103/PhysRevLett.24.611}, \href {https://ui.adsabs.harvard.edu/abs/1970PhRvL..24..611C} {24, 611}

\bibitem[\protect\citeauthoryear{{Chatterjee}, {Novak}  \& {Oertel}}{{Chatterjee} et~al.}{2021}]{chatterjee2021}
{Chatterjee} D.,  {Novak} J.,   {Oertel} M.,  2021, \mn@doi [European Physical Journal A] {10.1140/epja/s10050-021-00525-5}, \href {https://ui.adsabs.harvard.edu/abs/2021EPJA...57..249C} {57, 249}

\bibitem[\protect\citeauthoryear{{Chen}, {Yuan}  \& {Vasilopoulos}}{{Chen} et~al.}{2020}]{Chen_2020}
{Chen} A.~Y.,  {Yuan} Y.,   {Vasilopoulos} G.,  2020, \mn@doi [\apjl] {10.3847/2041-8213/ab85c5}, \href {https://ui.adsabs.harvard.edu/abs/2020ApJ...893L..38C} {893, L38}

\bibitem[\protect\citeauthoryear{{Choudhuri} \& {Konar}}{{Choudhuri} \& {Konar}}{2002}]{konar2002}
{Choudhuri} A.~R.,  {Konar} S.,  2002, \mn@doi [\mnras] {10.1046/j.1365-8711.2002.05362.x}, \href {https://ui.adsabs.harvard.edu/abs/2002MNRAS.332..933C} {332, 933}

\bibitem[\protect\citeauthoryear{{Coburn}, {Heindl}, {Rothschild}, {Gruber}, {Kreykenbohm}, {Wilms}, {Kretschmar}  \& {Staubert}}{{Coburn} et~al.}{2002}]{Coburn_2002}
{Coburn} W.,  {Heindl} W.~A.,  {Rothschild} R.~E.,  {Gruber} D.~E.,  {Kreykenbohm} I.,  {Wilms} J.,  {Kretschmar} P.,   {Staubert} R.,  2002, \mn@doi [\apj] {10.1086/343033}, \href {https://ui.adsabs.harvard.edu/abs/2002ApJ...580..394C} {580, 394}

\bibitem[\protect\citeauthoryear{{Cumming}, {Zweibel}  \& {Bildsten}}{{Cumming} et~al.}{2001}]{cumming2001magnetic}
{Cumming} A.,  {Zweibel} E.,   {Bildsten} L.,  2001, \mn@doi [\apj] {10.1086/321658}, \href {https://ui.adsabs.harvard.edu/abs/2001ApJ...557..958C} {557, 958}

\bibitem[\protect\citeauthoryear{{Cutler}}{{Cutler}}{2002}]{cutler2002}
{Cutler} C.,  2002, \mn@doi [\prd] {10.1103/PhysRevD.66.084025}, \href {https://ui.adsabs.harvard.edu/abs/2002PhRvD..66h4025C} {66, 084025}

\bibitem[\protect\citeauthoryear{{Das}, {Porth}  \& {Watts}}{{Das} et~al.}{2022}]{pushpitadas}
{Das} P.,  {Porth} O.,   {Watts} A.~L.,  2022, \mn@doi [\mnras] {10.1093/mnras/stac1817}, \href {https://ui.adsabs.harvard.edu/abs/2022MNRAS.515.3144D} {515, 3144}

\bibitem[\protect\citeauthoryear{{Fattoyev}, {Horowitz}  \& {Lu}}{{Fattoyev} et~al.}{2018}]{fattoyev2018}
{Fattoyev} F.~J.,  {Horowitz} C.~J.,   {Lu} H.,  2018, \mn@doi [arXiv e-prints] {10.48550/arXiv.1804.04952}, \href {https://ui.adsabs.harvard.edu/abs/2018arXiv180404952F} {p. arXiv:1804.04952}

\bibitem[\protect\citeauthoryear{{Friedman} \& {Schutz}}{{Friedman} \& {Schutz}}{1978}]{friedman1978}
{Friedman} J.~L.,  {Schutz} B.~F.,  1978, \mn@doi [\apj] {10.1086/156143}, \href {https://ui.adsabs.harvard.edu/abs/1978ApJ...222..281F} {222, 281}

\bibitem[\protect\citeauthoryear{{Fujisawa}, {Kisaka}  \& {Kojima}}{{Fujisawa} et~al.}{2022}]{fujisawa2022magneticallymulti}
{Fujisawa} K.,  {Kisaka} S.,   {Kojima} Y.,  2022, \mn@doi [\mnras] {10.1093/mnras/stac2585}, \href {https://ui.adsabs.harvard.edu/abs/2022MNRAS.516.5196F} {516, 5196}

\bibitem[\protect\citeauthoryear{{Geppert}}{{Geppert}}{2017}]{Geppert2017}
{Geppert} U.,  2017, \mn@doi [Journal of Astrophysics and Astronomy] {10.1007/s12036-017-9460-y}, \href {https://ui.adsabs.harvard.edu/abs/2017JApA...38...46G} {38, 46}

\bibitem[\protect\citeauthoryear{{Giliberti} \& {Cambiotti}}{{Giliberti} \& {Cambiotti}}{2022}]{giliberti2022}
{Giliberti} E.,  {Cambiotti} G.,  2022, \mn@doi [\mnras] {10.1093/mnras/stac245}, \href {https://ui.adsabs.harvard.edu/abs/2022MNRAS.511.3365G} {511, 3365}

\bibitem[\protect\citeauthoryear{{Goedbloed} \& {Poedts}}{{Goedbloed} \& {Poedts}}{2004}]{2004goed}
{Goedbloed} J.~P.~H.,  {Poedts} S.,  2004, {Principles of Magnetohydrodynamics}

\bibitem[\protect\citeauthoryear{{Haensel}, {Potekhin}  \& {Yakovlev}}{{Haensel} et~al.}{2007}]{haensel2007neutron}
{Haensel} P.,  {Potekhin} A.~Y.,   {Yakovlev} D.~G.,  2007, {Neutron Stars 1 : Equation of State and Structure}.
~ Vol. 326, Springer New York, NY

\bibitem[\protect\citeauthoryear{{Hameury}, {Bonazzola}, {Heyvaerts}  \& {Lasota}}{{Hameury} et~al.}{1983}]{hameury1983magnetohydrostatics}
{Hameury} J.~M.,  {Bonazzola} S.,  {Heyvaerts} J.,   {Lasota} J.~P.,  1983, \aap, \href {https://ui.adsabs.harvard.edu/abs/1983A&A...128..369H} {128, 369}

\bibitem[\protect\citeauthoryear{{Haskell}, {Antonelli}  \& {Pizzochero}}{{Haskell} et~al.}{2022}]{haskell2022}
{Haskell} B.,  {Antonelli} M.,   {Pizzochero} P.,  2022, \mn@doi [Universe] {10.3390/universe8120619}, \href {https://ui.adsabs.harvard.edu/abs/2022Univ....8..619H} {8, 619}

\bibitem[\protect\citeauthoryear{{Istomin} \& {Semerikov}}{{Istomin} \& {Semerikov}}{2016}]{istomin2016}
{Istomin} Y.~N.,  {Semerikov} I.~A.,  2016, \mn@doi [\mnras] {10.1093/mnras/stv2473}, \href {https://ui.adsabs.harvard.edu/abs/2016MNRAS.455.1938I} {455, 1938}

\bibitem[\protect\citeauthoryear{{Jahan Miri} \& {Bhattacharya}}{{Jahan Miri} \& {Bhattacharya}}{1994}]{jahanmiri}
{Jahan Miri} M.,  {Bhattacharya} D.,  1994, \mn@doi [\mnras] {10.1093/mnras/269.2.455}, \href {https://ui.adsabs.harvard.edu/abs/1994MNRAS.269..455J} {269, 455}

\bibitem[\protect\citeauthoryear{{Jones}}{{Jones}}{2010}]{jones2010}
{Jones} D.~I.,  2010, \mn@doi [\mnras] {10.1111/j.1365-2966.2009.16059.x}, \href {https://ui.adsabs.harvard.edu/abs/2010MNRAS.402.2503J} {402, 2503}

\bibitem[\protect\citeauthoryear{{Kalapotharakos}, {Wadiasingh}, {Harding}  \& {Kazanas}}{{Kalapotharakos} et~al.}{2021}]{Kalapotharakos_2021}
{Kalapotharakos} C.,  {Wadiasingh} Z.,  {Harding} A.~K.,   {Kazanas} D.,  2021, \mn@doi [\apj] {10.3847/1538-4357/abcec0}, \href {https://ui.adsabs.harvard.edu/abs/2021ApJ...907...63K} {907, 63}

\bibitem[\protect\citeauthoryear{{Kerin} \& {Melatos}}{{Kerin} \& {Melatos}}{2022}]{kerin2022}
{Kerin} A.~D.,  {Melatos} A.,  2022, \mn@doi [\mnras] {10.1093/mnras/stac1351}, \href {https://ui.adsabs.harvard.edu/abs/2022MNRAS.514.1628K} {514, 1628}

\bibitem[\protect\citeauthoryear{{Konar}}{{Konar}}{2017}]{konar2017magnetic}
{Konar} S.,  2017, \mn@doi [Journal of Astrophysics and Astronomy] {10.1007/s12036-017-9467-4}, \href {https://ui.adsabs.harvard.edu/abs/2017JApA...38...47K} {38, 47}

\bibitem[\protect\citeauthoryear{{Konar} \& {Bhattacharya}}{{Konar} \& {Bhattacharya}}{1997}]{konar97}
{Konar} S.,  {Bhattacharya} D.,  1997, \mn@doi [\mnras] {10.1093/mnras/284.2.311}, \href {https://ui.adsabs.harvard.edu/abs/1997MNRAS.284..311K} {284, 311}

\bibitem[\protect\citeauthoryear{{Konar} \& {Bhattacharya}}{{Konar} \& {Bhattacharya}}{1999a}]{konar99a}
{Konar} S.,  {Bhattacharya} D.,  1999a, \mn@doi [\mnras] {10.1046/j.1365-8711.1999.02287.x}, \href {https://ui.adsabs.harvard.edu/abs/1999MNRAS.303..588K} {303, 588}

\bibitem[\protect\citeauthoryear{{Konar} \& {Bhattacharya}}{{Konar} \& {Bhattacharya}}{1999b}]{konar99b}
{Konar} S.,  {Bhattacharya} D.,  1999b, \mn@doi [\mnras] {10.1046/j.1365-8711.1999.02781.x}, \href {https://ui.adsabs.harvard.edu/abs/1999MNRAS.308..795K} {308, 795}

\bibitem[\protect\citeauthoryear{{Konar} \& {Choudhuri}}{{Konar} \& {Choudhuri}}{2004}]{konar2004}
{Konar} S.,  {Choudhuri} A.~R.,  2004, \mn@doi [\mnras] {10.1111/j.1365-2966.2004.07397.x}, \href {https://ui.adsabs.harvard.edu/abs/2004MNRAS.348..661K} {348, 661}

\bibitem[\protect\citeauthoryear{{Kulkarni} \& {Romanova}}{{Kulkarni} \& {Romanova}}{2013}]{kulkarni2013}
{Kulkarni} A.~K.,  {Romanova} M.~M.,  2013, \mn@doi [\mnras] {10.1093/mnras/stt945}, \href {https://ui.adsabs.harvard.edu/abs/2013MNRAS.433.3048K} {433, 3048}

\bibitem[\protect\citeauthoryear{{Kulsrud} \& {Sunyaev}}{{Kulsrud} \& {Sunyaev}}{2020}]{Kulsrud_Sunyaev_2020}
{Kulsrud} R.~M.,  {Sunyaev} R.,  2020, \mn@doi [Journal of Plasma Physics] {10.1017/S0022377820001026}, \href {https://ui.adsabs.harvard.edu/abs/2020JPlPh..86f9002K} {86, 905860602}

\bibitem[\protect\citeauthoryear{{Lattimer} \& {Prakash}}{{Lattimer} \& {Prakash}}{2004}]{lattimer}
{Lattimer} J.~M.,  {Prakash} M.,  2004, \mn@doi [Science] {10.1126/science.1090720}, \href {https://ui.adsabs.harvard.edu/abs/2004Sci...304..536L} {304, 536}

\bibitem[\protect\citeauthoryear{{Lindblom} \& {Mendell}}{{Lindblom} \& {Mendell}}{1995}]{lindblom1995}
{Lindblom} L.,  {Mendell} G.,  1995, \mn@doi [\apj] {10.1086/175653}, \href {https://ui.adsabs.harvard.edu/abs/1995ApJ...444..804L} {444, 804}

\bibitem[\protect\citeauthoryear{{Long}, {Romanova}  \& {Lovelace}}{{Long} et~al.}{2005}]{Long_2005}
{Long} M.,  {Romanova} M.~M.,   {Lovelace} R.~V.~E.,  2005, \mn@doi [\apj] {10.1086/497000}, \href {https://ui.adsabs.harvard.edu/abs/2005ApJ...634.1214L} {634, 1214}

\bibitem[\protect\citeauthoryear{{Mastrano}, {Suvorov}  \& {Melatos}}{{Mastrano} et~al.}{2015}]{mastrano2015}
{Mastrano} A.,  {Suvorov} A.~G.,   {Melatos} A.,  2015, \mn@doi [\mnras] {10.1093/mnras/stu2671}, \href {https://ui.adsabs.harvard.edu/abs/2015MNRAS.447.3475M} {447, 3475}

\bibitem[\protect\citeauthoryear{{Melatos} \& {Payne}}{{Melatos} \& {Payne}}{2005}]{Melatos_2005}
{Melatos} A.,  {Payne} D.~J.~B.,  2005, \mn@doi [\apj] {10.1086/428600}, \href {https://ui.adsabs.harvard.edu/abs/2005ApJ...623.1044M} {623, 1044}

\bibitem[\protect\citeauthoryear{{Melatos} \& {Phinney}}{{Melatos} \& {Phinney}}{2001}]{melatos2001hydromagnetic}
{Melatos} A.,  {Phinney} E.~S.,  2001, \mn@doi [\pasa] {10.1071/AS01056}, \href {https://ui.adsabs.harvard.edu/abs/2001PASA...18..421M} {18, 421}

\bibitem[\protect\citeauthoryear{{Melatos}, {Douglass}  \& {Simula}}{{Melatos} et~al.}{2015}]{melatos2015}
{Melatos} A.,  {Douglass} J.~A.,   {Simula} T.~P.,  2015, \mn@doi [\apj] {10.1088/0004-637X/807/2/132}, \href {https://ui.adsabs.harvard.edu/abs/2015ApJ...807..132M} {807, 132}

\bibitem[\protect\citeauthoryear{{Mukherjee}}{{Mukherjee}}{2017}]{muk2017revisit}
{Mukherjee} D.,  2017, \mn@doi [Journal of Astrophysics and Astronomy] {10.1007/s12036-017-9465-6}, \href {https://ui.adsabs.harvard.edu/abs/2017JApA...38...48M} {38, 48}

\bibitem[\protect\citeauthoryear{{Mukherjee} \& {Bhattacharya}}{{Mukherjee} \& {Bhattacharya}}{2012}]{mukherjee2012phase}
{Mukherjee} D.,  {Bhattacharya} D.,  2012, \mn@doi [\mnras] {10.1111/j.1365-2966.2011.20085.x}, \href {https://ui.adsabs.harvard.edu/abs/2012MNRAS.420..720M} {420, 720}

\bibitem[\protect\citeauthoryear{{Mukherjee}, {Bhattacharya}  \& {Mignone}}{{Mukherjee} et~al.}{2013a}]{mukherjee2013mhd2d}
{Mukherjee} D.,  {Bhattacharya} D.,   {Mignone} A.,  2013a, \mn@doi [\mnras] {10.1093/mnras/stt020}, \href {https://ui.adsabs.harvard.edu/abs/2013MNRAS.430.1976M} {430, 1976}

\bibitem[\protect\citeauthoryear{{Mukherjee}, {Bhattacharya}  \& {Mignone}}{{Mukherjee} et~al.}{2013b}]{mukherjee2013mhd3d}
{Mukherjee} D.,  {Bhattacharya} D.,   {Mignone} A.,  2013b, \mn@doi [\mnras] {10.1093/mnras/stt1344}, \href {https://ui.adsabs.harvard.edu/abs/2013MNRAS.435..718M} {435, 718}

\bibitem[\protect\citeauthoryear{{Mukherjee}, {Bult}, {van der Klis}  \& {Bhattacharya}}{{Mukherjee} et~al.}{2015}]{Mukherjee15a}
{Mukherjee} D.,  {Bult} P.,  {van der Klis} M.,   {Bhattacharya} D.,  2015, \mn@doi [\mnras] {10.1093/mnras/stv1542}, \href {https://ui.adsabs.harvard.edu/abs/2015MNRAS.452.3994M} {452, 3994}

\bibitem[\protect\citeauthoryear{{Muslimov} \& {Tsygan}}{{Muslimov} \& {Tsygan}}{1985}]{muslimov}
{Muslimov} A.~G.,  {Tsygan} A.~I.,  1985, Soviet Astronomy Letters, \href {https://ui.adsabs.harvard.edu/abs/1985SvAL...11...80M} {11, 80}

\bibitem[\protect\citeauthoryear{{Mytidis}, {Coughlin}  \& {Whiting}}{{Mytidis} et~al.}{2015}]{mytidis2015}
{Mytidis} A.,  {Coughlin} M.,   {Whiting} B.,  2015, \mn@doi [\apj] {10.1088/0004-637X/810/1/27}, \href {https://ui.adsabs.harvard.edu/abs/2015ApJ...810...27M} {810, 27}

\bibitem[\protect\citeauthoryear{{N{\"a}ttil{\"a}}, {Cho}, {Skinner}, {Most}  \& {Ripperda}}{{N{\"a}ttil{\"a}} et~al.}{2024}]{Nättilä_2024}
{N{\"a}ttil{\"a}} J.,  {Cho} J. Y.~K.,  {Skinner} J.~W.,  {Most} E.~R.,   {Ripperda} B.,  2024, \mn@doi [\apj] {10.3847/1538-4357/ad54c2}, \href {https://ui.adsabs.harvard.edu/abs/2024ApJ...971...37N} {971, 37}

\bibitem[\protect\citeauthoryear{{Nazari} \& {Roshan}}{{Nazari} \& {Roshan}}{2020}]{nazari2020}
{Nazari} E.,  {Roshan} M.,  2020, \mn@doi [\mnras] {10.1093/mnras/staa2322}, \href {https://ui.adsabs.harvard.edu/abs/2020MNRAS.498..110N} {498, 110}

\bibitem[\protect\citeauthoryear{{Newcomb}}{{Newcomb}}{1961}]{newcomb1961}
{Newcomb} W.~A.,  1961, \mn@doi [Physics of Fluids] {10.1063/1.1706342}, \href {https://ui.adsabs.harvard.edu/abs/1961PhFl....4..391N} {4, 391}

\bibitem[\protect\citeauthoryear{{Owen}}{{Owen}}{2010}]{owen2010}
{Owen} B.~J.,  2010, \mn@doi [\prd] {10.1103/PhysRevD.82.104002}, \href {https://ui.adsabs.harvard.edu/abs/2010PhRvD..82j4002O} {82, 104002}

\bibitem[\protect\citeauthoryear{{Paczynski}}{{Paczynski}}{1983}]{paczynski1983models}
{Paczynski} B.,  1983, \mn@doi [\apj] {10.1086/160870}, \href {https://ui.adsabs.harvard.edu/abs/1983ApJ...267..315P} {267, 315}

\bibitem[\protect\citeauthoryear{{Pandharipande}, {Pines}  \& {Smith}}{{Pandharipande} et~al.}{1976}]{pandharipande1976}
{Pandharipande} V.~R.,  {Pines} D.,   {Smith} R.~A.,  1976, \mn@doi [\apj] {10.1086/154637}, \href {https://ui.adsabs.harvard.edu/abs/1976ApJ...208..550P} {208, 550}

\bibitem[\protect\citeauthoryear{{Parfrey} \& {Tchekhovskoy}}{{Parfrey} \& {Tchekhovskoy}}{2017}]{parfrey17b}
{Parfrey} K.,  {Tchekhovskoy} A.,  2017, \mn@doi [\apjl] {10.3847/2041-8213/aa9c85}, \href {https://ui.adsabs.harvard.edu/abs/2017ApJ...851L..34P} {851, L34}

\bibitem[\protect\citeauthoryear{{Parfrey}, {Spitkovsky}  \& {Beloborodov}}{{Parfrey} et~al.}{2017}]{parfrey}
{Parfrey} K.,  {Spitkovsky} A.,   {Beloborodov} A.~M.,  2017, \mn@doi [\mnras] {10.1093/mnras/stx950}, \href {https://ui.adsabs.harvard.edu/abs/2017MNRAS.469.3656P} {469, 3656}

\bibitem[\protect\citeauthoryear{{Patruno}}{{Patruno}}{2012}]{Patruno_2012}
{Patruno} A.,  2012, \mn@doi [\apjl] {10.1088/2041-8205/753/1/L12}, \href {https://ui.adsabs.harvard.edu/abs/2012ApJ...753L..12P} {753, L12}

\bibitem[\protect\citeauthoryear{{Payne} \& {Melatos}}{{Payne} \& {Melatos}}{2004}]{payne2004burial}
{Payne} D.~J.~B.,  {Melatos} A.,  2004, \mn@doi [\mnras] {10.1111/j.1365-2966.2004.07798.x}, \href {https://ui.adsabs.harvard.edu/abs/2004MNRAS.351..569P} {351, 569}

\bibitem[\protect\citeauthoryear{{Payne} \& {Melatos}}{{Payne} \& {Melatos}}{2007}]{payne2007burial}
{Payne} D.~J.~B.,  {Melatos} A.,  2007, \mn@doi [\mnras] {10.1111/j.1365-2966.2007.11451.x}, \href {https://ui.adsabs.harvard.edu/abs/2007MNRAS.376..609P} {376, 609}

\bibitem[\protect\citeauthoryear{{Potekhin} \& {Chabrier}}{{Potekhin} \& {Chabrier}}{2000}]{potekhin2000}
{Potekhin} A.~Y.,  {Chabrier} G.,  2000, \mn@doi [\pre] {10.1103/PhysRevE.62.8554}, \href {https://ui.adsabs.harvard.edu/abs/2000PhRvE..62.8554P} {62, 8554}

\bibitem[\protect\citeauthoryear{{Press}, {Teukolsky}, {Vetterling}  \& {Flannery}}{{Press} et~al.}{1992}]{PresTeukVettFlan92}
{Press} W.~H.,  {Teukolsky} S.~A.,  {Vetterling} W.~T.,   {Flannery} B.~P.,  1992, {Numerical recipes in C. The art of scientific computing}.
Cambridge University Press

\bibitem[\protect\citeauthoryear{{Priymak}, {Melatos}  \& {Payne}}{{Priymak} et~al.}{2011}]{priymak2011quadrupole}
{Priymak} M.,  {Melatos} A.,   {Payne} D.~J.~B.,  2011, \mn@doi [\mnras] {10.1111/j.1365-2966.2011.19431.x}, \href {https://ui.adsabs.harvard.edu/abs/2011MNRAS.417.2696P} {417, 2696}

\bibitem[\protect\citeauthoryear{{Priymak}, {Melatos}  \& {Lasky}}{{Priymak} et~al.}{2014}]{priymak2014cyclotron}
{Priymak} M.,  {Melatos} A.,   {Lasky} P.~D.,  2014, \mn@doi [\mnras] {10.1093/mnras/stu1825}, \href {https://ui.adsabs.harvard.edu/abs/2014MNRAS.445.2710P} {445, 2710}

\bibitem[\protect\citeauthoryear{{Psaltis} \& {Chakrabarty}}{{Psaltis} \& {Chakrabarty}}{1999}]{Psaltis_1999}
{Psaltis} D.,  {Chakrabarty} D.,  1999, \mn@doi [\apj] {10.1086/307525}, \href {https://ui.adsabs.harvard.edu/abs/1999ApJ...521..332P} {521, 332}

\bibitem[\protect\citeauthoryear{{Radhakrishnan} \& {Srinivasan}}{{Radhakrishnan} \& {Srinivasan}}{1984}]{radhakrishnan1984}
{Radhakrishnan} V.,  {Srinivasan} G.,  1984, in {Hidayat} B.,  {Feast} M.~W.,  eds, Second Asian-Pacific Regional Meeting on Astronomy. p.~423

\bibitem[\protect\citeauthoryear{{Rigoselli} \& {Mereghetti}}{{Rigoselli} \& {Mereghetti}}{2018}]{refId0}
{Rigoselli} M.,  {Mereghetti} S.,  2018, \mn@doi [\aap] {10.1051/0004-6361/201732408}, \href {https://ui.adsabs.harvard.edu/abs/2018A&A...615A..73R} {615, A73}

\bibitem[\protect\citeauthoryear{{Riles}}{{Riles}}{2023}]{riles2023}
{Riles} K.,  2023, \mn@doi [Living Reviews in Relativity] {10.1007/s41114-023-00044-3}, \href {https://ui.adsabs.harvard.edu/abs/2023LRR....26....3R} {26, 3}

\bibitem[\protect\citeauthoryear{{Riley} et~al.,}{{Riley} et~al.}{2021}]{Riley_2021}
{Riley} T.~E.,  et~al., 2021, \mn@doi [\apjl] {10.3847/2041-8213/ac0a81}, \href {https://ui.adsabs.harvard.edu/abs/2021ApJ...918L..27R} {918, L27}

\bibitem[\protect\citeauthoryear{{Romani}}{{Romani}}{1990}]{Romani1990}
{Romani} R.~W.,  1990, \mn@doi [\nat] {10.1038/347741a0}, \href {https://ui.adsabs.harvard.edu/abs/1990Natur.347..741R} {347, 741}

\bibitem[\protect\citeauthoryear{{Romanova}, {Kulkarni}  \& {Lovelace}}{{Romanova} et~al.}{2008}]{Romanova_2008}
{Romanova} M.~M.,  {Kulkarni} A.~K.,   {Lovelace} R. V.~E.,  2008, \mn@doi [\apjl] {10.1086/527298}, \href {https://ui.adsabs.harvard.edu/abs/2008ApJ...673L.171R} {673, L171}

\bibitem[\protect\citeauthoryear{{Rossetto}, {Frauendiener}, {Brunet}  \& {Melatos}}{{Rossetto} et~al.}{2023}]{rossetto23}
{Rossetto} P. H.~B.,  {Frauendiener} J.,  {Brunet} R.,   {Melatos} A.,  2023, \mn@doi [\mnras] {10.1093/mnras/stad2850}, \href {https://ui.adsabs.harvard.edu/abs/2023MNRAS.526.2058R} {526, 2058}

\bibitem[\protect\citeauthoryear{{Rossetto}, {Frauendiener}  \& {Melatos}}{{Rossetto} et~al.}{2025}]{Rossetto_2025}
{Rossetto} P. H.~B.,  {Frauendiener} J.,   {Melatos} A.,  2025, \mn@doi [\apj] {10.3847/1538-4357/ada276}, \href {https://ui.adsabs.harvard.edu/abs/2025ApJ...979...10R} {979, 10}

\bibitem[\protect\citeauthoryear{{Ruderman}}{{Ruderman}}{1969}]{ruderman1969}
{Ruderman} M.,  1969, \mn@doi [\nat] {10.1038/223597b0}, \href {https://ui.adsabs.harvard.edu/abs/1969Natur.223..597R} {223, 597}

\bibitem[\protect\citeauthoryear{{Ruderman}}{{Ruderman}}{1991a}]{ruderman1}
{Ruderman} M.,  1991a, \mn@doi [\apj] {10.1086/169558}, \href {https://ui.adsabs.harvard.edu/abs/1991ApJ...366..261R} {366, 261}

\bibitem[\protect\citeauthoryear{{Ruderman}}{{Ruderman}}{1991b}]{ruderman2}
{Ruderman} R.,  1991b, \mn@doi [\apj] {10.1086/170744}, \href {https://ui.adsabs.harvard.edu/abs/1991ApJ...382..576R} {382, 576}

\bibitem[\protect\citeauthoryear{{Ruderman} \& {Sutherland}}{{Ruderman} \& {Sutherland}}{1975}]{ruderman1975}
{Ruderman} M.~A.,  {Sutherland} P.~G.,  1975, \mn@doi [\apj] {10.1086/153393}, \href {https://ui.adsabs.harvard.edu/abs/1975ApJ...196...51R} {196, 51}

\bibitem[\protect\citeauthoryear{{Singh}, {Haskell}, {Mukherjee}  \& {Bulik}}{{Singh} et~al.}{2020}]{singh_2020}
{Singh} N.,  {Haskell} B.,  {Mukherjee} D.,   {Bulik} T.,  2020, \mn@doi [\mnras] {10.1093/mnras/staa442}, \href {https://ui.adsabs.harvard.edu/abs/2020MNRAS.493.3866S} {493, 3866}

\bibitem[\protect\citeauthoryear{{Smarr} \& {Blandford}}{{Smarr} \& {Blandford}}{1976}]{smarr1976}
{Smarr} L.~L.,  {Blandford} R.,  1976, \mn@doi [\apj] {10.1086/154524}, \href {https://ui.adsabs.harvard.edu/abs/1976ApJ...207..574S} {207, 574}

\bibitem[\protect\citeauthoryear{{Srinivasan}, {Bhattacharya}, {Muslimov}  \& {Tsygan}}{{Srinivasan} et~al.}{1990}]{srinivasan1990}
{Srinivasan} G.,  {Bhattacharya} D.,  {Muslimov} A.~G.,   {Tsygan} A.~J.,  1990, Current Science, \href {https://ui.adsabs.harvard.edu/abs/1990CSci...59...31S} {59, 31}

\bibitem[\protect\citeauthoryear{{Sur} \& {Haskell}}{{Sur} \& {Haskell}}{2021a}]{surhaskell2021a}
{Sur} A.,  {Haskell} B.,  2021a, \mn@doi [\pasa] {10.1017/pasa.2021.39}, \href {https://ui.adsabs.harvard.edu/abs/2021PASA...38...43S} {38, e043}

\bibitem[\protect\citeauthoryear{{Sur} \& {Haskell}}{{Sur} \& {Haskell}}{2021b}]{sur_2021b}
{Sur} A.,  {Haskell} B.,  2021b, \mn@doi [\mnras] {10.1093/mnras/stab307}, \href {https://ui.adsabs.harvard.edu/abs/2021MNRAS.502.4680S} {502, 4680}

\bibitem[\protect\citeauthoryear{{Suvorov} \& {Melatos}}{{Suvorov} \& {Melatos}}{2019}]{suvorov2019relaxation}
{Suvorov} A.~G.,  {Melatos} A.,  2019, \mn@doi [\mnras] {10.1093/mnras/sty3518}, \href {https://ui.adsabs.harvard.edu/abs/2019MNRAS.484.1079S} {484, 1079}

\bibitem[\protect\citeauthoryear{{Suvorov} \& {Melatos}}{{Suvorov} \& {Melatos}}{2020}]{suvorov2020multi}
{Suvorov} A.~G.,  {Melatos} A.,  2020, \mn@doi [\mnras] {10.1093/mnras/staa3132}, \href {https://ui.adsabs.harvard.edu/abs/2020MNRAS.499.3243S} {499, 3243}

\bibitem[\protect\citeauthoryear{{Taam} \& {van den Heuvel}}{{Taam} \& {van den Heuvel}}{1986}]{1986taam}
{Taam} R.~E.,  {van den Heuvel} E.~P.~J.,  1986, \mn@doi [\apj] {10.1086/164243}, \href {https://ui.adsabs.harvard.edu/abs/1986ApJ...305..235T} {305, 235}

\bibitem[\protect\citeauthoryear{Trottenberg, Oosterlee  \& Schuller}{Trottenberg et~al.}{2000}]{trottenberg2000multigrid}
Trottenberg U.,  Oosterlee C.~W.,   Schuller A.,  2000, Multigrid.
Elsevier

\bibitem[\protect\citeauthoryear{{Urpin} \& {Geppert}}{{Urpin} \& {Geppert}}{1995}]{urpin95}
{Urpin} V.,  {Geppert} U.,  1995, \mn@doi [\mnras] {10.1093/mnras/275.4.1117}, \href {https://ui.adsabs.harvard.edu/abs/1995MNRAS.275.1117U} {275, 1117}

\bibitem[\protect\citeauthoryear{{Urpin} \& {Konenkov}}{{Urpin} \& {Konenkov}}{1997}]{urpin97}
{Urpin} V.,  {Konenkov} D.,  1997, \mn@doi [\mnras] {10.1093/mnras/284.3.741}, \href {https://ui.adsabs.harvard.edu/abs/1997MNRAS.284..741U} {284, 741}

\bibitem[\protect\citeauthoryear{{Vigelius} \& {Melatos}}{{Vigelius} \& {Melatos}}{2008}]{vigelius2008three}
{Vigelius} M.,  {Melatos} A.,  2008, \mn@doi [\mnras] {10.1111/j.1365-2966.2008.13139.x}, \href {https://ui.adsabs.harvard.edu/abs/2008MNRAS.386.1294V} {386, 1294}

\bibitem[\protect\citeauthoryear{{Vigelius} \& {Melatos}}{{Vigelius} \& {Melatos}}{2009a}]{vigelius2009a}
{Vigelius} M.,  {Melatos} A.,  2009a, \mn@doi [\mnras] {10.1111/j.1365-2966.2009.14690.x}, \href {https://ui.adsabs.harvard.edu/abs/2009MNRAS.395.1972V} {395, 1972}

\bibitem[\protect\citeauthoryear{{Vigelius} \& {Melatos}}{{Vigelius} \& {Melatos}}{2009b}]{vigelius2009resistive}
{Vigelius} M.,  {Melatos} A.,  2009b, \mn@doi [\mnras] {10.1111/j.1365-2966.2009.14698.x}, \href {https://ui.adsabs.harvard.edu/abs/2009MNRAS.395.1985V} {395, 1985}

\bibitem[\protect\citeauthoryear{{Wette}, {Vigelius}  \& {Melatos}}{{Wette} et~al.}{2010}]{vigelius2010sinking}
{Wette} K.,  {Vigelius} M.,   {Melatos} A.,  2010, \mn@doi [\mnras] {10.1111/j.1365-2966.2009.15937.x}, \href {https://ui.adsabs.harvard.edu/abs/2010MNRAS.402.1099W} {402, 1099}

\bibitem[\protect\citeauthoryear{{Zanni} \& {Ferreira}}{{Zanni} \& {Ferreira}}{2009}]{zanni08}
{Zanni} C.,  {Ferreira} J.,  2009, \mn@doi [\aap] {10.1051/0004-6361/200912879}, \href {https://ui.adsabs.harvard.edu/abs/2009A&A...508.1117Z} {508, 1117}

\bibitem[\protect\citeauthoryear{{Zimmermann}}{{Zimmermann}}{1978}]{zimmermann1978}
{Zimmermann} M.,  1978, \mn@doi [\nat] {10.1038/271524a0}, \href {https://ui.adsabs.harvard.edu/abs/1978Natur.271..524Z} {271, 524}

\bibitem[\protect\citeauthoryear{{{\c{C}}{\i}k{\i}nto{\u{g}}lu}}{{{\c{C}}{\i}k{\i}nto{\u{g}}lu}}{2023}]{cikintoglu}
{{\c{C}}{\i}k{\i}nto{\u{g}}lu} S.,  2023, \mn@doi [\mnras] {10.1093/mnras/stad2164}, \href {https://ui.adsabs.harvard.edu/abs/2023MNRAS.524.3846C} {524, 3846}

\makeatother
\end{thebibliography}




\appendix

\section{Numerical Method}\label{app1}
For the logarithmic coordinates $(y,\mu)$ used here, $dr$ (spherical coordinates $(r,\theta)$) is given by
\begin{equation*}
    dr = (r-aR_{\ast})dy
\end{equation*}
The radial extent of the grid lies between $R_{\ast}$ and $R_{\rm out}$ i.e. $y_{1}=0$ and $y_{2}=\log((R_{\rm out}-aR_{\ast})/(R_{\ast}(1-a)))$. Let $N_{y}$ be the number of points between $y_{1}$ and $y_{2}$. Thus $dy = (y_{2}-y_{1})/(N_{y}+1)$. 
\begin{figure}
\vbox{
\includegraphics[width=\columnwidth]{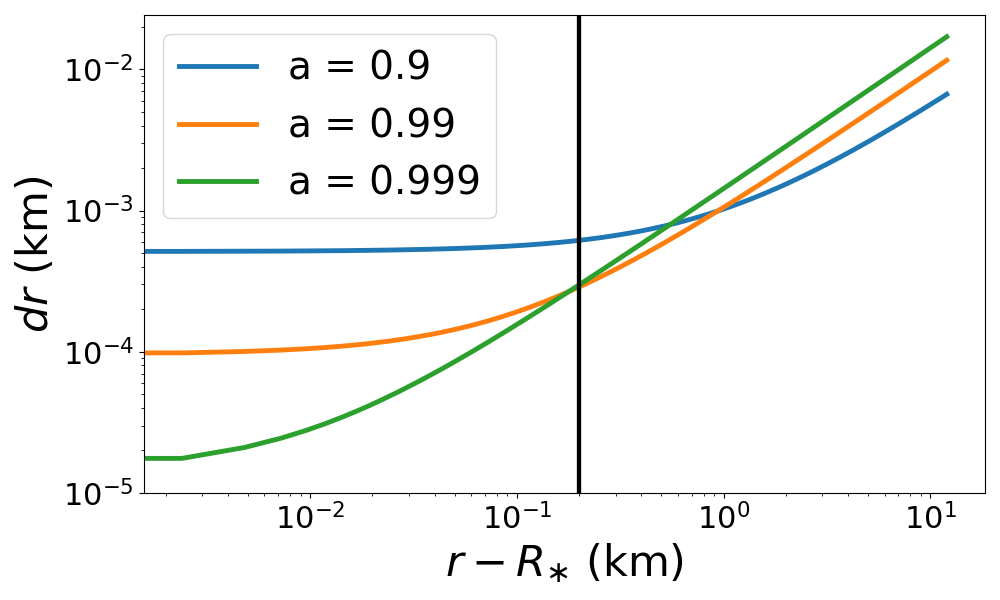}
}
\caption{The value of $dr$ is plotted as a function of $r-R_{\ast}$, both in km for three values of $a$ and for the logarithmic radial coordinates used in this work. The vertical line is at $200$ m above the neutron star surface i.e approximately the maximum heights of mounds used in this work.}
\label{dr}
\end{figure}
Therefore, 
\begin{align}
    dr = \frac{(r-aR_{\ast})}{(N_{y}+1)} \log\left(\frac{R_{\rm out}-aR_{\ast}}{R_{\ast}(1-a)}\right).
\end{align}
For fixed $R_{\ast}$, $R_{\rm out}$ and $N_{y}$, the value of $dr$ depends on $r$ and $a$. This dependence has been plotted in Figure \ref{dr} for $R_{\ast}=10$ km, $R_{\rm out}=22$ km, $N_{y}=5000$ and three values of $a$. From Figure \ref{dr}, we can see that as the value of $a$ increases, the stretching of the radial grid increases. The vertical line in figure \ref{dr} has been plotted for a height $200$ m above the surface. The heights less than $200$ m are better resolved by $a=0.999$ relative to the smaller values of $a$. Since the mounds in this work are close to $200$ m or do not exceed $200$ m, we selected an optimal value of $a=0.999$ for all calculations. For $a=0.999$ and for a resolution of $5000$ points, the value of $dr$ varies from $0.01419$ m at the inner boundary to $17$ m at the outer radial boundary. The typical scale heights of the solutions (for example $1-1000$ m for the solution in subsection \ref{rout_effect}) are resolved by the selected grid.

Equation \ref{GSeq} can be written in second order finite differenced form as follows
\begin{align}\label{aeq:1}
    & \notag y_{c}^{+} \psi[y+\Delta y;\mu] + y_{c}^{-} \psi[y-\Delta y;\mu] + e_{c} \psi[y;\mu] \\ & \notag + \mu_{c} (\psi[y;\mu+\Delta \mu] + \psi[y;\mu - \Delta \mu]) \\ & =  src_{\mbox{coeff}}\;  K(\psi[y;\mu],y,\mu)
\end{align}
where
\begin{equation*}
    y_{c}^{+} = (\Delta \mu)^{2} - \frac{(\Delta \mu)^{2} (\Delta y)}{2}
\end{equation*}
\begin{equation*}
    y_{c}^{-} = (\Delta \mu)^{2} + \frac{(\Delta \mu)^{2} (\Delta y)}{2}
\end{equation*}
\begin{equation*}
    \mu_{c} = \frac{(1-\mu^{2}) (\Delta y)^{2}}{\left(1 + \frac{a}{e^{y}(1-a)}\right)^{2}}
\end{equation*}
\begin{equation*}
    e_{c} = \left(-2(\Delta \mu)^{2} - 2(\Delta y)^{2}\frac{(1-\mu^{2})}{\left(1 + \frac{a}{e^{y}(1-a)}\right)^{2}}  \right)
\end{equation*}
\begin{equation*}
    src_{\mbox{coeff}} = (\Delta y)^{2}(\Delta \mu)^{2}R_{\ast}^{2}(1-a)^{2}e^{2y}.
\end{equation*}
and $K(\psi,y,\mu)$ is the source function from Equation \ref{source_func}.

A dipolar solution is used as an initial guess and red black successive over-relaxation (SOR) algorithm is used to iteratively update $\psi$ to solve the equation \ref{aeq:1}. The R.H.S or the source term of equation \ref{aeq:1} is updated at each step using the value of the solution at the previous SOR step. All boundary conditions are updated at each SOR step using the solution at the previous step. 
Earlier studies have often used a two-layer relaxation method with a first layer of successive over-relaxation (SOR) having the source constant for each SOR iteration, followed by an under-relaxation with an update of the RHS, and a repetition of these steps till convergence. However, in this study we have employed a local relaxation method called the SOR-Picard method \citep{trottenberg2000multigrid} without under-relaxation, which is faster. In the new method, $\psi$ is relaxed by SOR and the non linear RHS term is updated at each SOR iteration using the $\psi$ from the previous iteration. The iteration is stopped after the values of $\mbox{anorm}_{n}$ and $\mbox{relp}_{n}$ ($n=$Number of the iteration step) flatten out or remain unchanged. This criteria for convergence, although more restrictive, is more robust, than stopping at an arbitrary value. Their values are calculated by 

\begin{align}
    \mbox{anorm}_{n} = \sum_{i,j} & |y_{c}^{+} \psi[y_{i}+\Delta y;\mu_{j}] + y_{c}^{-} \psi[y_{i}-\Delta y;\mu_{j}] \nonumber\\ & + (e_{c})_{i,j} (\psi[y_{i};\mu_{j}]) \nonumber \\ & + (\mu_{c})_{i,j} (\psi[y_{i};\mu_{j}+\Delta \mu] + \psi[y_{i};\mu_{j} - \Delta \mu]) \nonumber \\ & -  (src_{\mbox{coeff}})_{i,j} K(\psi[y_{i};\mu_{j}],y_{i},\mu_{j})|
\end{align}

\begin{equation*}
    \mbox{relp}_{n} = MAX\left(\frac{|\psi_{n+1}(y_{i},\mu_{j})-\psi_{n}(y_{i},\mu_{j})|}{\psi_{n+1}(y_{i},\mu_{j})}\right).
\end{equation*}

\begin{figure}
\vbox{
\includegraphics[width=\columnwidth]{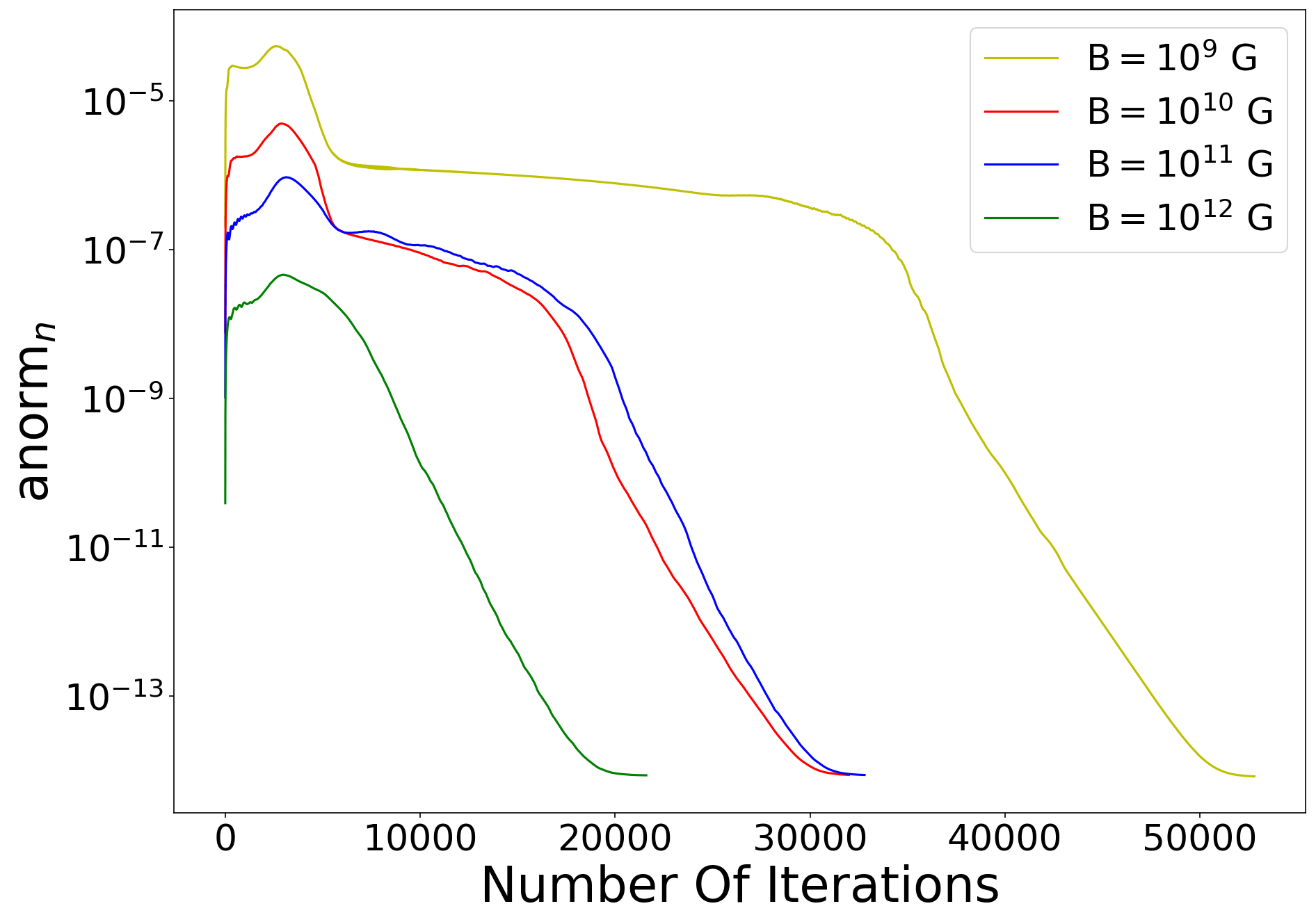}
\includegraphics[width=\columnwidth]{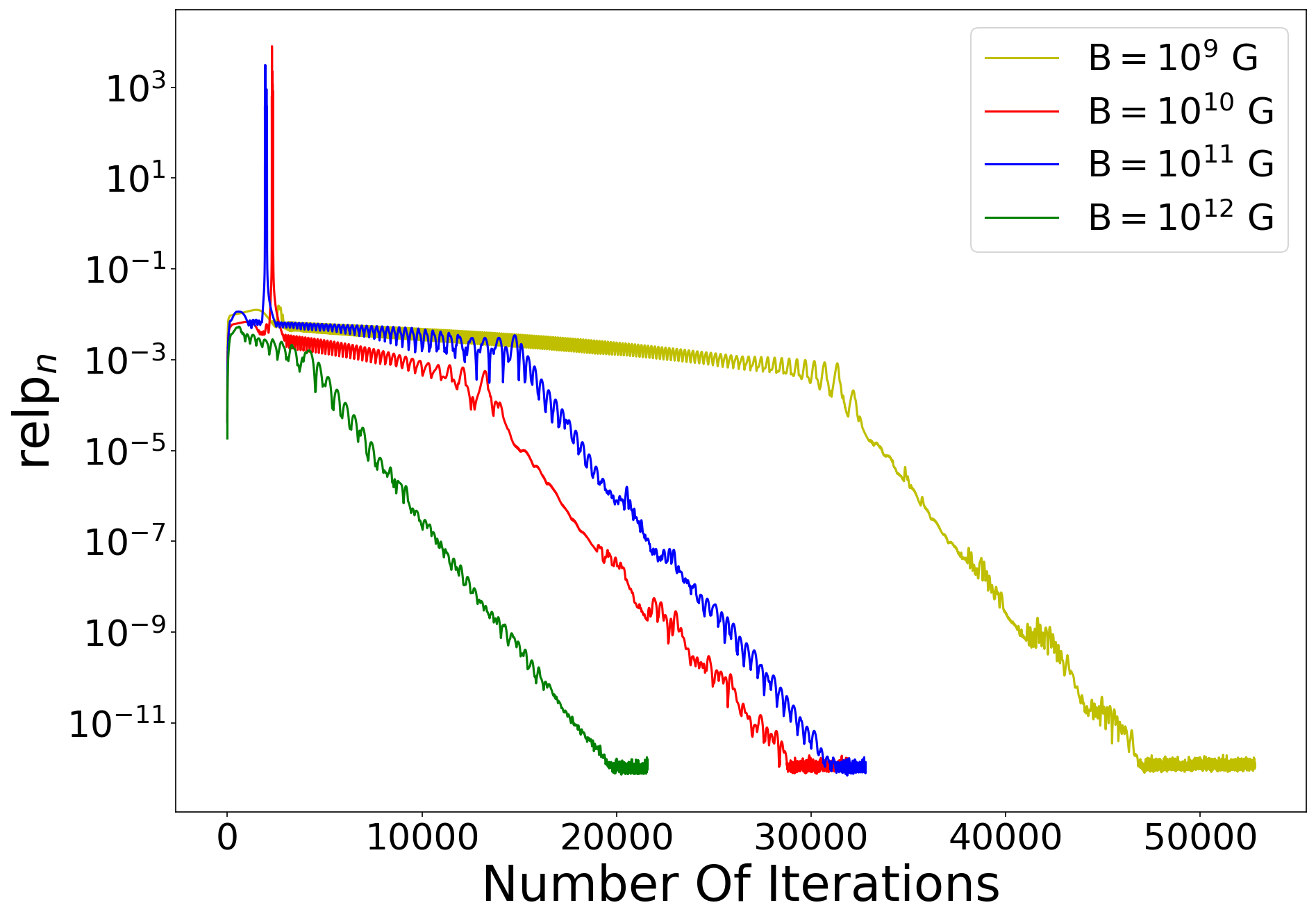}
}
\caption{Change in the value of $anorm_{n}$ (top) and in the value of $relp_{n}$ (bottom) with the number of iteration steps for 4 magnetic field strength cases have been plotted. Simulations are stopped after both show evidence of flattening out. All cases have been plotted for Maximum Mass solutions.}
\label{fig-a1}
\end{figure}

An example of the change in $\mbox{anorm}_{n}$ and $\mbox{relp}_{n}$ with the number of iteration steps is shown in Figure \ref{fig-a1}. The examples plotted here are for the numerical maximum mass (NMM) (refer to Appendix \ref{numapp}) of the solutions, with high gradients at some points.

\subsection{Optimal \texorpdfstring{$\omega$}{omega} for SOR}

From \cite{PresTeukVettFlan92}, the optimal choice of $\omega$ (overrelaxation parameter) for SOR is
\begin{equation}
    \omega = \frac{2}{1+\sqrt{1-\rho_{\mbox{Jacobi}}^{2}}}
\end{equation}
where $\rho_{\mbox{Jacobi}}$ is the spectral radius of the Jacobi iteration. $\rho_{\mbox{Jacobi}}$ for the 2D Poisson equation on a rectangular $J \times L$ grid is \citep{PresTeukVettFlan92}
\begin{equation}\label{aeq:2}
    \rho_{\mbox{Jacobi}} = \frac{\cos{\frac{\pi}{J}}+\left(\frac{\Delta x}{\Delta y}\right)^{2}\cos{\frac{\pi}{L}}}{1 + \left(\frac{\Delta x}{\Delta y}\right)^{2}}.
\end{equation}

Calculating $\rho_{\mbox{Jacobi}}$ for a general elliptic equation is non-trivial since the eigenvalues of the iteration matrix for such equations have spatial dependence. Empirical testing of different values of $\omega$ showed that the form of the spectral radius in equation \eqref{aeq:2} is optimal and adequate for the Grad Shafranov equation \ref{aeq:1}.

\subsection{Checking with global SOR-underrelaxation method}\label{A2}

\begin{figure}
{
\includegraphics[width=\columnwidth]{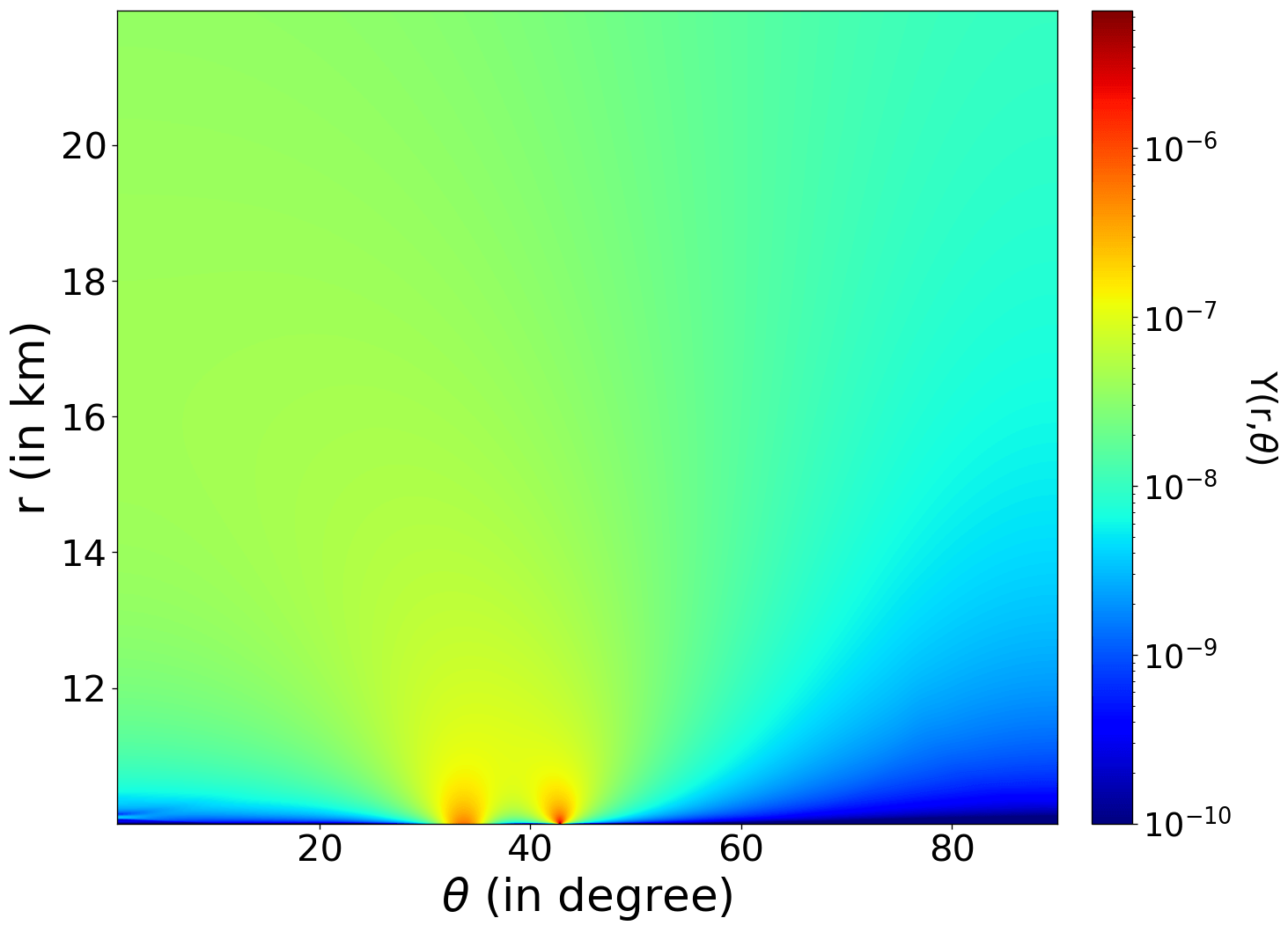}
}
\caption{Absolute relative difference between the solution using our current local SOR-Picard method and the solution using the old global SOR-underrelaxation method.}
\label{fig-a2}
\end{figure}

 For global SOR-underrelaxation method (earlier method), Chebyshev acceleration is used for SOR while an underrelaxation parameter of $0.01$ is used \citep{mukherjee2012phase}. Figure \ref{fig-a2} represents the colormap of the absolute relative difference between the solution from global SOR-underrelaxation method and the solution from local relaxation SOR Picard method. The plot shows that the solutions match by up to $10^{-5}$. Hence, the new method gives solutions that match well with those from the earlier method.

\section{Outer radial current free boundary condition}\label{app2}

The derivation of the current free boundary condition is presented in the first subsection \ref{derivn}. In the subsections following the first, a series of tests were performed on the new current free boundary condition (for the dipolar case) to optimize it and test its validity.
\subsection{Derivation}\label{derivn}
If $L$ is the Grad Shafranov operator and $L^{\ast}$ is the adjoint operator, then we need to solve the problem
\begin{equation*}
    L\psi = f
\end{equation*}
where $f$ is the source function. Using Green's theorem, and from equation (A5) in Appendix A1 in \cite{payne2004burial}, we know that 
\begin{equation}
    \psi = \int (G^{\ast}f) dV + \int \hat{n}.(-G^{\ast}(\Vec{\nabla}\psi) + \psi(\Vec{\nabla}G^{\ast}) - G^{\ast}\psi\Vec{b}) dS
\end{equation}
where $G^{\ast}$ is the greens function for the adjoint operator, and $\Vec{b}$ is calculated using $L^{\ast}$. We need to find the value of $\psi$ at the outer boundary ($R_{\rm out}$) at each step of the iteration. The above solution is valid for all regions. In order to evaluate the outer boundary independently at each iteration step of the SOR solver, we choose a source-free region ($f=0$) above the mound. This implies a domain extending radially from an inner boundary $r=R_{\rm in}$ to $r=\infty$ and $\theta \in (0^{0},90^{0})$. Appropriate constraints on the choice of $r=R_{\rm in}$ are discussed at the end of this subsection. Therefore, only the boundary terms contribute and we can write
\begin{equation*}
    \psi = \int \hat{n}.(-G^{\ast}(\Vec{\nabla}\psi) + \psi(\Vec{\nabla}G^{\ast}) - G^{\ast}\psi\Vec{b}) dS.
\end{equation*}
\begin{align*}
    \psi = \int  [ & - (\hat{r}.(-G^{\ast}(\Vec{\nabla}\psi) + \psi(\Vec{\nabla}G^{\ast}) - G^{\ast}\psi\Vec{b}))_{R_{\rm in}} \\ & + (\hat{r}.(-G^{\ast}(\Vec{\nabla}\psi) + \psi(\Vec{\nabla}G^{\ast}) - G^{\ast}\psi\Vec{b}))_{\infty} ]dS.
\end{align*}
At $r=R_{\rm in}$, $\psi=\psi(R_{\rm in},\theta)$ (value of $\psi$ from the previous iteration) and $G^{\ast} = 0$, \\
At $r=\infty$, $\psi=0$ and $G^{\ast} = 0$ \\
Thus,
\begin{equation*}
    \psi = \int \left(-\psi(R_{\rm in},\theta)\left(\frac{\partial G^{\ast}}{\partial r} \right)_{R_{\rm in}} \right).dS.
\end{equation*}
\begin{equation*}
    \psi(r^{\prime},\theta^{\prime}) = \int \left(-\psi(R_{\rm in},\theta)\left(\frac{\partial G^{\ast}(r,\theta,r^{\prime},\theta^{\prime})}{\partial r} \right)_{R_{\rm in}} \right).R_{\rm in}^{2} \sin \theta d\theta d\phi.
\end{equation*}
Differentiating $G^{\ast}$ previously calculated (Equation (A16) in Appendix A1) by \cite{payne2004burial} and writing its value at $R_{\rm in}$, we get
\begin{align*}
    \psi(r^{\prime},\theta^{\prime}) = \int_{0}^{\pi} & \psi(R_{\rm in},\theta) R_{\rm in}^{2} \sin \theta \\ & \times \sum_{\ell=1}^{\infty} \frac{(2\ell+1)R_{\rm in}^{\ell-2} P_{\ell}^{1}(\cos \theta) P_{\ell}^{1}(\cos \theta^{\prime}) \sin \theta^{\prime}}{2\ell(\ell+1)r^{\prime(\ell)} \sin \theta} d\theta.
\end{align*}
\begin{align}
    \psi(r^{\prime},\theta^{\prime}) = \sin \theta^{\prime} \sum_{\ell=1}^{\infty} & \left( \frac{(2\ell+1) P_{\ell}^{1}(\cos \theta^{\prime})}{2\ell(\ell+1)} \left(\frac{R_{\rm in}}{r^{\prime}}\right)^{\ell} \right. \nonumber \\ & \left. \quad \times \int_{0}^{\pi} \psi(R_{\rm in},\theta) P_{\ell}^{1}(\cos \theta) d\theta \right).
\end{align}

The above equation is used to calculate $\psi$ at $R_{\rm out}$, which is the outer radius of the Grad Shafranov simulation domain. When $\psi$ is symmetric about $\theta = \pi/2$ (not true for quadrudipolar field) i.e for $\psi(R_{\rm in},\theta) = \psi(R_{\rm in},\pi - \theta)$, equation can be simplified to 
\begin{align}
    \psi(r^{\prime},\theta^{\prime}) = \sin \theta^{\prime} \sum_{\ell=1}^{\infty} & \left( \frac{(2\ell+1) P_{\ell}^{1}(\cos \theta^{\prime})}{2\ell(\ell+1)} \left(\frac{R_{\rm in}}{r^{\prime}}\right)^{\ell} \right. \nonumber \\ & \left. \quad \times \int_{0}^{\frac{\pi}{2}} \psi(R_{\rm in},\theta) P_{\ell}^{1}(\cos \theta) (1 + (-1)^{\ell+1}) d\theta \right).
\end{align}
For this case, series in $\psi$ admits only odd '$\ell$' values ($\ell=1,3,5,7.....$). To find $\psi$ numerically, we need to choose an $\ell_{\rm max}$ up to which the series calculation should be done based on desired accuracy. Thus for dipolar inner boundary case, this equation can be written as
\begin{align}
    \psi(r^{\prime},\theta^{\prime}) = \sum_{\ell=1}^{\ell_{\rm max}} & \left( \frac{(2\ell+1) P_{\ell}^{1}(\cos \theta^{\prime})}{2\ell(\ell+1)} \left(\frac{R_{\rm in}}{r^{\prime}}\right)^{\ell} \sin \theta^{\prime} \right. \nonumber \\ & \left. \quad \times \int_{0}^{1} \frac{\psi(R_{\rm in},\theta) P_{\ell}^{1}(\cos \theta) (1 + (-1)^{\ell+1})}{\sqrt{1-\cos^{2}\theta}} d(\cos \theta) \right).
\end{align}

To calculate the solution at the outer boundary, value of $r^{\prime}=R_{\rm out}$. Different value of $R_{\rm in}$ needs to be selected as far away from $R_{\rm out}$ as possible, to make sure the term $\left(\frac{R_{\rm in}}{R_{\rm out}}\right)^{\ell}$ reduces drastically for values of $\ell>\ell_{\rm max}$, increasing the accuracy of this calculation. Thus, we suggest choosing $R_{\rm in}$ near the top of the mound in vacuum.

\subsection{Selection of \texorpdfstring{$\ell_{\rm max}$}{ell\_max}}\label{lmx}
A value of $\ell_{\rm max}$ needs to be selected to truncate the series calculation for the current free boundary condition. To decide $\ell_{\rm max}$, multiple solutions are calculated with the series for current free boundary condition truncated at multiple values of $\ell$. $\psi_{\ell}$ is thus the solution calculated with the series for the current free boundary condition truncated at $\ell$. $Z_{\ell}$ is a parameter used to compare solutions and finalize the value of $\ell_{\rm max}$. 
\begin{equation}\label{b1}
    Z_{\ell} = \sum_{i,j} |\psi_{\ell}(r_{i},\theta_{j})-\psi_{\ell-2}(r_{i},\theta_{j})|
\end{equation}

\subsubsection{Different value of $R_{\rm in}$}
\begin{figure}
{
\includegraphics[width=\columnwidth]{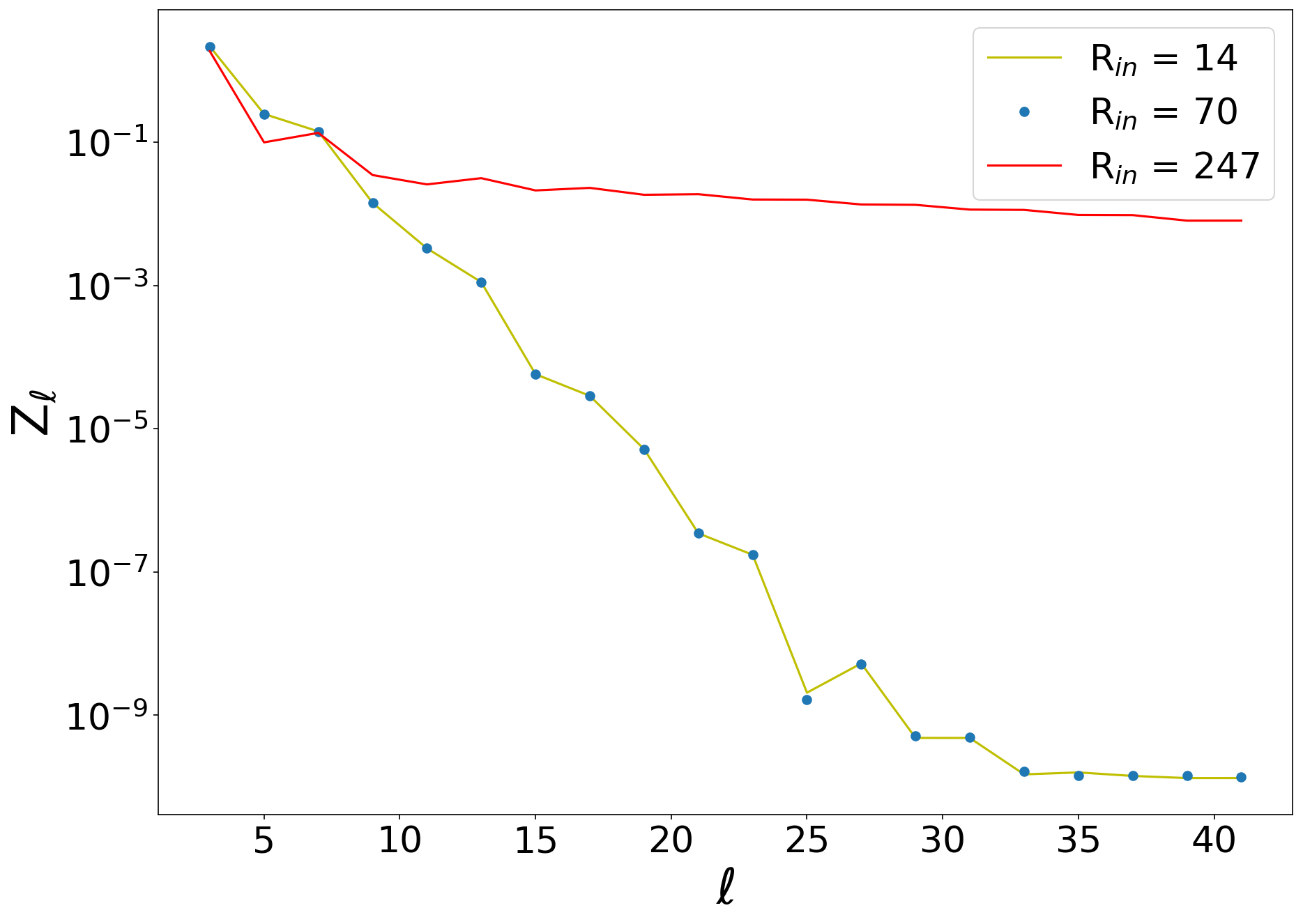}
}
\caption{Plot of $Z_{\ell}$ (Equation \ref{b1}) vs $\ell$ for different values of inner radius. Here the outer boundary is at $250$ cell number. Unless the inner radius is selected closer to the outer boundary, higher multipole moments do not contribute significantly to the series solution for current free boundary.}
\label{fig-a3}
\end{figure}

To test the effect of different values of the parameter $R_{\rm in}$ (radius at which solution is used to calculate and update the outer radial boundary) on $\ell_{\rm max}$, multiple simulations are setup for a $4$ m accretion mound with surface magnetic field strength $10^{9}$ G, truncation angle $45.5^{0}$, number of points $250\times250$ and outer radius above the surface ($R_{\rm out}-R_{\ast}$) $10$ km. Plots of $Z_{\ell}$ for three different values of $R_{\rm in}$ i.e. $14$ ($R_{\rm in}-R_{\ast}$ $=$ $4.3$ m)($14$ means $14$th point), $70$ ($R_{\rm in}-R_{\ast}$ $=$ $56.8$ m), $247$ ($R_{\rm in}-R_{\ast}$ $=$ $8.7$ km) have been represented in Figure \ref{fig-a3}. 
It is apparent that unless $R_{\rm in}$ is selected near the outer radial boundary (like $247$), $Z_{\ell}$ converges at $\ell=33$ for these parameters.

\subsubsection{Effect Of Resolution}
\begin{figure}
{
\includegraphics[width=\columnwidth]{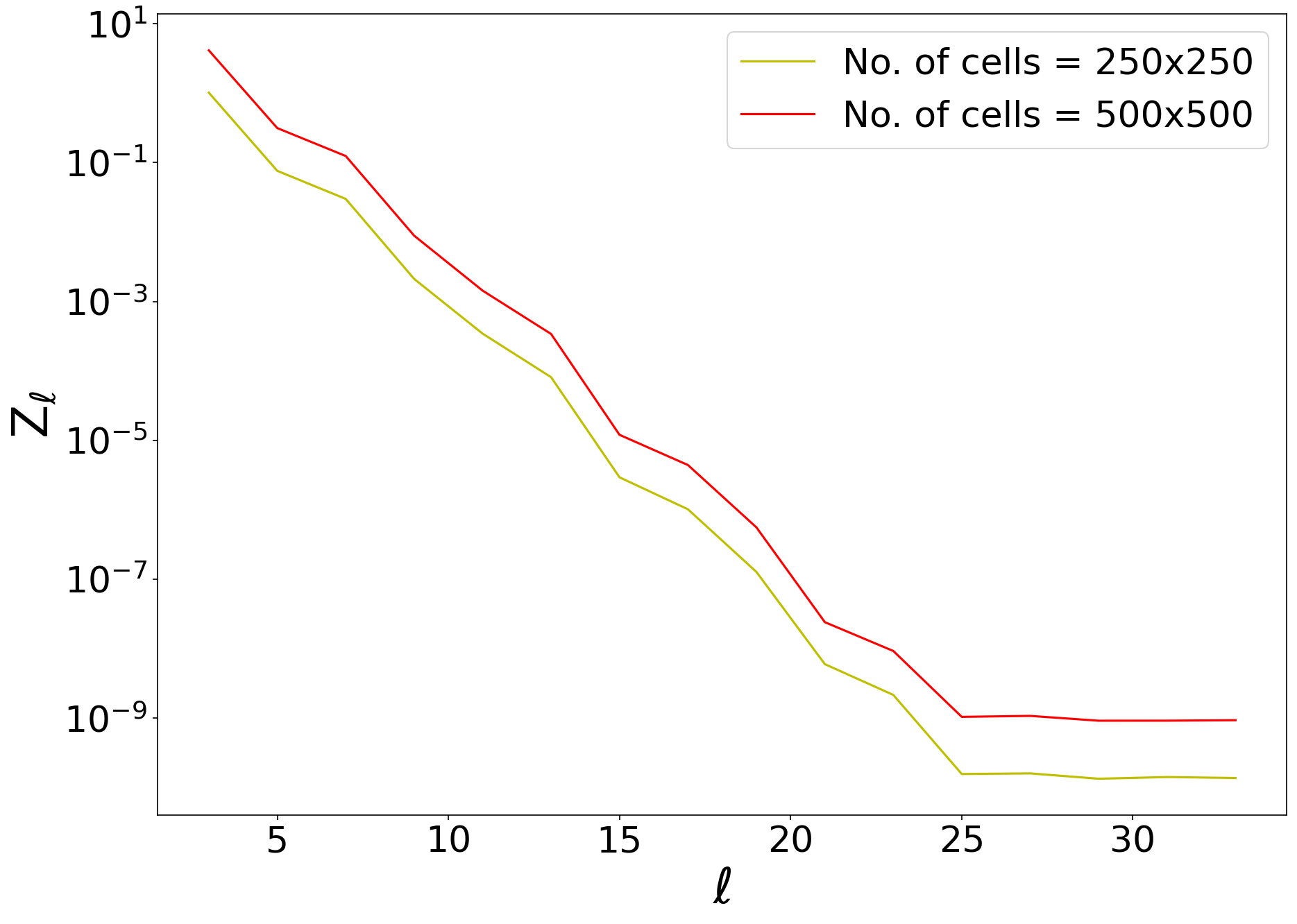}
}
\caption{Plot of $Z_{\ell}$ (Equation \ref{b1}) vs $\ell$ for different values of resolution. Result is not affected by resolution.}
\label{fig-a4}
\end{figure}

To test the effect of resolution on $\ell_{\rm max}$, multiple simulations are setup for a $4$ m accretion mound with surface magnetic field strength $10^{9}$ G, truncation angle $45.5^{0}$, $R_{\rm in}$ close to the top of the mound and outer radius above the surface ($R_{\rm out}-R_{\ast}$) $14$ km. Behaviour of $Z_{\ell}$ for two different resolutions have been plotted in Figure \ref{fig-a4}.
Value of $Z_{\ell}$ is higher for a higher resolution as the addition takes place over larger number of cells. From Figure \ref{fig-a4}, we can conclude that resolution affects the value of $Z_{\ell}$ but not the value of $\ell$ at which $Z_{\ell}$ converges. 

\subsubsection{Effect of changing outer boundary}\label{app-b1.3}
\begin{figure}
{
\includegraphics[width=\columnwidth]{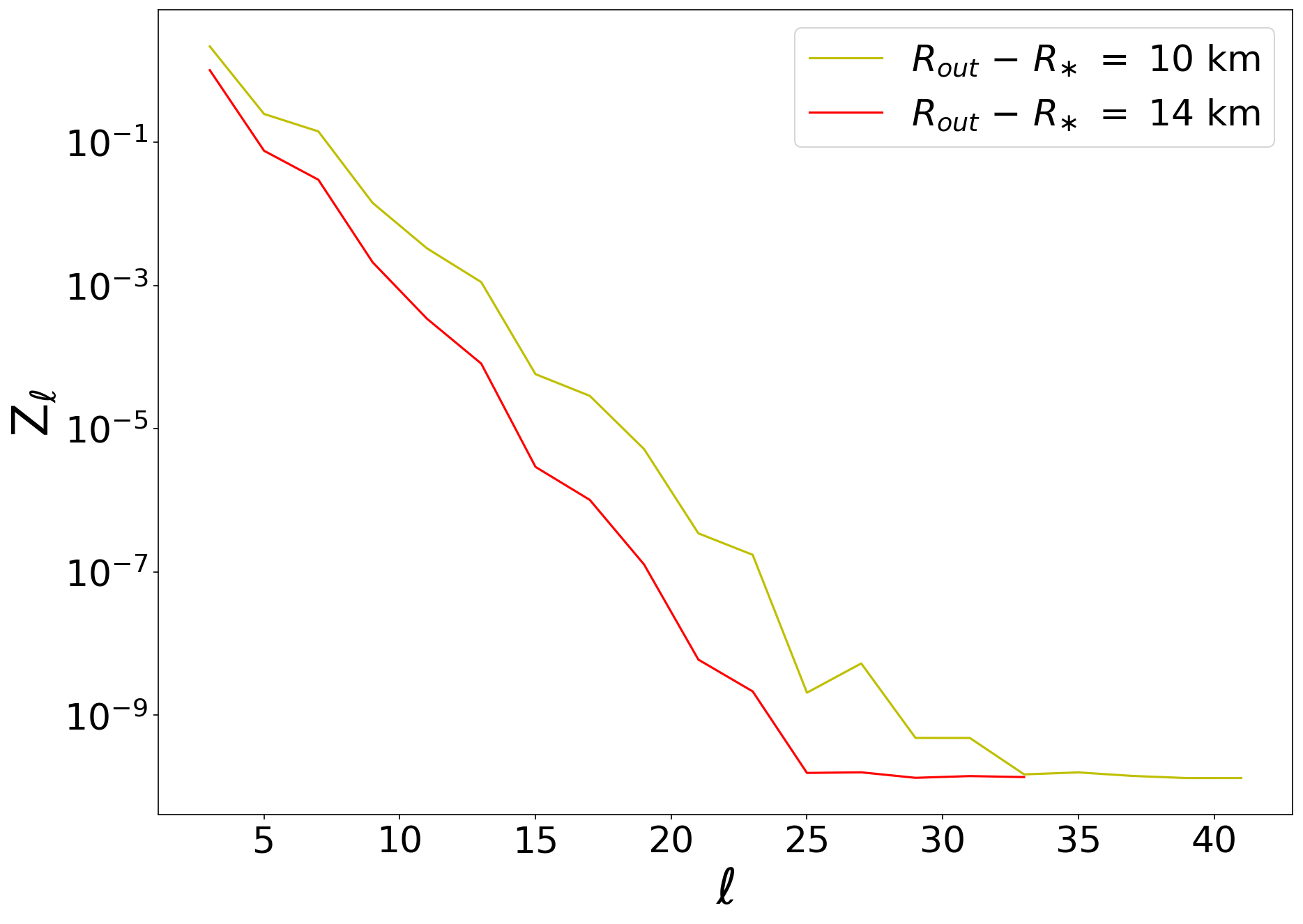}
}
\caption{Plot of $Z_{\ell}$ (Equation \ref{b1}) vs $\ell$ for different values of outer boundary. A larger outer boundary requires lesser multipole terms in the series solutions for the same accuracy.}
\label{fig-a5}
\end{figure}

To test the effect of different values of the outer radius on $\ell_{\rm max}$, multiple simulations are setup for a $4$ m accretion mound with surface magnetic field strength $10^{9}$ G, truncation angle $45.5^{0}$, number of points $250\times250$ and $R_{\rm in}$ close to the top of the mound. Plots of $Z_{\ell}$ for two different values of outer radius above the surface ($R_{\rm out}-R_{\ast}$) i.e $10$ km and $14$ km are represented in Figure \ref{fig-a5}.
We can conclude that a larger value of outer radius makes $Z_{\ell}$ converge at a lower value of $\ell$. A larger mass or a lower dipole moment does not effect the behaviour of $Z_{\ell}$ with $\ell_{\rm max}$. For almost all the simulations performed in this work (unless explicitly mentioned), the outer radius above the surface ($R_{\rm out}-R_{\ast}$) is $12$ km and $R_{\rm in}$ is close to the top of the mound, thus from all the tests performed above, $\ell_{\rm max}$ was chosen to be $33$ (convergence for outer radius above the surface ($R_{\rm out}-R_{\ast}$) $10$ km). 

\subsection{Checking with Greens function volume boundary condition}\label{greensvol}
Instead of using an arbitrary $R_{\rm in}$ to evaluate the outer boundary, we can use the Greens function formalism to evaluate the outer boundary using all the cells with non zero density or non zero source function. \cite{fujisawa2022magneticallymulti} have used such an integral form to solve for the whole simulation domain. $\psi$ at the outer boundary is calculated using $\psi$ values of the whole simulation domain as follows

\begin{align}
    \quad & \quad \psi(R_{\rm out},\theta^{\prime}) \nonumber \\ & = \frac{\psi^{\ast} \sin^{2}\theta^{\prime} R_{\ast}}{R_{\rm out}} \nonumber \\ & + \sum_{\ell=1,3,5...}^{\infty} \frac{\sin \theta^{\prime} P_{\ell}^{1}(\cos \theta^{\prime})}{\ell(\ell+1)} R_{\rm out}^{-\ell} \nonumber \\ 
    & \times \left( \int_{0}^{\frac{\pi}{2}} \int_{R_{\ast}}^{R_{\rm out}} r^{\ell+1} \left(\left(\frac{R_{\ast}}{r}\right)^{2\ell+1} - 1\right) P_{\ell}^{1}(\cos \theta) K(\psi,r,\theta) dr d\theta \right).
\end{align}

\begin{figure}
{
\includegraphics[width=\columnwidth]{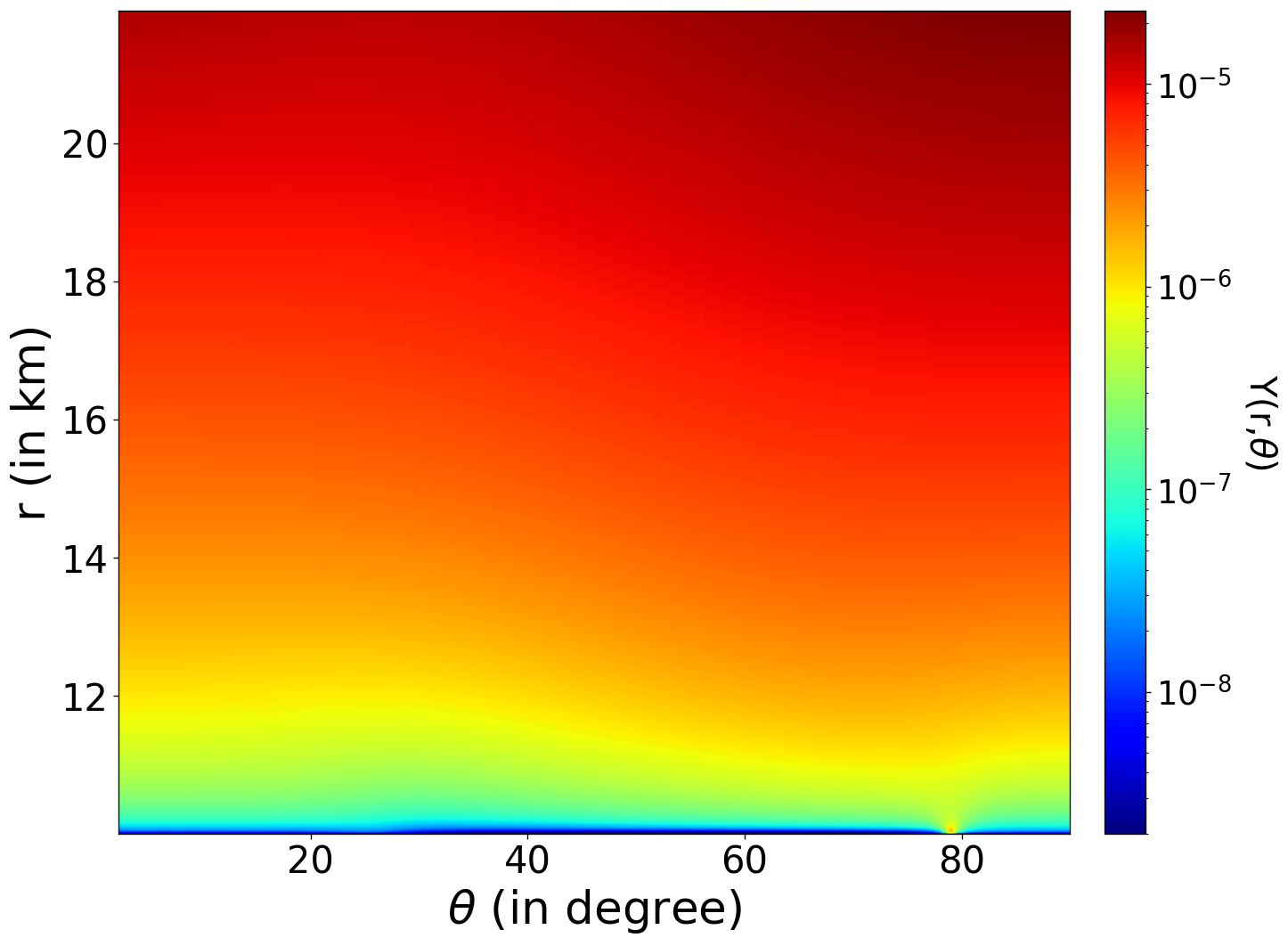}
}
\caption{Relative difference between the solution using our new current free boundary condition and a Greens function based boundary condition. The maximum relative difference is lower than $10^{-4}$.}
\label{fig-a6}
\end{figure}

Comparing the solution by applying the above boundary condition, with a solution with our simpler boundary condition with an arbitrary $R_{\rm in}$ closer to the top of the mound, we find that the maximum absolute relative difference $Y(r,\theta)$ between the solutions is lower than $5 \times 10^{-5}$, for resolution $600\times600$ and normalized outer dipole moment $0.83$. The absolute relative difference $Y(r,\theta)$ between the two solutions is plotted in Figure \ref{fig-a6} ($r_{c}=10^{2}$ m, $B_{d}=10^{12}$ G, $\theta_{t}=84^{0}$). In the case where the whole simulation domain includes significant screening currents or toroidal fields, the volume terms with the source function need to be included in the greens function formalism along with the boundary terms to calculate the solution at the outer radial boundary.

\section{Numerical Maximum Mass (NMM)}\label{numapp}
\subsection{NMM for fixed \texorpdfstring{$\theta_{t}$}{theta\_t} and fixed \texorpdfstring{$B_{d}$}{B\_d}} \label{appnew}

\begin{figure*}
\begin{multicols}{2}
    \includegraphics[width=\columnwidth]{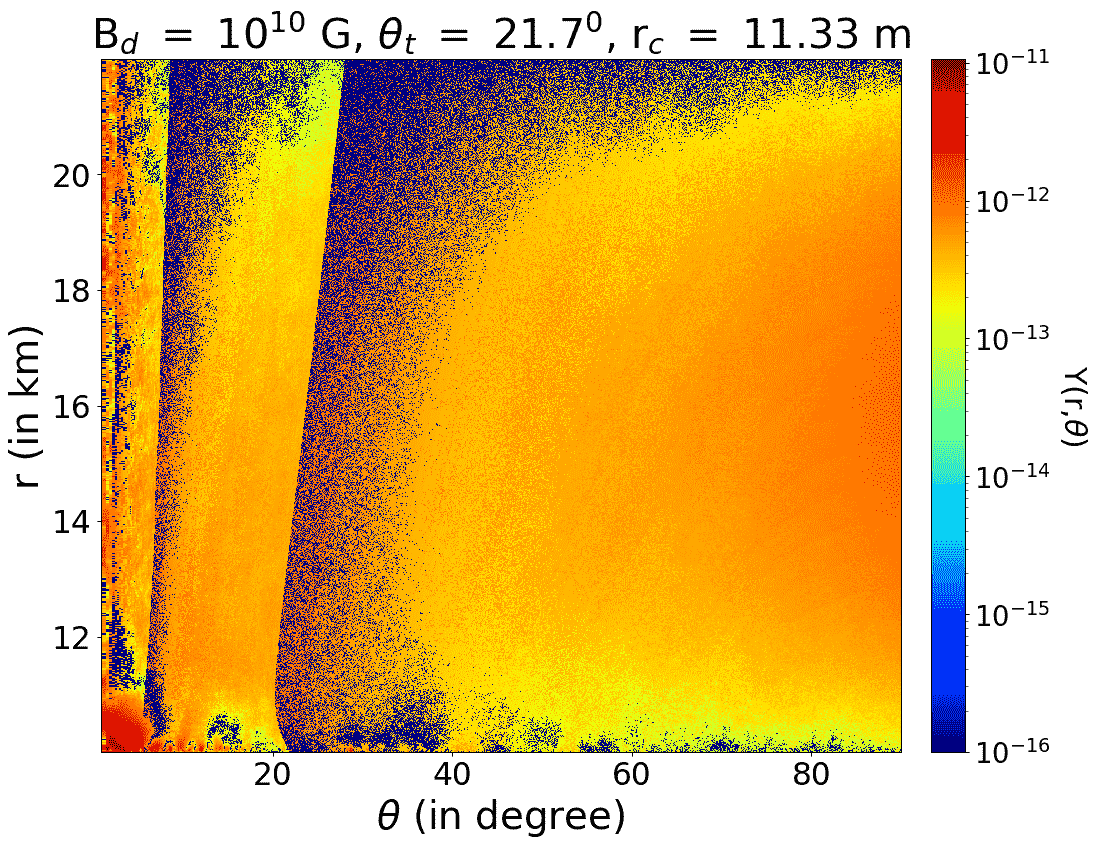}\par 
    \includegraphics[width=\columnwidth]{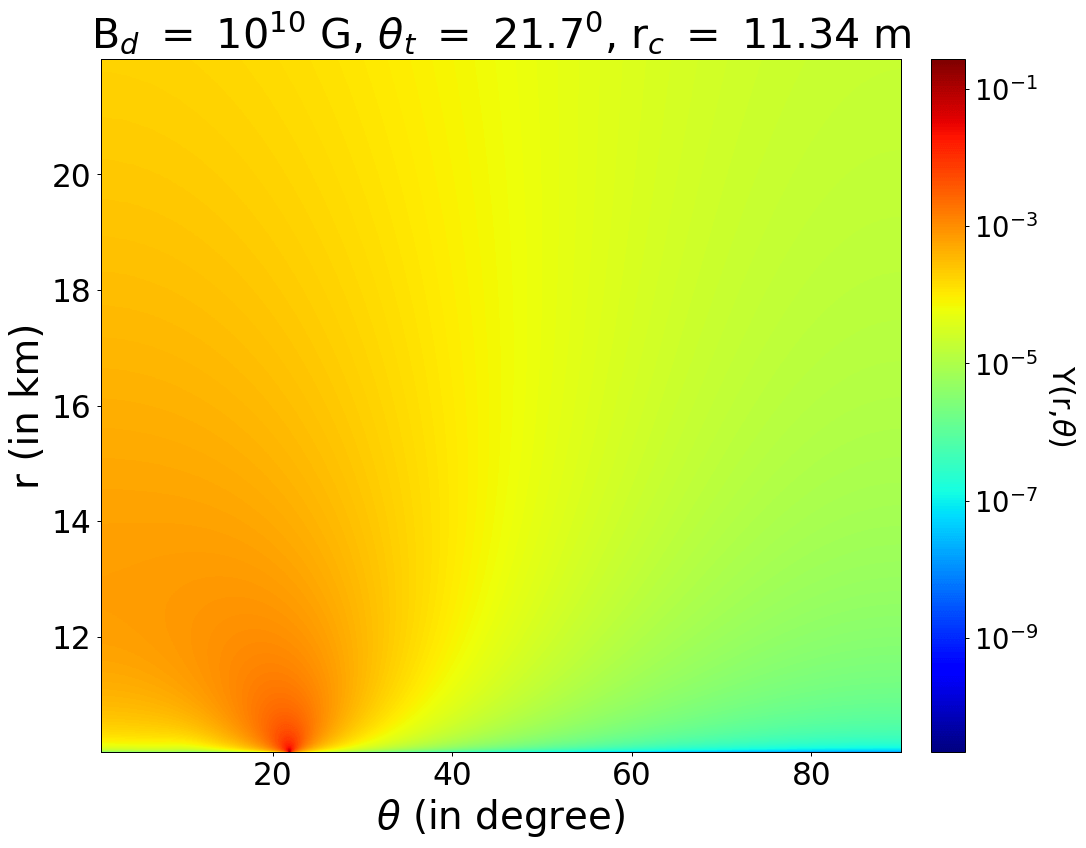}\par 
    \end{multicols}
\begin{multicols}{2}
    \includegraphics[width=\columnwidth]{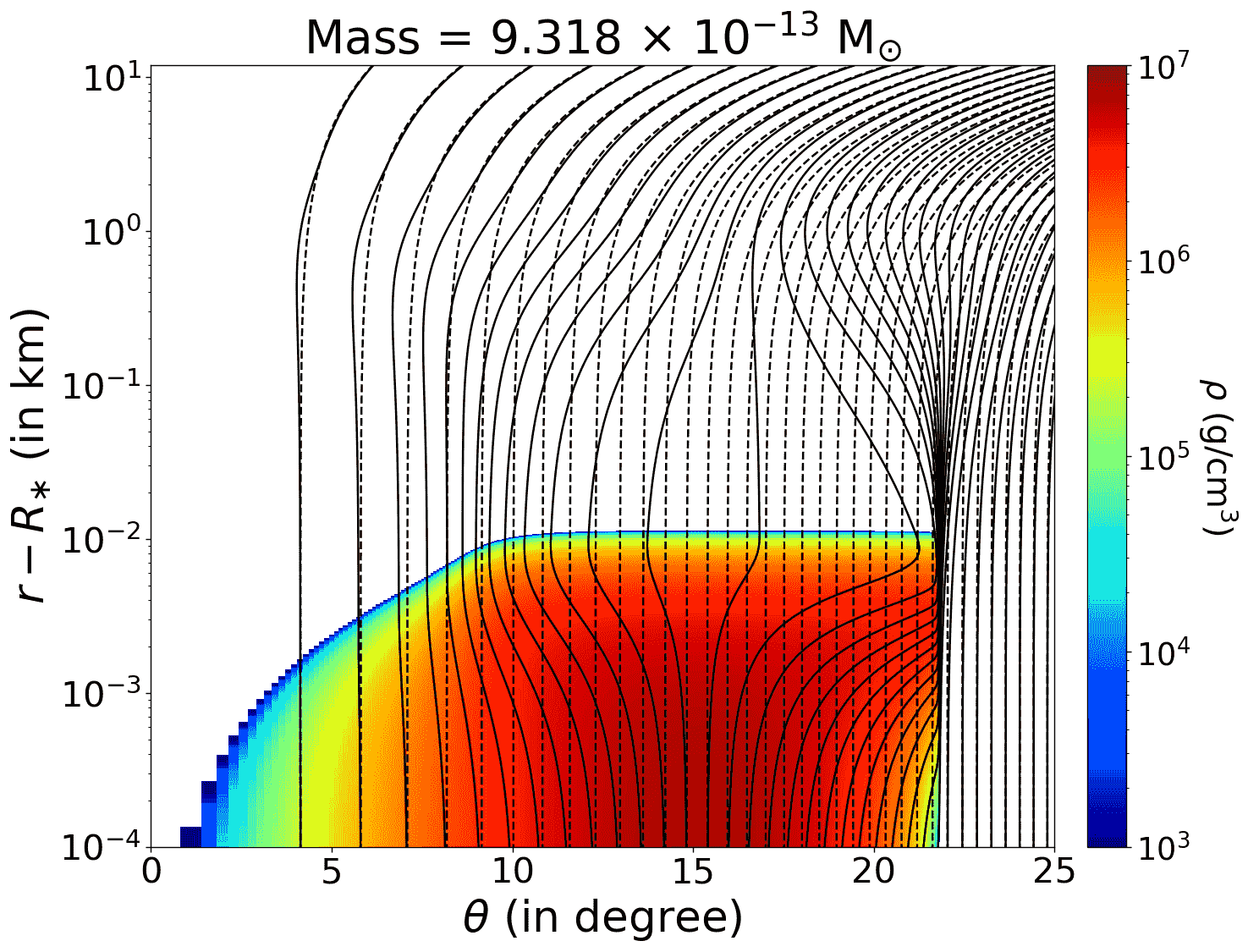}\par
    \includegraphics[width=\columnwidth]{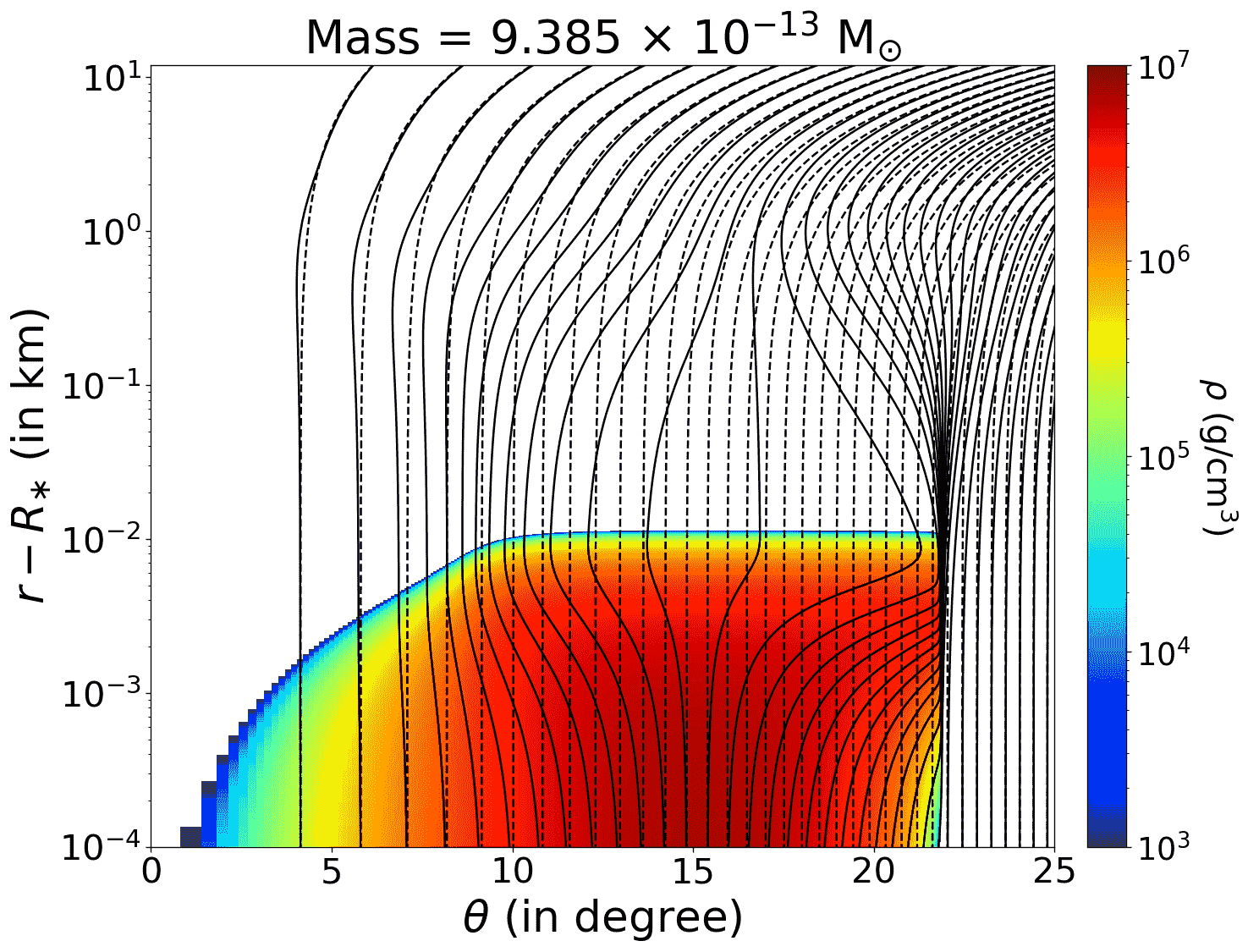}\par
\end{multicols}
\caption{Colormap of $Y(r,\theta)$ between perturbed and unperturbed solution for parameters $B_{d}$ $=$ $10^{10}$ G, $\theta_{t}$ $=$ $21.7^{0}$, $r_{c}$ $=$ $11.33$ m, Mass $=$ $9.318 \times 10^{-13}$ M$_{\odot}$ (top left) and Density profile, magnetic field lines (solid) and dipolar magnetic field lines (dashed) for the same parameters (bottom left). Maximum $Y(r,\theta)$ is lower than $10^{-10}$, thus even with the perturbed guess, the code converges to the same solution. Colormap of $Y(r,\theta)$ between perturbed and unperturbed solution for parameters $B_{d}$ $=$ $10^{10}$ G, $\theta_{t}$ $=$ $21.7^{0}$, $r_{c}$ $=$ $11.34$ m, Mass $=$ $9.385 \times 10^{-13}$ M$_{\odot}$ (top right), Density profile, magnetic field lines (solid) and dipolar magnetic field lines (dashed) for the same parameters (bottom right).Maximum $Y(r,\theta)$ is greater than $0.1$, i.e the code converges to a different solution for a perturbed guess.}
\label{maxmass}
\end{figure*}

\begin{figure}
\includegraphics[width=\columnwidth]{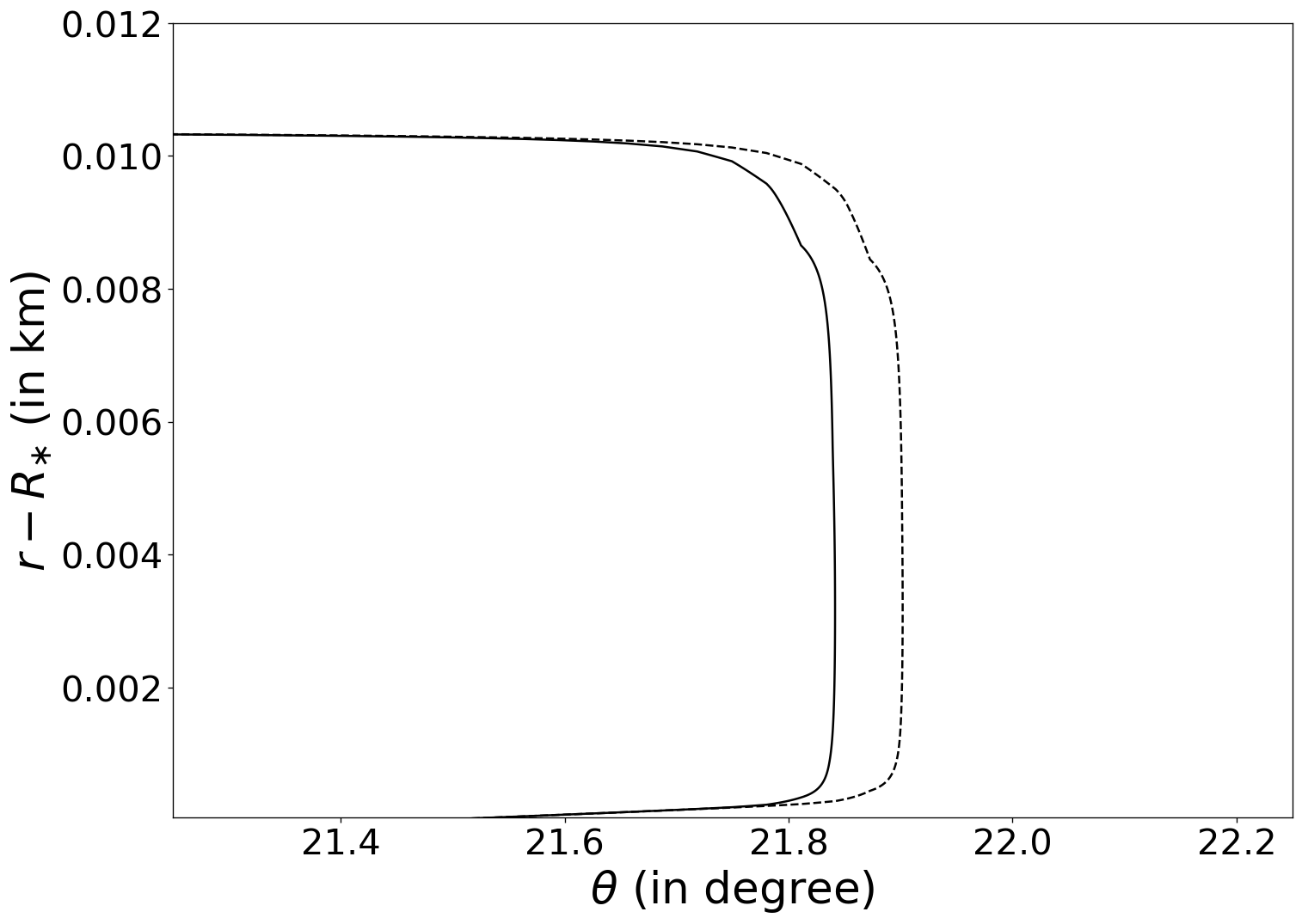}
\caption{Contours for density: $\rho = 10^{5}$ g$/$cm$^{3}$ for GS run with parameters corresponding to the rightmost plot in Figure \ref{maxmass}. The solution $\psi$ with a dipolar guess is presented in a solid line and the solution $\psi_{\mbox{pert}}$ with a perturbed guess in a dashed line. The perturbed input leads to spreading of matter beyond the first solution, resulting in strong deviation between the two solutions.
}
\label{diffuse}
\end{figure}

Earlier studies \cite{hameury1983magnetohydrostatics,payne2004burial,mukherjee2012phase} have shown that numerically converged GS solutions can be obtained up to a maximum confined mass, beyond which closed field lines start to appear, resulting in magnetic loops. However, it remains unclear whether such occurrences are due to the specific choice of profile functions or lack of resolution. In order to test such scenarios we have run a suite of models with fixed $B_{d}$ and $\theta_{t}$, but varying $r_{c}$. Solutions for ring-shaped mound profile keep spreading as shown in subsection \ref{3.4}. It is necessary to check the validity of these solutions. We apply a perturbation based test, as discussed below. 
    \begin{itemize}
        \item Use an initial dipolar guess and obtain a solution $\psi(r,\theta)$ from our prescribed numerical method.
        \item Now, use $\psi(r,\theta)\times(1.0+q\times u(r,\theta))$ ($q$ is assumed to be $10$, while $u(r,\theta)$ is a random number from a uniform distribution between 0 and 1 drawn for each point of the grid) as an initial guess and obtain a solution $\psi_{\mbox{pert}}(r,\theta)$ from the same prescribed numerical method. 
        \item Compute
   \begin{equation}\label{Yeq}
    Y(r,\theta) = \frac{|\psi(r,\theta) - \psi_{\mbox{pert}}(r,\theta)|}{\psi(r,\theta)},
\end{equation}     
        the difference of the perturbed solution from the original to compare $\psi(r,\theta)$ and $\psi_{\mbox{pert}}(r,\theta)$.

        \item If $MAX(Y(r,\theta))$ is larger than $10^{-3}$, then a non unique solution has been obtained by using a different guess but evaluated with the same method. Thus, this solution is deemed invalid. 
    \end{itemize}
In general, we find that with an increase in $r_{c}$ for fixed $\theta_{t}$ and $B_{d}$, after a certain height, $MAX(Y(r,\theta))$ transitions from a low value (e.g. $\sim 10^{-11}$) to a high value (viz. $ > 10^{-3}$). This indicates a strong deviation from the original solution, implying multiple branches or families of potential solutions, resulting in non-uniqueness. We deem such solutions invalid.
    
An example of this behaviour has been plotted in Figure \ref{maxmass}. The left panels show the $Y(r,\theta)$ and the solutions for a mound with $r_{c}=$ $11.33$ m. However, increasing $r_{c}$ by $0.01$ m, the perturbation analysis shows very large deviations from the unperturbed solution. We call the last valid solution mass (i.e. in this case for $r_{c}=$ $11.33$ m) as the numerical maximum mass. All the solutions presented in this paper are numerically valid.
    
Now, the interesting question is "Why does the numerical method converge, but presents a non-unique solution for a different guess ?". This is because there are large gradients in $\psi$ near the truncation angle $\theta_{t}$, which increase with increasing $r_{c}$. Eventually, mounds with a perturbed starting guess solution tend to diffuse beyond the solutions obtained from a dipolar guess value, making the solution invalid, as presented in Figure $\ref{diffuse}$. This is likely mediated by numerical resistivity. For low resolutions such as $600\times600$, numerical instabilities make closed loops of magnetic field lines near the truncation angle for a starting solution with a perturbed $\psi$, as presented in figure \ref{balloon} in Appendix \ref{app3}. Closed loops have also been reported in earlier studies \citep{hameury1983magnetohydrostatics,payne2004burial,mukherjee2012phase}. However, such cases were either due to lack of resolution (Appendix \ref{app3}) or the choice of the profile function (i.e, closed loops in Figure \ref{cosh}). For higher resolutions, the large gradients in $\psi$ near $\theta_{t}$ are well resolved keeping the solution valid, allowing the matter to continue to spread towards the equator with an increase in $r_{c}$ (Subsection \ref{3.4}).
    
Since diffusion of matter or formation of closed loops near $\theta_{t}$ deems the solution invalid, and is dependent on resolution, the value of Numerical Maximum Mass is resolution dependent. Table \ref{apptab1} lists down the numerical maximum mass for different simulation domain resolutions.

Since the latitudinal direction is represented in the $\cos{\theta}$ coordinates, the regions near the equator are better resolved in $\theta$ than the poles. Thus, for the same resolution, a solution with a larger $\theta_{t}$ naturally favors a larger numerical maximum mass for our choice of coordinates, because the resolution in $\theta$ increases as the value of $\theta$ increases from $0^{0}$ to $90^{0}$. Once a solution is deemed numerically valid, dynamical stability of the solution needs to be probed.

\subsection{NMM resolution dependence}\label{app3}

It has been established in the above subsection \ref{appnew} and by Table \ref{apptab1} that NMM of the mounds defined by our criteria depends on resolution.

\begin{table}
\caption{Resolution vs Maximum Mass for parameters $B_{d}=10^{9}$ G and $\theta_{t}=45.5^{0}$}
\label{apptab1}
\begin{center}
\begin{tabular}{|c|c|}
\hline
 Resolution & Maximum Mass (M$_{\odot}$) \\ 
 \hline\hline
 $700 \times 700$ & $3.3363 \times 10^{-13}$ M$_{\odot}$ \\ 
 \hline
 $800 \times 800$ & $3.4282 \times 10^{-13}$ M$_{\odot}$ \\ 
 \hline
 $1000 \times 1000$ & $3.4421 \times 10^{-13}$ M$_{\odot}$ \\ 
 \hline 
 $2000 \times 2000$ & $3.6073 \times 10^{-13}$ M$_{\odot}$ \\ 
 \hline
 $3000 \times 3000$ & $4.1566 \times 10^{-13}$ M$_{\odot}$ \\ 
 \hline
 $5000 \times 5000$ & $5.65 \times 10^{-13}$ M$_{\odot}$ \\ 
 \hline
\end{tabular}
\end{center}
\end{table}

\begin{figure}
\vbox{
\includegraphics[width=0.8\columnwidth]{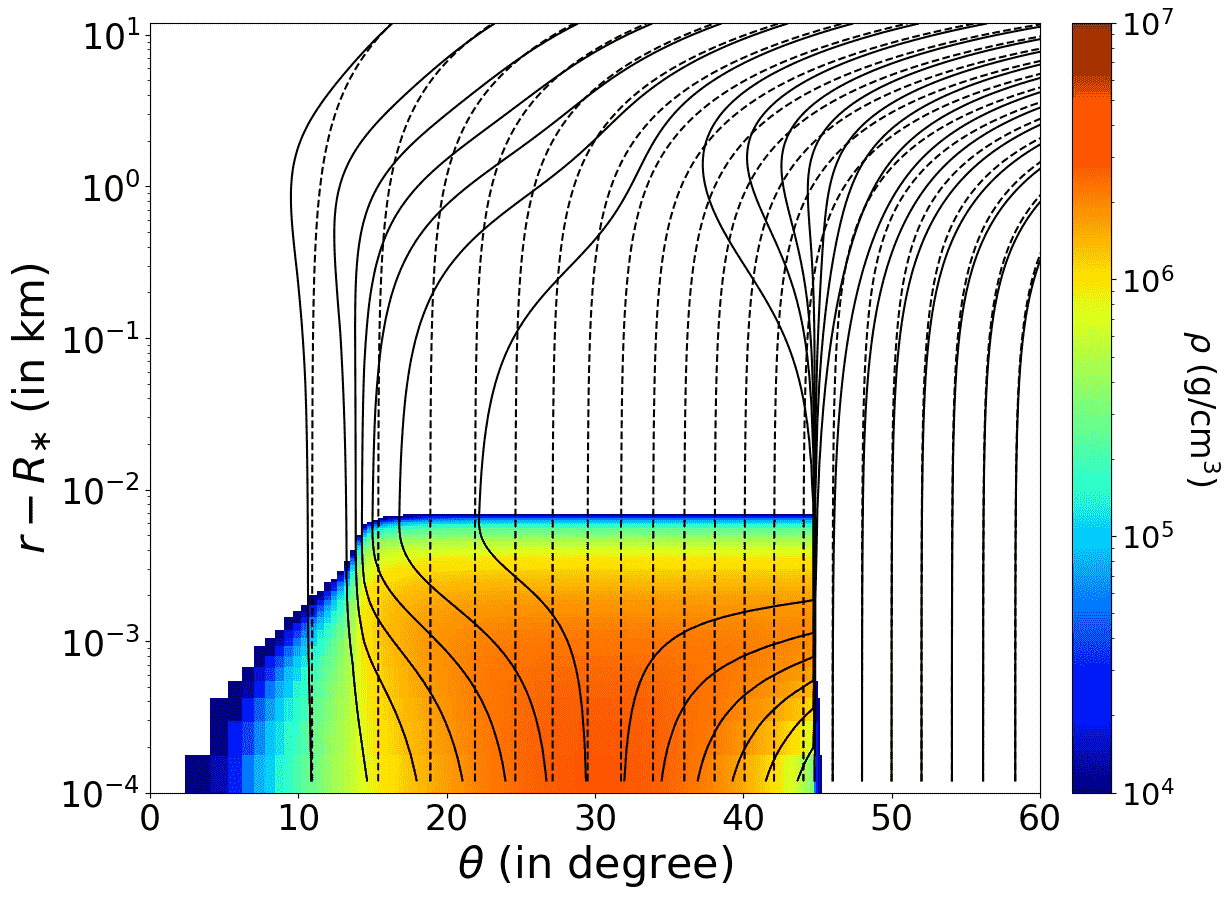}
\includegraphics[width=0.8\columnwidth]{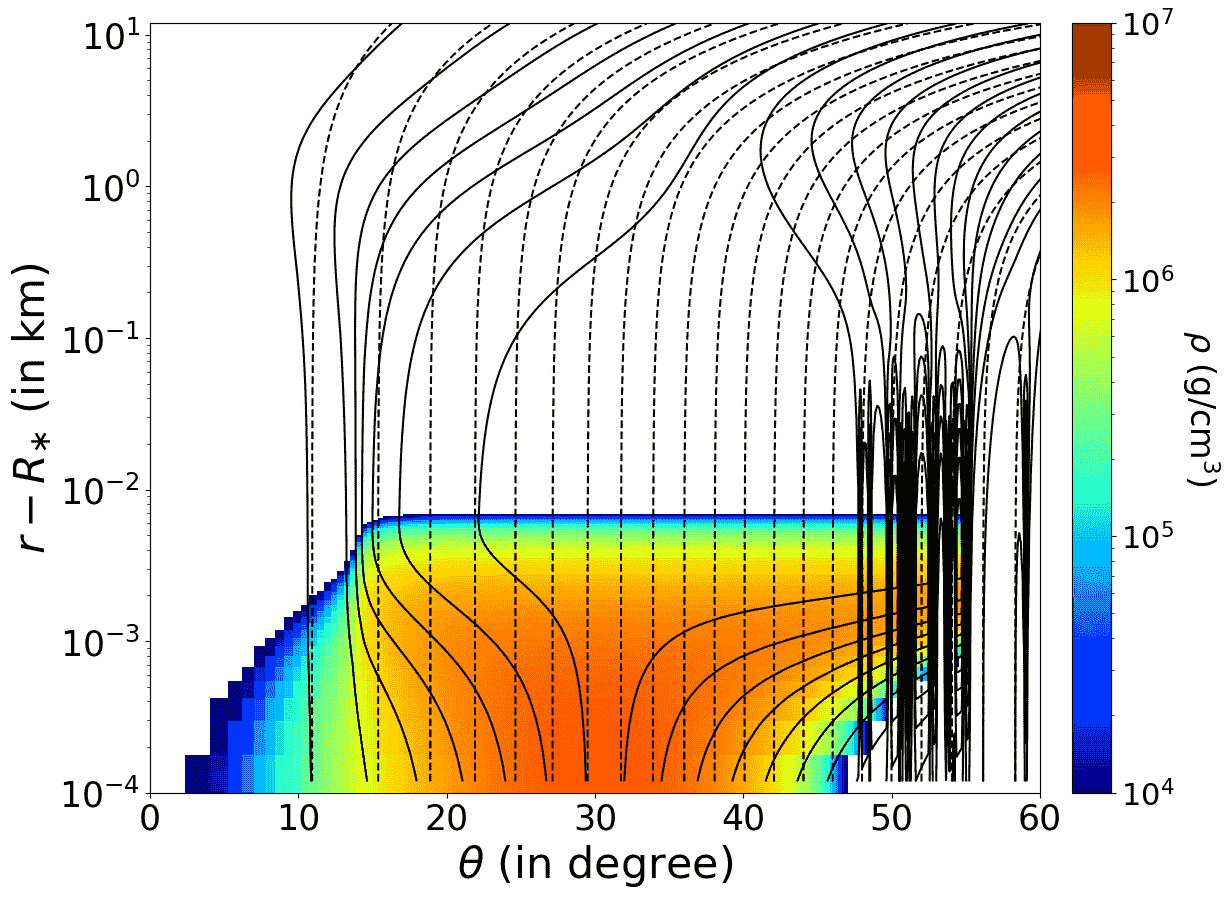}
\includegraphics[width=0.8\columnwidth]{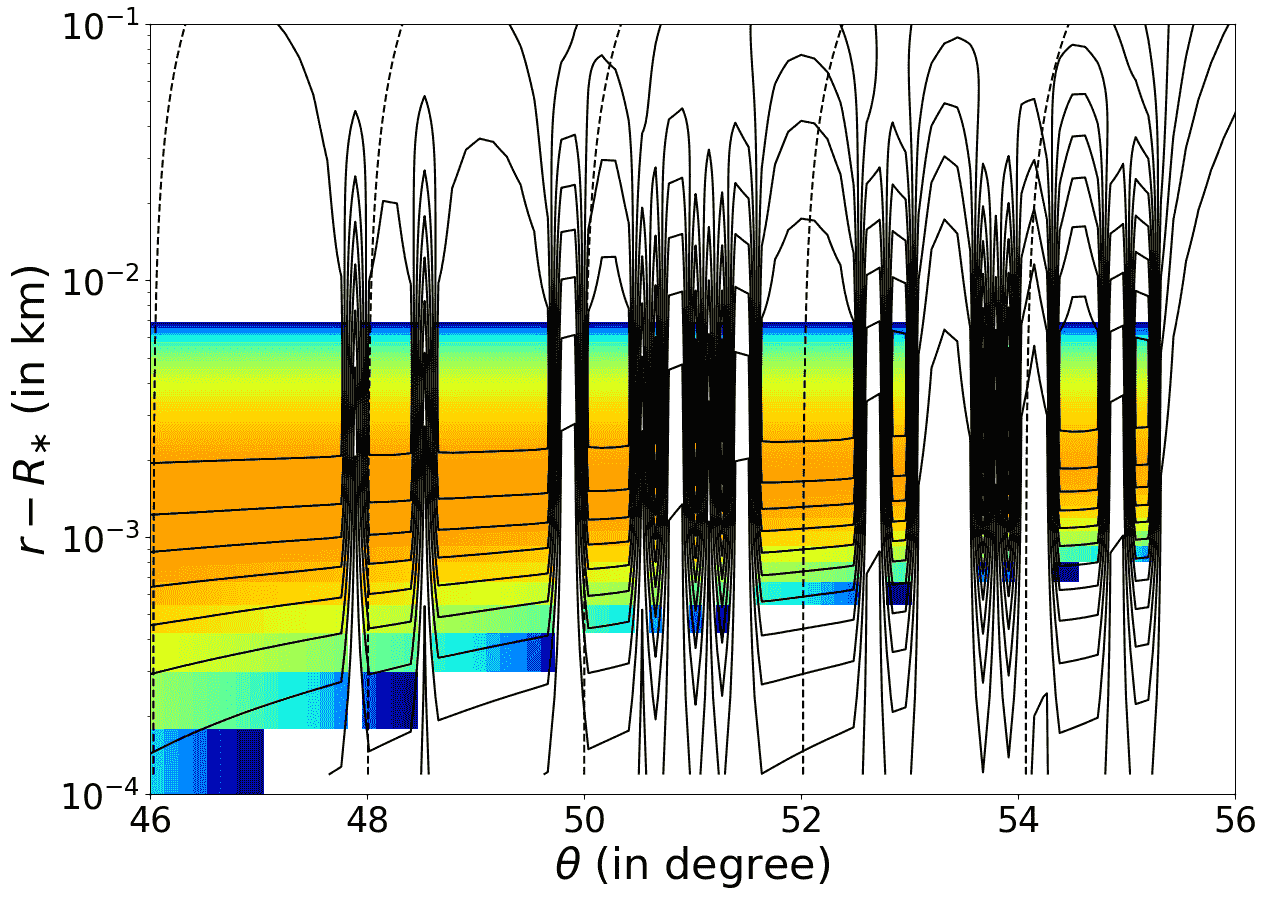}
}
\caption{Plot of Density Profile and Magnetic field lines (solid) for resolution $600\times600$, parameters $B_{d}$ $=$ $10^{9}$ G, $r_{c}=7$ m and $\theta_{t}=45.5^{0}$. Solution from a dipolar guess (top) and the Solution from a perturbed guess (middle) are plotted here. Closed magnetic loops are clearly visible in the bottom plot (close-up of the middle plot).}
\label{balloon}
\end{figure}

For low resolutions, closed loops form after solving with a perturbed guess near the truncation angle due to numerical resistivity. For a $600\times600$ simulation, Figure \ref{balloon} shows the solution derived from a dipolar guess before perturbing and the solution derived from a perturbed guess of the previous solution. Multiple closed magnetic loops at the right of the truncation angle are visible in the bottom plot of Figure \ref{balloon}. 

\bsp	
\label{lastpage}
\end{document}